\documentclass[12pt]{article}
\usepackage{graphicx}
\usepackage{bm}
\usepackage{amsmath}
\usepackage{amssymb}
\usepackage{amsfonts}
\usepackage{mathtools}
\usepackage{hyperref}
\usepackage{braket}
\usepackage{anyfontsize}
\usepackage[normalem]{ulem}
\usepackage{wrapfig}
\usepackage[left=1in,right=1in,top=1in,bottom=1in]{geometry}

\hypersetup{colorlinks=true,urlcolor=[rgb]{0,0,0.5},citecolor=[rgb]{0.5,0,0},linkcolor=[rgb]{0,0,0.4}}

\renewcommand{\title}[1]{\vbox{\center\bf{\Large{#1}}}\vspace{5mm}}
\renewcommand{\author}[1]{\vbox{\center#1}\vspace{5mm}}
\newcommand{\address}[1]{\vbox{\center\em#1}}

\newcommand\emails[1]{\begingroup
	\renewcommand\thefootnote{}\footnote{#1}
	\addtocounter{footnote}{-1}\endgroup}

\def\Tr{{\rm Tr}}

\def\op{{\cal O}}
\def\C{\mathbb{C}}
\def\R{\mathbb{R}}

\def\vev#1{\langle{#1}\rangle}

\def\1den{\hbox{$1\hskip -1.2pt\vrule depth 0pt height 1.53ex width 0.7pt
                  \vrule depth 0pt height 0.3pt width 0.12em$}}

\def\where{\quad {\rm where} \quad}
\def\for{\quad {\rm for} \quad}
\def\and{\quad {\rm and} \quad}

\def\ni{\noindent}
\def\nn{\nonumber\\}

\def\ra{\rightarrow}
\def\ie{{\rm i.e.\ }}

\def\CA{{\cal A}}

\def\CE{{\cal E}}
\def\CF{{\cal F}}
\def\CH{{\cal H}}

\def\CR{{\cal R}}
\def\CW{{\cal W}}

\def\Wg{{\cal W}\! g}

\begin{document}

\begin{titlepage}
\hfill {\footnotesize CALT-TH-2017-047}\\
\null\hfill {\footnotesize SU-ITP-17/10}

\begin{center}
\vspace*{1.5cm}
\title{Chaos, Complexity, and Random Matrices}
\author{Jordan Cotler,${}^a$ Nicholas Hunter-Jones,${}^b$ Junyu Liu,${}^b$ and Beni Yoshida${}^c$}
\address{
${}^a$ Stanford Institute for Theoretical Physics,\\ Stanford University, Stanford, California 94305, USA\\
\vspace*{2mm}
${}^b$ Institute for Quantum Information and Matter \&\\ Walter Burke Institute for Theoretical Physics,\\ California Institute of Technology,\\ Pasadena, California 91125, USA\\
\vspace*{2mm}
${}^c$ Perimeter Institute for Theoretical Physics,\\ Waterloo, Ontario N2L 2Y5, Canada
}
\emails{ \hspace*{-8mm}
\href{mailto:jcotler@stanford.edu}{\tt jcotler@stanford.edu},
\href{mailto: nickrhj@caltech.edu}{\tt nickrhj@caltech.edu},
\href{mailto:jliu2@caltech.edu}{\tt jliu2@caltech.edu},
\href{mailto:byoshida@perimeterinstitute.ca}{\tt byoshida@perimeterinstitute.ca}
}

\end{center}

\begin{abstract}
Chaos and complexity entail an entropic and computational obstruction to describing a system, and thus are intrinsically difficult to characterize.  In this paper, we consider time evolution by Gaussian Unitary Ensemble (GUE) Hamiltonians and analytically compute out-of-time-ordered correlation functions (OTOCs) and frame potentials to quantify scrambling, Haar-randomness, and circuit complexity.  While our random matrix analysis gives a qualitatively correct prediction of the late-time behavior of chaotic systems, we find unphysical behavior at early times including an $\mathcal{O}(1)$ scrambling time and the apparent breakdown of spatial and temporal locality.  The salient feature of GUE Hamiltonians which gives us computational traction is the Haar-invariance of the ensemble, meaning that the ensemble-averaged dynamics look the same in any basis.  Motivated by this property of the GUE, we introduce $k$-invariance as a precise definition of what it means for the dynamics of a quantum system to be described by random matrix theory. We envision that the dynamical onset of approximate $k$-invariance will be a useful tool for capturing the transition from early-time chaos, as seen by OTOCs, to late-time chaos, as seen by random matrix theory.
\end{abstract}

\end{titlepage}

\tableofcontents
\newpage

\section{Introduction}

Quantum chaos is a general feature of strongly-interacting systems and has recently provided new insight into both strongly-coupled many-body systems and the quantum nature of black holes. Even though a precise definition of quantum chaos is not at hand, understanding how chaotic dynamics process quantum information has proven valuable.  For instance, Hayden and Preskill \cite{HaydenPreskill} considered a simple model of random unitary evolution to show that black holes rapidly process and scramble information. The suggestion that black holes are the fastest scramblers in nature \cite{SekinoSusskind,FastScrambling} has led to a new probe of chaos in quantum systems, namely the $4$-point out-of-time-order correlation function (OTOC). Starting with the work of Shenker and Stanford \cite{SSbutterfly,SSstringy}, it was shown \cite{MSSbound} that black holes are maximally chaotic in the sense that a bound on the early time behavior of the OTOC is saturated.  Seperately, Kitaev proposed a soluble model of strongly-interacting Majorana fermions \cite{Kitaev14,SachdevYe}, which reproduces many features of gravity and black holes, including the saturation of the chaos bound \cite{Kitaev15,MS_SYK}. The Sachdev-Ye-Kitaev model (SYK) has since been used as a testing ground for questions about black hole information loss and scrambling.

In recent work, \cite{BHRMT16} found evidence that the late time behavior of the SYK model can be described by random matrix theory, emphasizing a dynamical perspective on more standard notions of quantum chaos. Random matrix theory (RMT) has its roots in nuclear physics \cite{Wigner55, Dyson62} as a statistical approach to understand the spectra of heavy atomic nuclei, famously reproducing the distribution of nearest neighbor eigenvalue spacings of nuclear resonances. Random matrix theory's early success was later followed by its adoption in a number of subfields, including large $N$ quantum field theory, string theory, transport in disordered quantum systems, and quantum chaos.  Indeed, random matrix eigenvalue statistics have been proposed as a defining characteristic of quantum chaos, and it is thought that a generic classically chaotic system, when quantized, has the spectral statistics of a random matrix ensemble consistent with its symmetries \cite{BGSchaos}.

Current thinking holds that both spectral statistics and the behavior of the OTOC serve as central diagnostics of chaos, although the precise relation between the two is unclear. OTOCs have recently been studied using techniques from quantum
information theory, and it was found that their decay as a function of time quantifies scrambling \cite{ChaosChannels} and randomness \cite{ChaosDesign}. The goal of this paper is to connect various concepts as a step towards a quantum information-theoretic definition of quantum chaos that incorporates scrambling, chaotic correlation functions, complexity, approximate randomness, and random matrix universality.

As alluded to above, an important first step to bridge early-time chaos and late-time dynamics is to understand the relation between the OTOC and the spectral statistics. We derive an explicit analytical formula relating certain averages of OTOCs and spectral form factors which holds for arbitrary quantum mechanical systems. A simple corollary is that spectral form factors can be approximated by OTOCs defined with respect to random (typically non-local) operators, highlighting the fact that spectral statistics are good probes of macroscopic thermodynamic properties, but may miss important microscopic physics such as early-time chaos. We also compute correlation functions for an ensemble of Hamiltonians given by the Gaussian Unitary Ensemble (GUE), and find that $4$-point OTOCs decay faster than $2$-point correlators contrary to findings for local quantum Hamiltonians \cite{MSSbound}. Due to the basis independence of the GUE, averaged correlation functions do not depend on sizes of operators, and thus can be expressed solely in terms of spectral form factors. Furthermore, we find that correlators for GUE Hamiltonians do not even depend on the time-ordering of operators. These results imply that the GUE ignores not only spatial but also temporal locality.

Another important question is to understand the approach to entropic (as well as quantum complexity) equilibrium via pseudorandomization at late times in strongly coupled systems.  We consider the ensemble of unitaries generated by fixed GUE Hamiltonians, namely
\begin{equation}
\CE_t^{\rm GUE} = \big\lbrace e^{-iHt}, ~{\rm for}~ H\in {\rm GUE} \big\rbrace\,,
\end{equation}
and study its approach to Haar-randomness by computing frame potentials which quantify the ensemble's ability to reproduce Haar moments. We find that the ensemble forms an approximate $k$-design at an intermediate time scale, but then deviates from a $k$-design at late times. These results highlight that the $k$-design property fails to capture late time behavior of correlation functions. An interesting application of unitary $k$-designs is that Haar-randomness is a probe of quantum complexity. We apply techniques from~\cite{ChaosDesign} to lower bound the quantum circuit complexity of time evolution by GUE Hamiltonians and find a quadratic growth in time.

In order to make precise claims about the behavior of OTOCs and frame potentials for GUE Hamiltonians, we need explicit expressions for certain spectral quantities.  Accordingly, we compute the $2$-point and $4$-point spectral form factors for the GUE at infinite temperature, as well as the $2$-point form factor at finite temperature. We then use these expressions to discuss time scales for the frame potentials. We also analytically compute the late-time value of the $k$-th frame potential for arbitrary $k$.

Under time evolution by strongly-coupled systems, correlations are spread throughout the system and the locality of operators as well as time-ordering appear to be lost from the viewpoint of correlation functions, as implied by the late-time universality of random matrix theory. Also motivated by the $k$-design property's failure to capture late-time chaos (i.e., $\mathcal{E}_t^{\text{GUE}}$ fails to be Haar-random at late times), we propose a new property called $k$-invariance, which may provide a better probe of chaos at both early and late times. The property of $k$-invariance characterizes the degree to which an ensemble is Haar-invariant, meaning that the ensemble is invariant under a change of basis. When the dynamics becomes approximately Haar-invariant, correlation functions can be captured solely in terms of spectral functions, which signifies the onset of an effective random matrix theory description.  We thus provide an information theoretically precise definition of what it means for a system's dynamics to 	be described by random matrix theory. Specifically, we say that an ensemble of Hamiltonian time evolutions $\mathcal{E}_{t}$ is described by random matrix theory at times greater than or equal to $t$ with respect to $2k$-point OTOCs when $\mathcal{E}_{t}$ is approximately $k$-invariant with respect to its symmetry class, for example the symmetry class of either the unitary, orthogonal, or symplectic groups.

The paper is organized as follows: In Section \ref{sec:FFRMT}, we provide a brief overview of random matrix theory and explicitly compute the spectral form factors for the GUE at infinite and finite temperature. In Section \ref{sec:OTOC}, we compute correlation functions for the GUE, including the OTOC, and demonstrate that they can be expressed in terms of spectral correlators as well. In Section \ref{sec:FPRMT}, we compute frame potentials for the GUE, and extract the timescales when it becomes an approximate $k$-design both at finite and infinite temperatures. We show that the frame potentials can be also expressed as products of sums of spectral correlators. In Section \ref{sec:comp}, we discuss complexity bounds and complexity growth for the GUE. In Section \ref{sec:haarinv}, we discuss Haar-invariance as a diagnostic of delocalization of spatial degrees of freedom and random matrix universality at late times. We conclude with a discussion in Section \ref{sec:discussion}. The appendices contain an review of various information-theoretic definitions of scrambling in the literature, a discussion of information scrambling in black holes, more details of our random matrix calculations, and numerics.

\section{Form factors and random matrices}
\label{sec:FFRMT}

\begin{figure}
\centering
\includegraphics[width=0.60\linewidth]{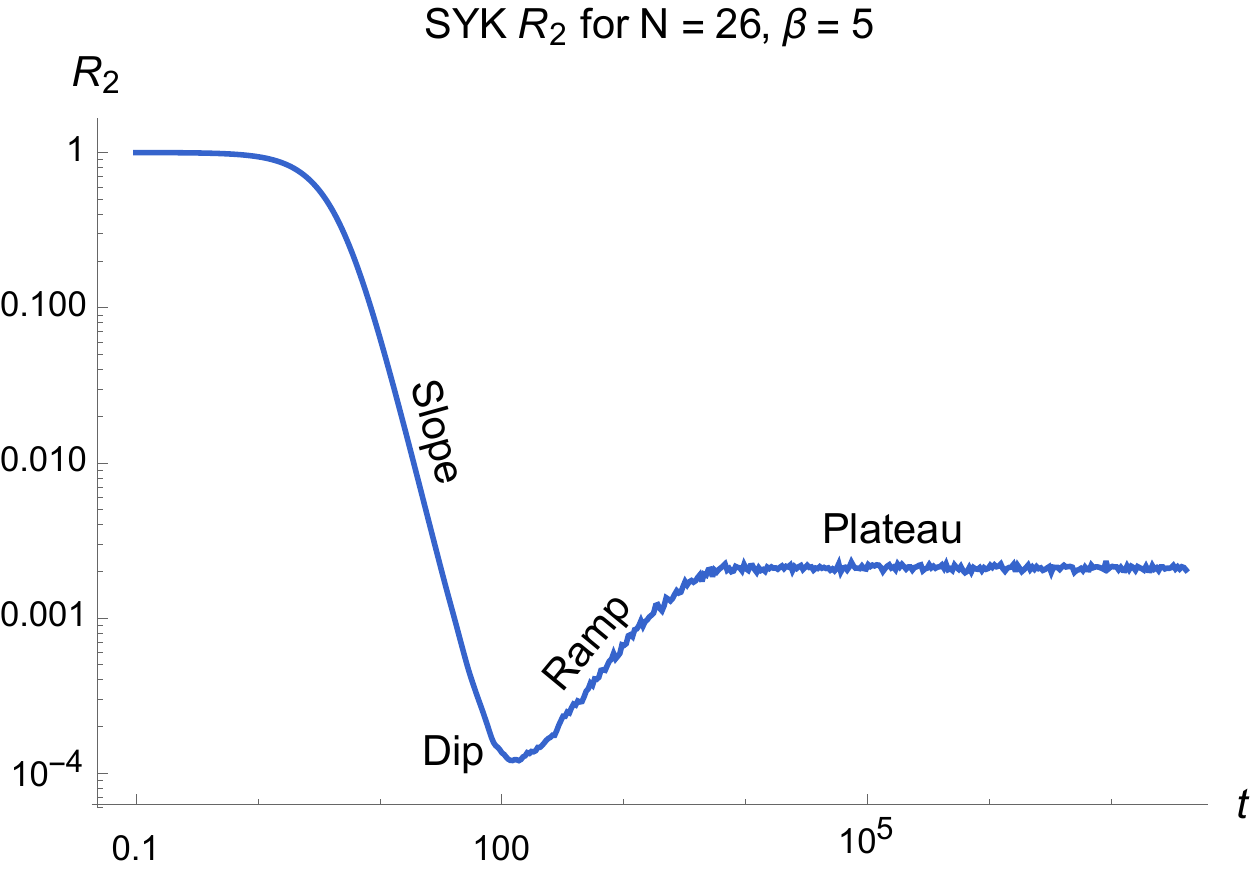}
\caption{The 2-point spectral form factor for SYK with $N=26$ Majoranas at inverse temperature $\beta=5$, computed for 1000 random samples. The slope, dip, ramp, and plateau are labeled.}
\label{fig:SYK_FF}
\end{figure}

For a long time, the spectral statistics of a random matrix were seen as a defining feature of quantum chaos.  More recently, it has been proposed that the late time behavior of certain strongly coupled theories with large numbers of degrees of freedom also exhibit a dynamical form of random matrix universality at late times \cite{BHRMT16}. The central object of study in this recent work is the $2$-point spectral form factor,\footnote{One motivation for studying this object is a simple version of the information loss problem in AdS/CFT~\cite{MaldacenaEternal}, where the apparent exponential decay of $2$-point correlation functions in bulk effective field theory contradicts the finite late-time value of $e^{-\mathcal{O}(S)}$ implied by the discreteness of the spectrum. As we shall see in the next section, the $2$-point form factor is equivalent to the average of $2$-point correlation functions. More recently, chaos and information loss in correlation functions and form factors have also been studied in holographic CFTs \cite{ChaosCFT,CFTinfoloss1,CFTinfoloss2,CFTlatetimes,D1D5chaos}.} which is defined in terms of the analytically continued partition function
\begin{equation}
\CR_{2}(\beta,t) \equiv \big\langle |Z(\beta, t)|^2\big\rangle, \where Z(\beta,t) \equiv \Tr \big( e^{-\beta H - iHt}\big)
\end{equation}
and where $\langle \, \cdot \, \rangle$ denotes the average over an ensemble of Hamiltonians.  In SYK as well as standard RMT ensembles, the $2$-point spectral form factor decays from its initial value and then climbs linearly back up to a floor value at late times. The early time decay of the form factor is called the slope, the small value at intermediate times is called the dip, the steady linear rise is called the ramp, and the late time floor is called the plateau. In Fig.~\ref{fig:SYK_FF} we observe these features in SYK with $N=26$ Majoranas, which has GUE statistics at late times.\footnote{For SYK with $N$ Majoranas, particle-hole symmetry dictates the symmetry class of the spectrum, where $N$ (mod $ 8) \equiv 2$ or $6$ corresponds to GUE statistics \cite{You16}. Furthermore, the spectral density of SYK and its relation to random matrices has also been discussed in \cite{Garcia16}.} Furthermore, it was found that in SYK, time scales and many features of the slope, dip, ramp and plateau agree with predictions from RMT.

In this section, we briefly review random matrix theory.  Further, we study the $2$-point spectral form factor for the GUE at both infinite and finite temperature, compute its analytic form, and extract its dip and plateau times and values.\footnote{We consider the GUE since it corresponds to the least restrictive symmetry class of Hamiltonians. The generalization of our analysis to the GOE or GSE is left for future work.} In addition, we compute the $4$-point form factor and extract relevant time scales and values. We find that the late-time rise in the $4$-point form factor is quadratic in $t$, in contrast to the linear rise in the $2$-point form factor. The expressions derived in this section will give us analytic control over the correlation functions and frame potentials discussed in later sections. For a detailed treatment of the random matrix ensembles, we refer the reader to \cite{MehtaRMT, TaoRMT, RMTphys}.

\subsection{Random matrix theory}

The Gaussian Unitary Ensemble GUE$(L,\mu,\sigma)$ is an ensemble of $L \times L$ random Hermitian matrices, where the off-diagonal components are independent complex Gaussian random variables  $N(\mu, \sigma)_\C$ with mean $\mu$ and variance $\sigma^2$, and the diagonal components are independent real Gaussian random variables $N(\mu,\sigma)_\R$. It is common in the math literature to work with GUE$(L,0,1)$ which has zero mean and unit variance, but we will instead use the normalization GUE$(L,0,1/\sqrt{L})$ so that the eigenvalues do not scale with the system size.\footnote{The reason for using the normalization GUE$(L,0,1/\sqrt{L})$ instead of GUE$(L,0,1)$ is as follows:  With the standard normalization GUE$(L,0,1)$, the energy spectrum ranges from $-2\sqrt{L}$ to $2 \sqrt{L}$. This implies that by applying a local operator, one may change the energy of the system by $\mathcal{O}(\sqrt{L})$. With the physical normalization GUE$(L,0,1/\sqrt{L})$, the energies lie within the range $-2$ to $2 $, and local operators act with $\mathcal{O}(1)$ energy. See~\cite{SecondLawComplexity} for discussions on normalizing $q$-local Hamiltonians.} The probability density function of the ensemble has a Gaussian form
\begin{equation}
P(H) \propto \, e^{-\frac{L}{2} \Tr H^2}\,,
\end{equation}
up to a normalizing factor. As the GUE is invariant under unitary conjugation $H \ra UHU^\dagger$, the integration measure $dH = d(UH U^\dagger)$ is likewise invariant. The probability measure $P(H)\,dH$ on the ensemble integrates to unity.

Instead of integrating over $dH$ directly, it is convenient to change variables to eigenvalues and diagonalizing unitaries.  Up to a normalizing constant $C$ defined in Eq.~\eqref{jointprobwithfactors} in App.~\ref{app:FFFP}, the measure becomes
\begin{equation}
dH = C\, |\Delta(\lambda)|^2 \prod_i d\lambda_i dU\,,
\end{equation}
where $dU$ is the Haar measure on the unitary group $U(L)$ and $\Delta(\lambda)$ is the Vandermonde determinant
\begin{equation}
\Delta(\lambda) = \prod_{i>j} (\lambda_i - \lambda_j)\,.
\end{equation}
The joint probability distribution of eigenvalues is
\begin{equation}
P(\lambda_1,\ldots,\lambda_L) = C e^{-\frac{L}{2} \sum_i \lambda_i^2} |\Delta(\lambda)|^2\,,
\end{equation}
and is symmetric under permutations of its variables.  For simplicity, we define a measure $D\lambda$ which absorbs the Gaussian weights, eigenvalue determinant, and constant factors. We integrate over the GUE in the eigenvalue basis as
\begin{equation}
\vev{O(\lambda)}_{\rm GUE} \equiv \int D \lambda \,O(\lambda) \where \int D\lambda = C \int \prod_i d\lambda_i  |\Delta(\lambda)|^2 e^{-\frac{L}{2} \sum_i \lambda_i^2} = 1\,.
\label{eq:GUEmeas}
\end{equation}
The probability density of eigenvalues $\rho(\lambda)$, where
\begin{equation}
 \int d\lambda\, \rho(\lambda) = 1\,,
\end{equation}
can be written in terms of the joint eigenvalue probability density by integrating over all but one argument
\begin{equation}
\rho(\lambda) = \int d\lambda_1\ldots d\lambda_{L-1} P(\lambda_1, \ldots, \lambda_{L-1},\lambda)\,.
\end{equation}
The spectral $n$-point correlation function, \ie the joint probability distribution of $n$ eigenvalues, $\rho^{(n)}$ is defined as
\begin{equation}
\rho^{(n)}(\lambda_1,\ldots,\lambda_n) \equiv \int d\lambda_{n+1}\ldots d\lambda_L P(\lambda_1,\ldots,\lambda_L)\,.
\end{equation}

With these definitions at hand, we quote a few central results. In the large $L$ limit, the density of states for the Gaussian ensembles gives Wigner's famous semicircle law,
\begin{equation}
\rho(\lambda) = \frac{1}{2\pi} \sqrt{4-\lambda^2} \quad{\rm as}\quad L\ra \infty\,,
\end{equation}
where the semicircle diameter is fixed by our chosen eigenvalue normalization. Also in the large $L$ limit, the spectral $2$-point function
\begin{equation}
\rho^{(n)}(\lambda_1,\lambda_2) =\int d\lambda_{3}\ldots d\lambda_L P(\lambda_1,\ldots,\lambda_L)\,,
\end{equation}
can be expressed in terms of a disconnected piece and a squared sine kernel as \cite{MehtaRMT}
\begin{equation}
\rho^{(2)} (\lambda_1,\lambda_2) = \frac{L^2}{L(L-1)}\, \rho(\lambda_1) \rho(\lambda_2) - \frac{L^2}{L(L-1)} \frac{\sin^2 \big(L (\lambda_1-\lambda_2)\big)}{\big(L\pi (\lambda_1-\lambda_2)\big)^2}\,.
\label{eq:sinek}
\end{equation}

\subsection{Spectral form factors}
\label{sec:FFs}

The $2$-point spectral form factor for a single Hamiltonian $H$ is given in terms of the analytically continued partition function $Z(\beta, t) = \Tr\, (e^{-\beta H-iHt})$ as
\begin{equation}
\CR_{2}^{H}(\beta,t) \equiv Z(\beta, t) Z^*(\beta, t) = \Tr\, (e^{-\beta H-iHt}) \Tr\, (e^{-\beta H+ iHt}) \,.
\end{equation}
Similarly, the spectral form factor averaged over the GUE is denoted by
\begin{equation}
\CR_2(\beta,t) \equiv \big\langle Z(\beta, t) Z^*(\beta, t) \big\rangle_{\rm GUE}  = \int D\lambda \sum_{i,j} e^{i(\lambda_i-\lambda_j)t} e^{-\beta(\lambda_i+\lambda_j)}\,,
\end{equation}
which is the Fourier transform of the spectral 2-point function. At infinite temperature $\beta = 0$, the Fourier transform of the density of states is just $Z(t) = \Tr\, (e^{-iHt})$, the trace of unitary time evolution. Using the semicircle law, we take the average of $Z(t)$ at large $L$
\begin{equation}
\vev{Z(t)}_{\rm GUE}  = \int D\lambda \sum_i e^{-i\lambda_i t} = L \int_{-2}^2 d\lambda\,\rho(\lambda) e^{-i\lambda t} = \frac{L J_1(2t)}{t}\,,
\label{eq:Zavg}
\end{equation}
where $J_{1}(t)$ is a Bessel function of the first kind. The function $J_1(2t)/t$ is one at $t=0$ and oscillates around zero with decreasing amplitude that goes as $\sim 1/t^{3/2}$, decaying at late times.  At infinite temperature, the $2$-point spectral form factor for the GUE is
\begin{equation}
\CR_2(t) = \big\langle Z(t)Z^*(t) \big\rangle_{\rm GUE}  = \int dH \,\Tr \big( e^{-iHt}\big)\,\Tr\big( e^{iHt}\big) = \int D\lambda \sum_{i,j} e^{i(\lambda_i-\lambda_j)t}\,.
\end{equation}
More generally, we will also be interested in computing $2k$-point spectral form factors
\begin{equation}
\CR_{2k}(t) = \Big\langle \big(Z(t)Z^*(t)\big)^k\Big\rangle_{\rm GUE}  = \int D\lambda \sum_{i{\rm 's}, j{\rm 's}} e^{i(\lambda_{i_1} + \ldots + \lambda_{i_k}-\lambda_{j_1} - \ldots - \lambda_{j_k})t}\,,
\end{equation}
the Fourier transform of the spectral $2k$-point function $\rho^{(2k)}$.\footnote{In the random matrix literature, the $2$-point form factor is often defined as the Fourier transform of the {\it connected} piece of the spectral $2$-point correlation function, where the connected piece of the spectral $2k$-point function is often referred to as the $2k$-level cluster function.  Our definition for the $2k$-point spectral form factor $\mathcal{R}_{2k}$ includes both connected \textit{and} disconnected pieces.} Although the form factors can be written exactly at finite $L$, our analysis will focus on analytic expressions that capture the large $L$ behavior.\footnote{In addition to relating the form factor to the fidelty of certain states, \cite{delCampo17} also studies the 2-point spectral form factor for the GUE, computing an analytic form at finite $L$ and discussing the dip and plateau.}

Note that in \cite{BHRMT16}, $2$-point form factors were normalized via dividing by $Z(\beta)^2$. At infinite temperature, this simply amounts to dividing by $L^2$, but at finite temperature the situation is more subtle. As we will comment on later, the correct object to study is the quenched form factor $\vev{Z(\beta,t)Z^*(\beta,t)/Z(\beta)^2}$, but since we only have analytic control over the numerator and denominator averaged separately, we instead work with the unnormalized form factor $\mathcal{R}_2$ as defined above.

\subsubsection{2-point spectral form factor at infinite temperature}
Here we calculate the $2$-point form factor at $\beta=0$. Working at large $L$, we can evaluate $\CR_2$ by first pulling out the contribution from coincident eigenvalues
\begin{equation}
\CR_2(t) = \int D\lambda \,\sum_{i,j} e^{i (\lambda_i - \lambda_j)t } = L + L(L-1)\int d\lambda_1 d\lambda_2 \, \rho^{(2)}(\lambda_1,\lambda_2)e^{i(\lambda_1-\lambda_2)t}\,.
\label{eq:spec2pt}
\end{equation}
In the large $L$ limit, we can make use of the sine kernel form of the $2$-point function Eq.~\eqref{eq:sinek}. Using Eq.~\eqref{eq:Zavg}, we integrate the first term, a product of $1$-point functions, and find
\begin{equation}
\int d\lambda_1d\lambda_2\, \rho(\lambda_1) \, \rho(\lambda_2)\, e^{i(\lambda_1-\lambda_2)t} = \frac{J_1^2(2t)}{t^2}\,.
\end{equation}
In order to integrate the sine kernel, we make the change of variables:
\begin{equation}
u_1 = \lambda_1 - \lambda_2 \and u_2 = \lambda_2\,,
\end{equation}
which allows us to rewrite the integral
\begin{equation}
L^2 \int d\lambda_1d\lambda_2 \frac{\sin^2 \big(L (\lambda_1-\lambda_2)\big)}{\big(L \pi (\lambda_1-\lambda_2)\big)^2}\, e^{i(\lambda_1-\lambda_2)t}  = L^2 \int du_2 \int du_1 \frac{\sin^2 (Lu_1)}{L\pi u_1^2}\, e^{i u_1 t}\,.
\end{equation}
Having decoupled the variables, in order to integrate over $u_1$ and $u_2$, we must employ a short distance cutoff. We develop a certain approximation method which we refer to as the `box approximation,' and explain its justification in App.~\ref{app:FFFP}. Specifically, we integrate  $u_1$ from $0$ to $u_2$, and integrate $u_2$ from $-\pi/2$ to $\pi/2$,
\begin{equation}
L^2 \int du_1 du_2 \frac{\sin^2(Lu_1)}{L\pi u_1^2}\, e^{i u_1 t} =L \begin{cases} 1-\frac{t}{2L}\,, &{\rm for}\quad t<2L\\ 0\,, &{\rm for}\quad t>2L\end{cases} \,.
\end{equation}
Note that in the random matrix theory literature, a common treatment \cite{BrezinHikami1} is to approximate the short-distance behavior of $\rho^{(2)}(\lambda_1,\lambda_2)$ by adding a delta function for coincident points $\lambda_1 = \lambda_2$ and inserting a $1$-point function into the sine kernel. For $\CR_2$ this gives the same result as the approximation above, but this short-distance approximation does not generalize to higher $k$-point form factors, as discussed in App.~\ref{app:FFFP}. The $2$-point form factor we compute is\footnote{We emphasize that this function relied on an approximation and while it captures certain desired behavior, it should not be viewed as exact. In App.~\ref{app:num} we provide numerical checks and discuss an improvement of the ramp function $r_2(t)$.}
\begin{equation}
\CR_2(t) = L^2 r_1^2(t) - Lr_2 (t) +L\,,
\label{eq:R2func}
\end{equation}
where we define the functions
\begin{equation}
r_1(t) \equiv \frac{J_1(2t)}{t}\,, \and r_2(t) \equiv \begin{cases} 1-\frac{t}{2L}\,, &{\rm for}\quad t<2L\\ 0\,, &{\rm for}\quad t>2L\end{cases} \,.
\end{equation}

As was discussed in \cite{BHRMT16}, we can extract the dip and plateau times and values from $\mathcal{R}_2$. From the ramp function $r_2$, we observe that the plateau time is given by
\begin{equation}
t_{p} = 2L\,
\end{equation}
where after the plateau time, the height of the function $\CR_2$ is the constant $L$.  This value can also be derived by taking the infinite time average of $\CR_2$.

\begin{figure}[htb!]
\centering
\includegraphics[width=0.5\linewidth]{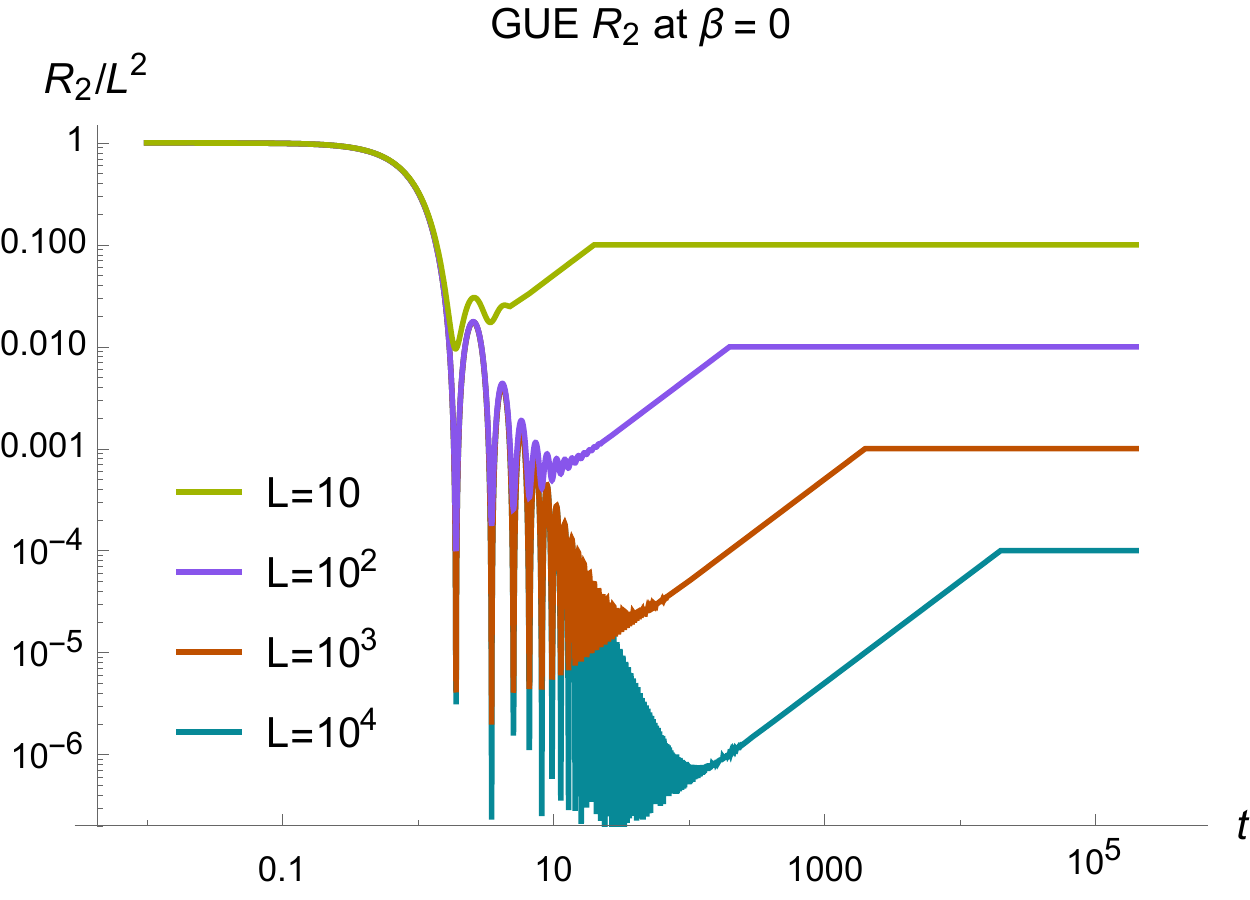}
\caption{The $2$-point spectral form factor at infinite temperature, as given in Eq.~\eqref{eq:R2func}, plotted for various values of $L$ and normalized by the initial value $L^2$. We observe the linear ramp and scaling of the dip and plateau with $L$.
}
\label{fig:2pt}
\end{figure}

The other important time scale is the dip time $t_d$, which we can estimate using the asymptotic form of the Bessel function at large $t$, which gives
\begin{equation}
r_1(t) \approx \frac{1}{t} \frac{\cos(2t-3\pi/4)}{\sqrt{\pi t}}\,,
\end{equation}
oscillating at times $\sim\mathcal{O}(1)$ with decaying envelope $\sim t^{-3/2}$. While the first dip time is $\mathcal{O}(1)$, we will be interested in the dip time as seen by the envelope, especially because the oscillatory behavior disappears at finite temperature (see Fig.~\ref{fig:2pttemp}). Solving for the minimum of the envelope of $\CR_2$, we find
\begin{equation}
t_d \approx\sqrt{L} \,,
\end{equation}
up to order one factors. The true minimum of the envelope and ramp is $(6/\pi)^{1/4} \sqrt{L} \approx 1.18 \sqrt{L} $, but in light of the approximations we made, and the fact that the precise ramp behavior is somewhat ambiguous, we simply quote the dip time as $t_d \approx \sqrt{L}$. At $t_d$, we find the dip value $\CR_2(t_d)\approx\sqrt{L}$. We plot the 2-point form factor for different dimensions $L$ in Fig.~\ref{fig:2pt}.

\noindent The oscillations in the early time slope behavior of the form factor simply arise from the oscillatory behavior of the Bessel function, \ie the zeros of $r_1(t)^2$.

\subsubsection{2-point spectral form factor at finite temperature}
Recall that spectral $2$-point function at finite temperature is defined as
\begin{equation*}
\CR_2(t,\beta) \equiv \big\langle Z(t,\beta)Z^*(t,\beta) \big\rangle_{\rm GUE} = \int D\lambda \sum_{i,j} e^{i(\lambda_i-\lambda_j)t} e^{-\beta(\lambda_i+\lambda_j)}\,.
\end{equation*}
As described in App.~\ref{app:FFFP}, we insert the spectral $2$-point function $\rho^{(2)}$ and, using the short-distance kernel, find $\CR_2(t,\beta)$ in terms of the above functions:
\begin{equation}
\CR_2(t,\beta) = L^2 r_1(t+i\beta ) r_1(-t+i\beta ) + L r_1(2i\beta ) -  L r_1(2i\beta ) r_2(t)\,.
\label{eq:R2beta}
\end{equation}

First we comment on the validity of the approximations used in the finite temperature case. The first and third terms of Eq.~\eqref{eq:R2beta}, dominating at early and late times respectively, are computed from the 1-point function. Therefore, the expression captures the early time, slope, and plateau behaviors. The dip and ramp behavior, encoded in the $r_2$ term, are more subtle. The expression correctly captures the slope of the ramp, but deviates from the true ramp at large $\beta$. We will discuss this more in App.~\ref{app:FFFP}, but here only discuss quantities around the dip for small $\beta$, where Eq.~\eqref{eq:R2beta} is a good approximation.

The ramp function $r_2$, which is the same as at infinite temperature, gives the plateau time
\begin{equation}
t_p = 2L\,.
\end{equation}
For convenience we define the function $h_1(\beta) \equiv J_1(2 i \beta)/i \beta$, which is real-valued in $\beta$.\footnote{For instance, to emphasize its real-valuedness, we could equivalently write $h_1(\beta)$ as a regularized hypergeometric function $h_1(\beta) \equiv  {}_0\widetilde F_1 (2;\beta^2)$.} The initial value and plateau value are thus given by
\begin{equation}
\CR_2(0) = (h_1(\beta))^2 L^2\,, \qquad \CR_2(t_p) = h_1(2\beta) L\,.
\end{equation}
To find the dip time, we make use of the asymptotic expansion of the Bessel function as
\begin{equation}
L^2 r_1(t+i\beta) r_1(-t+i\beta) \sim \frac{L^2}{2\pi t^3} \big( \cosh(4\beta) - \sin(4t)\big) \approx \frac{L^2}{\pi t^3} \cosh^2 (2\beta)\,.
\end{equation}
Finding the minimum of the expression gives the dip time
\begin{align}
t_d = h_2(\beta) \sqrt{L} \where h_2(\beta) \approx \left(1+ \frac{\beta^2}{2} + O(\beta^4)\right)\,,
\end{align}
and evaluating $\CR_2$ at the dip gives
\begin{equation}
\CR_2(t_d) \approx h_3(\beta) \sqrt{L} \where h_3(\beta) \approx \left( 1+ \frac{5\beta^2}{2} + \op(\beta^4) \right)\,,
\end{equation}
up to order one factors. While we could write down full expressions for the dip time $h_2$ and dip value $h_3$ in terms of the Bessel function, we only trust Eq.~\eqref{eq:R2beta} in this regime for small $\beta$, and thus report the functions perturbatively.

\begin{figure}[htb!]
\centering
\includegraphics[width=0.45\linewidth]{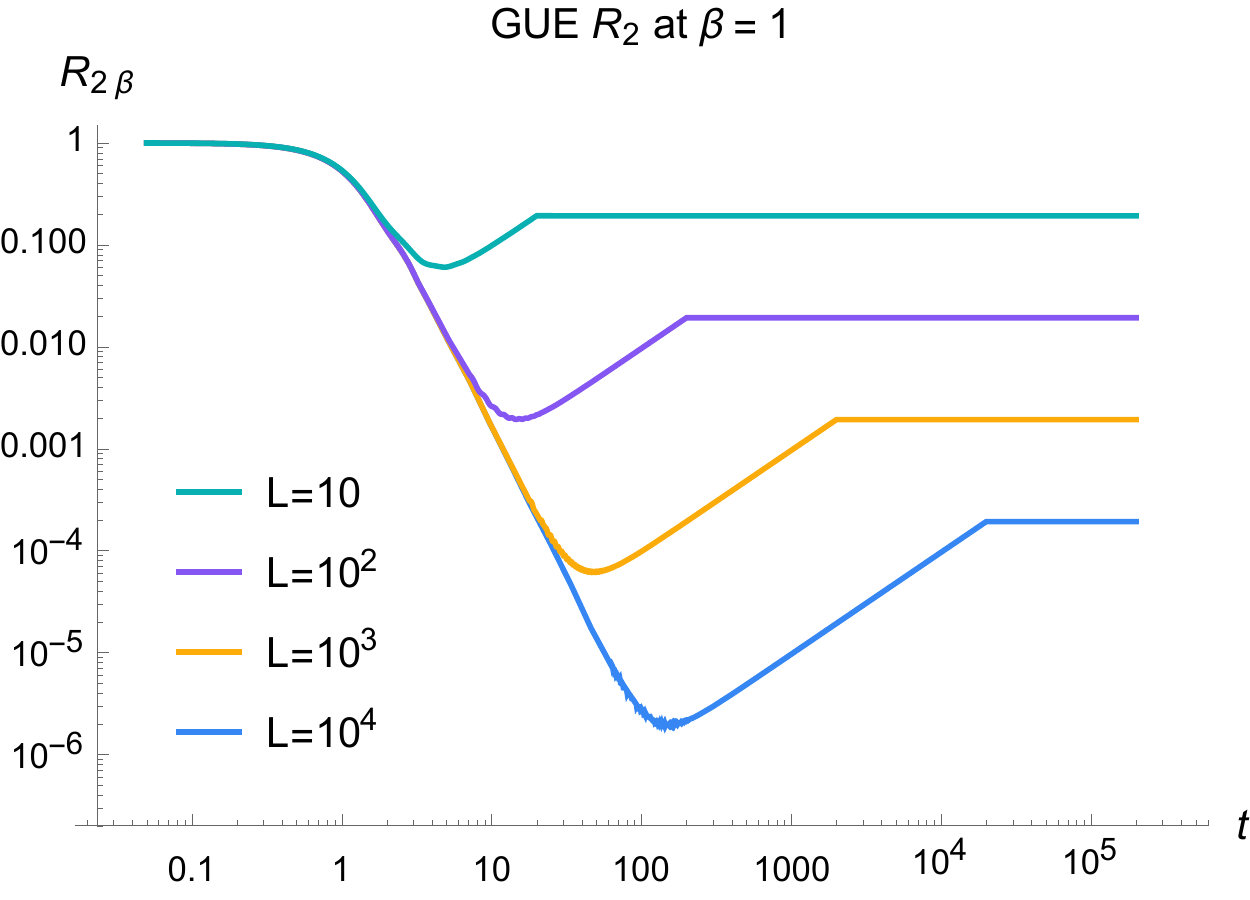}
\includegraphics[width=0.45\linewidth]{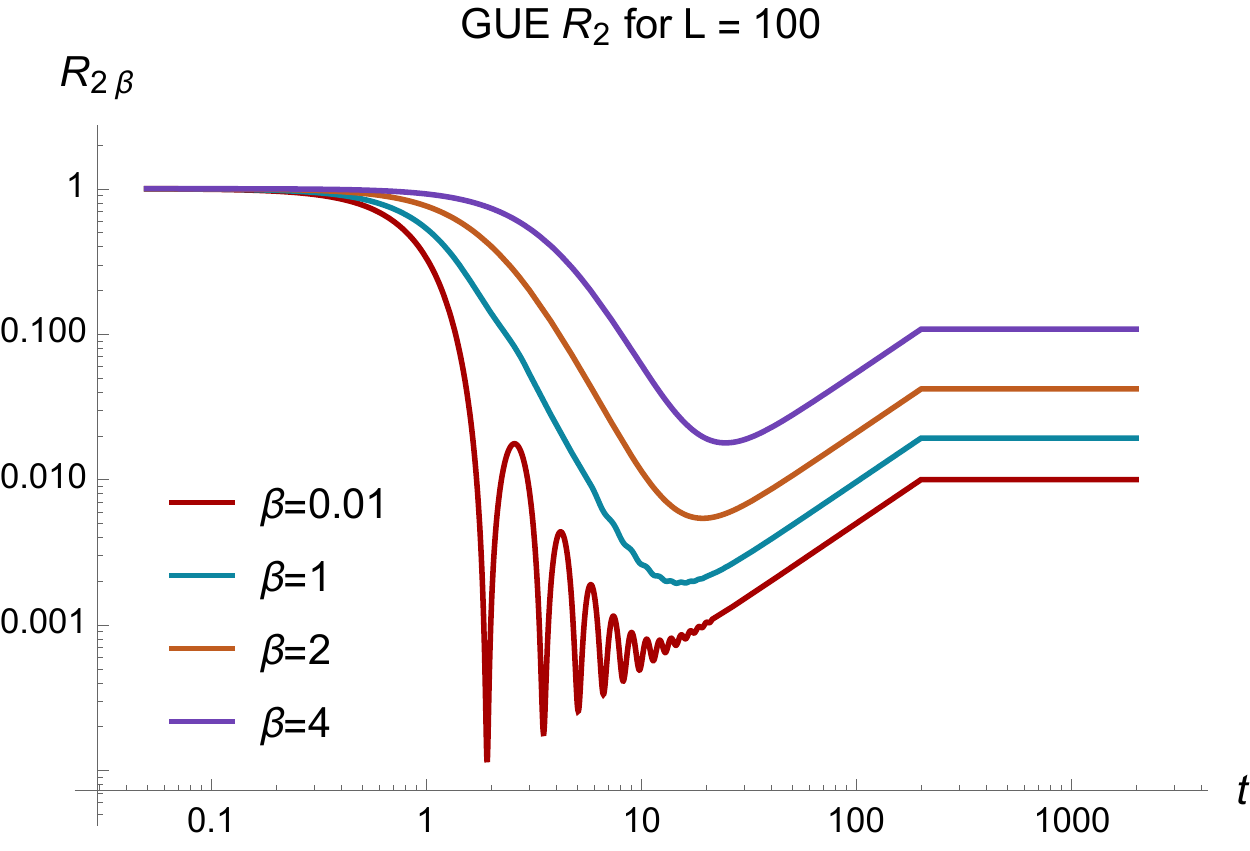}
\caption{The $2$-point spectral form factor at finite temperature as per Eq.~\eqref{eq:R2beta}, on the left plotted at different values of $L$, and on the right plotted at different temperatures, normalized by the initial value. We see that the dip and plateau both scale with $\beta$ and $L$ and that lowering the temperature smooths out the oscillations in $\CR_2$.
}
\label{fig:2pttemp}
\end{figure}

The 2-point form factor is plotted in Fig.~\ref{fig:2pttemp} for various values of $L$ and $\beta$. While increasing the dimension $L$ lowers the dip and plateau values and delays the dip and plateau times, decreasing temperature raises the dip and plateau values and delays the dip times. We also note that lowering the temperature smooths out oscillations from the Bessel function.\footnote{While the oscillatory behavior still persists at finite temperature, the width of the dips become very sharp as we increase $\beta$ and thus the oscillations are not observed when plotted. Furthermore, if we average over a small time window, the oscillations are also smoothed out.}  After normalizing $\CR_{2}(\beta,t)$ by its initial value, the late-time value is $\simeq 2^{-S^{(2)}}$ where $S^{(2)}$ is the thermal R\'{e}nyi-$2$ entropy.

\subsection{4-point spectral form factor at infinite temperature}
\begin{figure}[t]
\centering
\includegraphics[width=0.50\linewidth]{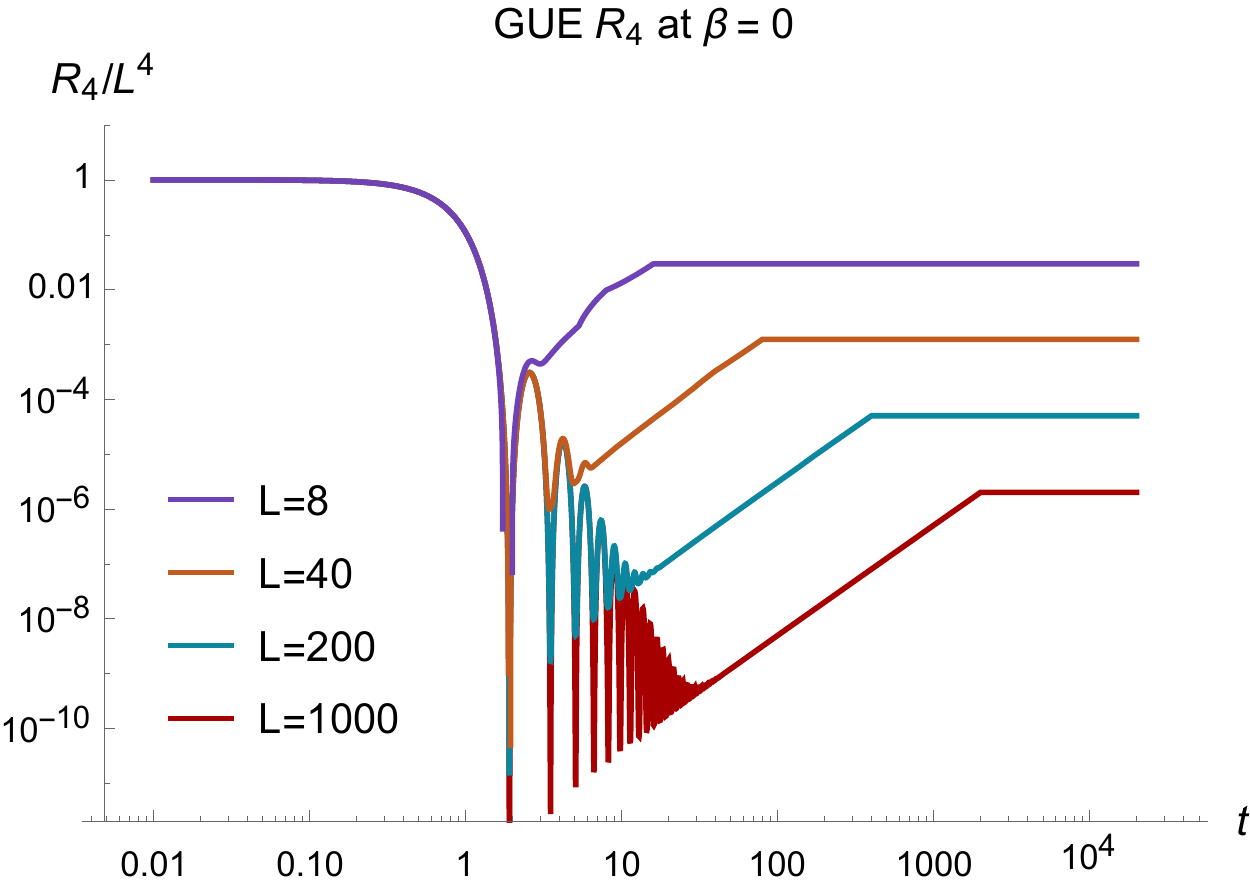}
\caption{The GUE $4$-point spectral form factor at infinite temperature, plotted for different values of $L$ and normalized by their initial values. We observe the scaling of the dip and plateau, and the quadratic rise $\sim t^2$.
}
\label{fig:4pt}
\end{figure}

We can also compute the $4$-point form factor at infinite temperature, defined as
\begin{equation}
\CR_4(t) \equiv \big\langle Z(t)Z(t)Z^*(t)Z^*(t) \big\rangle_{\rm GUE}  = \int D\lambda \,\sum_{i,j,k,\ell} e^{i (\lambda_i+\lambda_j - \lambda_k-\lambda_\ell)t }\,.
\label{eq:spec4pt}
\end{equation}
As we explain in App.~\ref{app:FFFP}, we compute $\CR_4$ by replacing $\rho^{(4)}$ by a determinant of sine kernels and carefully integrating each term using the box approximation. The result is
\begin{equation}
\CR_4(t) = L^4 r_1^4(t) + 2L^2 r_2^2 (t) - 4 L^2 r_2 (t) - 7L r_2 (2t) + 4Lr_2 (3t)+4L r_2 (t) + 2 L^2 -L\,,
\end{equation}
given in terms of the functions $r_1(t)$ and $r_2(t)$ defined above. The initial value of $\CR_4$ is $L^4$. Given the dependence on the ramp function, the plateau time is still $t_p =2L $. The plateau value $2 L^2-L$ matches the infinite time average of Eq.~\eqref{eq:spec4pt}. The dip time is found again by considering the leading behavior of $\CR_4$ and expanding the Bessel functions
\begin{equation}
\CR_4 \approx L^4\frac{J_1^4(2t)}{t^4} + \frac{t}{2}(t-2) \sim \frac{L^4}{t^6\pi^2} + \frac{t}{2}(t-2) \,.
\end{equation}
Solving for the minimum, we find the dip time
\begin{equation}
t_d \approx \sqrt{L}\,,
\end{equation}
where at the dip time $\CR_4(t_d) \approx L$. We plot the $\CR_4(t)$ for various values of $L$ in Fig.~\ref{fig:4pt}.

Let us summarize the time scales and values for the form factors considered above:
\begin{center}
\begin{tabular}{c|c|c|c}
form factor & time scale & time & value\\
\hline
$\CR_2(t)$
& initial & 0 & $L^2$\\
& dip & $\sqrt{L}$ & $ \sqrt{L}$\\
& plateau  & $2L$ & $L$\\
\hline
$\CR_2(t,\beta)$
& initial & 0 & $ h^2_1(\beta)L^2$\\
& dip &$h_2(\beta) \sqrt{L}$ & $h_3(\beta) \sqrt{L}$\\
& plateau & $2L$ & $h_1(2\beta) L$\\
\hline
$\CR_4(t)$
& initial & 0 & $L^4$\\
& dip & $\sqrt{L}$ & $ L$\\
& plateau  & $2L$ & $2L^2$\\
\end{tabular}
\end{center}
The $\beta$--dependent functions were defined above. 

With an understanding of the first few form factors, we briefly describe the expected behavior for $2k$-point form factors $\CR_{2k}(t)$ (with $k\ll L$). Initially, $\CR_{2k}$ decays from $L^{2k}$ as $\sim J_1^{2k}(2t)/t^{2k}$, reaching the dip at time $t_d\approx \sqrt{L}$ where $\CR_{2k}(t_d) \approx L^{k/2}$. The $\sim t^k$ growth after the dip levels off at the plateau time $2L$, with plateau value $\sim k L^k$. 

Given that we employed some approximation to compute the form factors, we perform numerical checks for the expressions above in App.~\ref{app:num}. At both infinite and finite temperature, we correctly capture the time scales, early time decay, dip behavior, and the late-time plateau, but find slight deviations from the analytic prediction for the ramp. We discuss this and possible improvements to the ramp function in App.~\ref{app:num}.

Later we will study frame potentials which diagnose whether an ensemble forms a $k$-design.  We will find that the frame potentials for the ensemble of unitaries generated by the GUE can be written in terms of the spectral form factors discussed here, thereby allowing us to extract important time scales pertaining to $k$-designs.

\section{Out-of-time-order correlation functions}
\label{sec:OTOC}

\subsection{Spectral form factor from OTOCs}
\label{sec:OTOCFF}

Although quantum chaos has traditionally focused on spectral statistics, recent developments from black hole physics and quantum information theory suggest an alternative way of characterizing quantum chaos via OTOCs \cite{HaydenPreskill, SSbutterfly, MSSbound, ChaosChannels}. In this subsection, we bridge the two notions by relating the average of $2k$-point OTOCs to spectral form factors. We work at infinite temperature ($\beta=0$), but note that by distributing operator insertions around the thermal circle, the generalization to finite temperature is straightforward. The results in this subsection are not specific to GUE and are applicable to any quantum mechanical system.

Consider some Hamiltonian $H$ acting on an $L=2^n$-dimensional Hilbert space, \ie consisting of $n$ qubits. We start by considering the $2$-point autocorrelation function $\langle A(0) A^{\dagger}(t) \rangle$, time evolved by $H$. We are interested in the averaged $2$-point function:
\begin{align}
\int dA \langle A(0) A^{\dagger}(t) \rangle \equiv \frac{1}{L} \int dA \  \text{Tr}( A  e^{-iHt} A^{\dagger} e^{iHt} ) \label{eq:2pt-average}
\end{align}
where $\int dA$ represents an integral with respect to a unitary operator $A$ over the Haar measure on $U(2^n)$. We note that since the 2-point Haar integral concerns only the first moment of the Haar ensemble, we can instead average over the ensemble of Pauli operators\footnote{This is because the Pauli operators form a 1-design.}
\begin{align}
\int dA \langle A(0) A^{\dagger}(t) \rangle = \frac{1}{L^3} \sum_{j=1}^{L^2} \text{Tr}( A_{j}  e^{-iHt} A_{j}^{\dagger} e^{iHt} )\,,
\end{align}
where $A_{j}$ are Pauli operators and $L^2=4^{n}$ is the number of total Pauli operators for a system of $n$ qubits. To derive the spectral form factor, we will need the first moment of the Haar ensemble
\begin{align}
\int dA\, A^j_k A^\dagger{}^\ell_m = \frac{1}{L} \delta^j_m \delta^\ell_k\,,\quad {\rm or~equivalently}\quad \int dA \ A O A^{\dagger} =\frac{1}{L} \text{Tr}(O) I. \label{eq:Haar1}
\end{align}
Applying Eq.~\eqref{eq:Haar1} to Eq.~\eqref{eq:2pt-average}, we obtain
\begin{align}
\int dA \langle A(0) A^{\dagger}(t) \rangle = \frac{| \text{Tr}(e^{-iHt}) |^2}{L^2} = \frac{\CR_{2}^H(t)}{L^2}. \label{eq:OTOC_factor_2}
\end{align}
where $\CR_{2k}^H(t) \equiv |\text{Tr}(e^{-i H t})|^{2k}$ is the same as $\CR_{2k}(t)$ from before, but written for a single Hamiltonian $H$ instead of averaged over the GUE. Thus, the $2$-point form factor is proportional to the averaged $2$-point function.

This formula naturally generalizes to $2k$-point OTOCs and $2k$-point form factors. Consider $2k$-point OTOCs with some particular ordering of operators
\begin{align}
\langle A_{1}(0)B_{1}(t)\cdots A_{k}(0)B_{k}(t) \rangle \where A_{1}B_{1}\cdots A_{k}B_{k}=I.
\end{align}
Operators which do not multiply to the identity have zero expectation value at $t=0$, and the value stays small as we time-evolve. We are interested in the average of such $2k$-point OTOCs. By using Eq.~\eqref{eq:Haar1} $2k-1$ times, we obtain
\begin{align}
\int dA_{1}\cdots dB_{k-1}dA_{k} \langle A_{1}(0)B_{1}(t)\cdots A_{k}(0)B_{k}(t) \rangle = \frac{|\text{Tr}(e^{-iHt})|^{2k}}{L^{2k}} = \frac{\mathcal{R}_{2k}^H(t)}{L^{2k}} \label{eq:OTOC_factor_2k}
\end{align}
where $B_{k}=A_{k}^{\dagger}\cdots B_{1}^{\dagger}A_{1}^{\dagger}$.
Thus, higher-point spectral form factors can be also computed from OTOCs. In fact, by changing the way we take an average, we can access various types of form factors. For instance, let us consider OTOCs $\langle A_{1}(0)B_{1}(t)\cdots A_{k}(0)B_{k}(t) \rangle$ with $B_{j}=A_{j}^{\dagger}$. We then have
\begin{align}
\int dA_{1}dA_{2}\cdots dA_{k} \langle A_{1}(0)A_{1}^{\dagger}(t)\cdots A_{k}(0)A_{k}^{\dagger}(t) \rangle  = \frac{\text{Tr}(e^{-iHt})^{k} \text{Tr}(e^{iHkt}) }{L^{k+1}}. \label{eq:asymmetric}
\end{align}
The fact that the expression on the right-hand side is asymmetric is because the operator $A_{1}(0)A_{1}^{\dagger}(t)\cdots A_{k}(0)A_{k}^{\dagger}(t)$ is not Hermitian.\footnote{BY learned Eq.~\eqref{eq:asymmetric} from Daniel Roberts.}

These expressions not only provides a direct link between spectral statistics and physical observables, but also give a practical way of computing the spectral form factor. If one wishes to compute or experimentally measure the $2$-point form factor $\CR_{2}(t)$, one just needs to pick a random unitary operator $A$ and study the behavior of the $2$-point correlator $\langle A(0) A^{\dagger}(t) \rangle$. In order to obtain the exact value of $\CR_{2}(t)$, we should measure $\langle A(0) A(t) \rangle$ for all possible Pauli operators and take their average. Yet, it is possible to obtain a pretty good estimate of $\CR_{2}(t)$ from $\langle A(0) A(t) \rangle$ with only a few instances of unitary operator $A$. Consider the variance of $\langle A(0) A(t) \rangle$,
\begin{align}
\Delta \langle A(0) A^{\dagger}(t) \rangle_{\rm avg}^2 \equiv  \int dA | \langle A(0) A^{\dagger}(t) \rangle |^2 - \Big| \int dA \langle A(0) A^{\dagger}(t)\rangle \Big|^2.
\end{align}
If the variance is small, then the estimation by a single $A$ would suffice to obtain a good estimate of $\CR_{2}(t)$. Computing this, we obtain
\begin{align}
\Delta \langle A(0) A^{\dagger}(t) \rangle_{\rm avg}^2 \sim \mathcal{O}\Big(\frac{1}{L^2}\Big).
\end{align}
This implies that the estimation error is suppressed by $1/L$. By choosing a Haar unitary operator $A$ (or $2$-design operator, such as a random Clifford operator), one can obtain a good estimate of $\CR_{2}(t)$.

\subsubsection*{A check in a non-local spin system}
To verify Eq.~\eqref{eq:OTOC_factor_2} and the claim that the variance of the 2-point functions is small, consider a random non-local (RNL) spin system with the Hamiltonian given as the sum over all 2-body operators with random Gaussian couplings $J_{ij\alpha \beta}$\,\cite{Erdos14}:
\begin{equation}
H_{\rm RNL} = -\sum_{i,j,\alpha,\beta} J_{ij\alpha\beta} S_i^\alpha S_j^\beta\,,
\label{eq:HRNL}
\end{equation}
where $i,j$ sum over the number of sites and $\alpha,\beta$ sum over the Pauli operators at a given site. Such Hamiltonians have a particularly useful property where locally rotating the spins of $H_{\rm RNL}$ with couplings $J_{ij\alpha\beta}$ creates another Hamiltonian $H_{\rm RNL}'$ with different couplings $J_{ij\alpha\beta}'$. More precisely, if we consider an ensemble of such $2$-local Hamiltonians;
\begin{align}
\mathcal{E}_{\rm RNL} = \{ H_{\rm RNL}, \ \text{for} \ J_{ij\alpha\beta}\in \text{Gaussian} \}
\end{align}
the ensemble is invariant under conjugation by any $1$-local Clifford operator
\begin{align}
\mathcal{E}_{\rm RNL} =V \mathcal{E}_{\rm RNL}V^{\dagger}\,,\qquad V \in \text{$1$-body Clifford}.
\end{align}
Here a Clifford operator refers to unitary operators which transform a Pauli operator to a Pauli operator. For this reason, the $2$-point correlation function $\langle A(0)A^{\dagger}(t) \rangle_{\mathcal{E}_{\rm RNL}}$ depends only of the weight of Pauli operator $A$:
\begin{align}
\langle A(0)A^{\dagger}(t) \rangle_{\mathcal{E}_{\rm RNL}} = c_{m}\,,\quad \text{where $A$ is an $m$-body Pauli operator}
\end{align}
and where $\langle \, \cdot \, \rangle_{\mathcal{E_{\rm RNL}}}$ denotes the ensemble (disorder) average. Thus, this system is desirable for studying the weight dependence of $2$-point correlation functions.

\begin{figure}[t]
\centering
\includegraphics[width=0.60\linewidth]{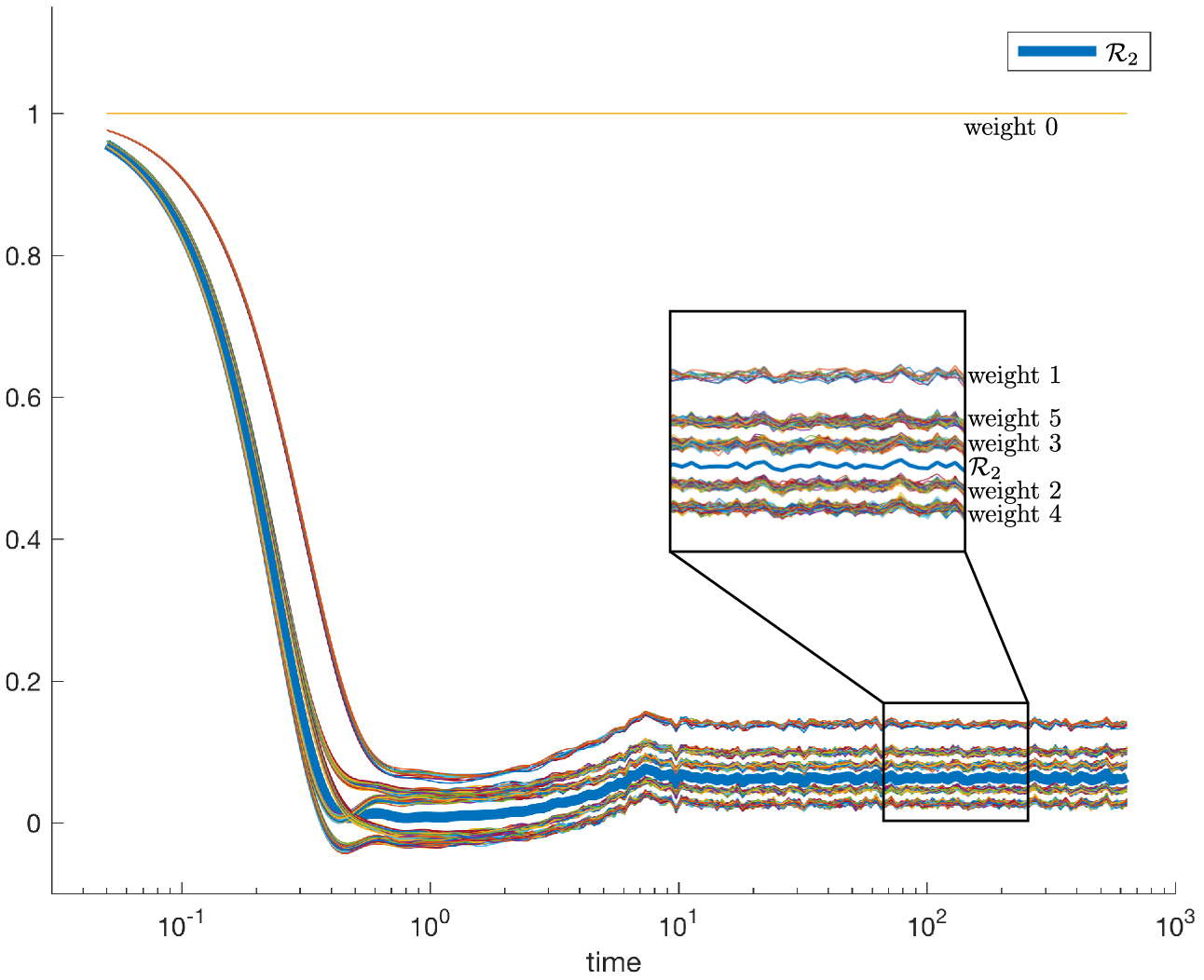}
\caption{The 2-point form factor and the 2-point functions $\vev{A_j A_j(t)}$ of Pauli operators for $H_{\rm RNL}$ for $n=5$ sites and averaged over $500$ samples. The thick blue line is $\CR_2/L^2$ and surrounding bands of lines are all 1024 Pauli 2-point functions of different weight.}
\label{fig:HRNL}
\end{figure}

As mentioned above, we can write the average over $2$-point correlation functions as the average over all Paulis as
\begin{equation}
\int dA \vev{A(0)A^\dagger(t)} = \frac{1}{4^n}\sum_{A\in {\rm Pauli}} \vev{A(0)A^\dagger(t)} = \frac{\CR_2^{H_{\rm RNL}} (t)}{L^2}\,,
\end{equation}
time evolving with $H_{\rm RNL}$. Numerically, for a single instance of $H_{\rm RNL}$, we find that the average over all 2-point functions of Pauli operators gives $\CR_2$ as expected. In Fig.~\ref{fig:HRNL}, for $n=5$ sites and averaged over 500 random instances of $H_{\rm RNL}$ to suppress fluctuations, we plot $\CR_2$ along side all 2-point functions of Pauli operators. We observe that correlation functions depend only on the weight of $A$, with the higher weight Pauli operators clustered around $\CR_2$. The arrangement of the 2-point functions for Paulis of different weight depends on the number of sites $n$. But for $n=5$, the even and odd weight Paulis are respectively below and above $\CR_2$ at later times and weight 2 and 3 Paulis are the closest to $\CR_2$. We will comment on the size dependence of correlators in Sec.~\ref{sec:haarinv}.

The conclusion is that we can choose a few random Paulis, and by computing 2-point functions, quickly approximate $\CR_2$. We also checked that by increasing the number of spins, the variance becomes small and 2-point functions become closer to $\CR_2$.

\subsubsection*{Operator averages and locality}

Let us pause for a moment and discuss the meaning of considering the operator average from the perspective of spatial locality in quantum mechanical systems. In deriving the above exact formulae relating the spectrum and correlators, we considered the average of OTOCs over all the possible Pauli operators. For a system of $n$ qubits, a typical Pauli operator has support on $\simeq 3n/4$ qubits because there are four one-body Pauli operators, $I,X,Y,Z$. It is essential to recognize that the average of correlation functions is dominated by correlations of non-local operators with big supports covering the whole system. Thus, the spectral statistics have a tendency to ignore the spatial locality of operators in correlation functions.\footnote{Signatures of the locality of an individual Hamiltonian may be seen in properties of its spectrum, as argued in~\cite{LocalitySpectrum}. }

In fact, the spectral statistics ignore not only spatial locality but also temporal locality of operators. Namely, similar formulas can be derived for correlation functions with various ordering of time. For instance, consider the following $4$-point correlation function:
\begin{align}
\langle A(0)B(t)C(2t)D(t) \rangle
\end{align}
where the $C$ operator acts at time $2t$ instead of $0$ such that the correlator is not out-of-time-ordered. Computing the average of the correlator with $ABCD=I$, we obtain
\begin{align}
\int dAdBdC \langle A(0)B(t)C(2t)D(t) \rangle = \frac{\CR_{4} (t)}{L^4}
\end{align}
which is exactly the same result as the average of $4$-point OTOCs in Eq.~\eqref{eq:OTOC_factor_2k}. Indeed, time-ordering is washed away since GUE Hamiltonians cause a system to rapidly delocalize, thus destroying all local temporal correlations. 

In strongly coupled systems with local Hamiltonians, correlation functions behave rather differently depending on the time ordering of operators, as long as the time gaps involved are small or comparable to the scrambling time~\cite{LarkinOv69, SSbutterfly, SSstringy, Kitaev15}. This observation hints that the spectral statistics are good probes of correlations at long time scales, but may miss some important physical signatures at shorter time scales, such as the exponential growth of OTOCs with some Lyapunov exponent.

\subsection{OTOCs in random matrix theory}
\label{sec:OTOCRMT}
Next, we turn our attention to correlators averaged over random matrices, analytically computing the $2$-point correlation functions and $4$-point OTOCs for the GUE. We begin with the $2$-point correlation functions for the GUE
\begin{align}
\langle A(0)B(t) \rangle_{\text{GUE}}\equiv \int dH \langle A(0)B(t) \rangle \where B(t)=e^{-iHt}B(0)e^{iHt}\,,
\end{align}
where $\int dH$ represents an integral over Hamiltonians $H$ drawn from the GUE. Since the GUE measure $dH$ is invariant under unitary conjugation $dH = d(UHU^{\dagger}) $ for all $U$, we can express the GUE average as
\begin{align}
\langle A(0)B(t) \rangle_{\text{GUE}}=\iint dH dU \big\langle A U e^{-iHt}U^{\dagger} B U e^{iHt}U^{\dagger} \big\rangle
\end{align}
by inserting $U,U^{\dagger}$ where $dU$ is the Haar measure. Haar integrating, we obtain
\begin{align}
\langle A(0)B(t) \rangle_{\text{GUE}} = \langle A \rangle \langle B \rangle + \frac{\CR_{2}(t)-1}{L^{2}-1} \langle \! \langle AB \rangle \!\rangle \,, \qquad  \langle \! \langle AB \rangle \!\rangle \equiv \langle AB \rangle -  \langle A \rangle \langle B \rangle
\end{align}
where $\langle \! \langle AB \rangle \!\rangle$ represents the connected correlator. If $A,B$ are non-identity Pauli operators, we have
\begin{equation}
\begin{split}
\langle A(0)B(t) \rangle_{\text{GUE}}  &= \frac{\CR_{2}(t)-1}{L^{2}-1}\qquad \ (A=B)\\
&= 0 \qquad \qquad \qquad (A\not=B)\,.
\end{split}
\end{equation}
If $\CR_{2}(t)\gg 1$, we have
\begin{align}
\langle A(0)A^{\dagger}(t) \rangle_{\text{GUE}} \simeq \frac{\CR_{2}(t)}{L^2} \label{eq:OTOC_GUE_2}
\end{align}
for any non-identity Pauli operator $A$. It is worth emphasizing the similarity between Eq.~\eqref{eq:OTOC_GUE_2} and Eq.~\eqref{eq:OTOC_factor_2}. Recall that Eq.~\eqref{eq:OTOC_factor_2} was derived by taking an average over all Pauli operators $A$ and is valid for any quantum mechanical system while Eq.~\eqref{eq:OTOC_GUE_2} was derived without any additional assumption on the locality of Pauli operator $A$. Namely, the key ingredient in deriving Eq.~\eqref{eq:OTOC_GUE_2} was the Haar-invariance of the GUE measure $dH$. The resemblance of Eq.~\eqref{eq:OTOC_GUE_2} and Eq.~\eqref{eq:OTOC_factor_2} implies that the GUE is suited for studying physical properties of chaotic Hamiltonians at macroscopic scales such as thermodynamic quantities.

Next, we compute the $4$-point OTOCs for the GUE
\begin{align}
\langle A(0)B(t)C(0)D(t) \rangle_{\text{GUE}}\,.
\end{align}
Inserting $U,U^{\dagger}$, we must compute the fourth Haar moment
\begin{align}
\langle A(0)B(t)C(0)D(t) \rangle_{\text{GUE}}
= \iint dH dU \big\langle AUe^{-iHt}U^{\dagger}B U e^{iHt}U^{\dagger} CUe^{-iHt}U^{\dagger}DU e^{iHt}U^{\dagger}\big\rangle\,.
\end{align}
We can avoid dealing directly with the $(4!)^2$ terms generated by integrating here and focus on the leading behavior. Assuming that $A,B,C,D$ are non-identity Pauli operators, we obtain
\begin{align}
\langle A(0)B(t)C(0)D(t) \rangle_{\text{GUE}}\simeq  \langle ABCD \rangle \frac{\CR_{4}(t)}{L^4}.
\label{eq:R4eq}
\end{align}
Thus, OTOCs are almost zero unless $ABCD=I$.\footnote{In fact, one can prove that the GUE averaged OTOCs are exactly zero if $ABCD$ is non-identity Pauli operator for all times.}\textsuperscript{,}\footnote{For analysis related to Eq.~\eqref{eq:R4eq} in the context of SYK, see \cite{Altland17}.} A similar analysis allows us to obtain the following result for $2k$-point OTOCs:
\begin{align}
\langle A_{1}(0)B_{1}(t)\ldots A_{k}(0)B_{k}(t) \rangle_{\text{GUE}}\simeq  \langle A_{1}B_{1}\ldots A_{k}B_{k} \rangle \frac{\CR_{2k}(t)}{L^{2k}}.\label{eq:OTOC_GUE_2k}
\end{align}
The above equation is nonzero when $A_{1}B_{1}\ldots A_{k}B_{k}=I$. Again, note the similarity between Eq.~\eqref{eq:OTOC_GUE_2k} and Eq.~\eqref{eq:OTOC_factor_2k}. Recall that in order to derive Eq.~\eqref{eq:OTOC_factor_2k}, we took an average over OTOCs with $A_{1}B_{1}\ldots A_{k}B_{k}=I$. This analysis also supports our observation that the GUE tends to capture global-scale physics very well.

Similar calculations can be carried out for correlation functions with arbitrary time-ordering. For $m$-point correlators, at the leading order, we have
\begin{align}
\langle A_{1}(t_{1}) A_{2}(t_{2}) \ldots A_{m}(t_{m}) \rangle_{\text{GUE}}\simeq \langle A_{1}\ldots A_{m} \rangle \frac{1}{L^{m}}\text{Tr}(e^{-it_{12}H})\text{Tr}(e^{-it_{23}H})\ldots\text{Tr}(e^{-it_{m1}H})
\end{align}
where $t_{ij}=t_{j}-t_{i}$. Namely, we have:
\begin{align}
\langle A(0)B(t)C(2t)D(t) \rangle_{\text{GUE}}\simeq  \langle ABCD \rangle \frac{\CR_{4}(t)}{L^4}\,.
\end{align}
So, for the GUE, $\langle A(0)B(t)C(2t)D(t) \rangle_{\text{GUE}}\simeq \langle A(0)B(t)C(0)D(t) \rangle_{\text{GUE}}$. This implies that the GUE does not care if operators in the correlator are out-of-time-ordered or not, ignoring both spatial and temporal locality.

Careful readers may have noticed that the only property we used in the above derivations is the unitary invariance of the GUE ensemble. If one is interested in computing correlation functions for an ensemble of Hamiltonians which are invariant under conjugation by unitary operators, then correlation functions can be expressed in terms of spectral form factors. Such techniques have been recently used to study thermalization in many-body systems, see~\cite{BrandaoHoro12} for instance. We discuss this point further in Sec.~\ref{sec:haarinv}.

\subsection{Scrambling in random matrices}

Finally, we discuss thermalization and scrambling phenomena in random matrices by studying the time scales for correlation functions to decay. 

We begin with $2$-point correlators and thermalization. In a black hole (or any thermal system), quantum information appears to be lost from the viewpoint of local observers. This apparent loss of quantum information is called \emph{thermalization}, and is often associated with the decay of $2$-point correlation functions $\langle A(0)B(t)\rangle$ where $A$ and $B$ are some local operators acting on subsystems $\mathcal{H}_{A}$ and $\mathcal{H}_{B}$ which local observers have access to. In the context of black hole physics, $\mathcal{H}_{A}$ and $\mathcal{H}_{B}$ correspond to infalling and outgoing Hawking radiation and such $2$-point correlation functions can be computed from the standard analysis of Hawking and Unruh~\cite{Hawking75, Unruh76}. $2$-point correlation functions of the form $\langle A(0)B(t)\rangle$ have an interpretation as how much information about initial perturbations on $\mathcal{H}_{A}$ can be detected from local measurements on $\mathcal{H}_{B}$ at time $t$. A precise and quantitative relation between quantum information (mutual information) and $2$-point correlation functions is derived in Appendix~\ref{app:QI}. The upshot is that the smallness of $\langle A(0)B(t)\rangle$ implies the information theoretic impossibility of reconstructing from Hawking radiation (defined on $\mathcal{H}_{B}$) an unknown quantum state (supported on $\mathcal{H}_{A}$) that has fallen into a black hole.

Is the GUE a good model for describing thermalization? For the GUE, we found $\langle A(0) B(t) \rangle \simeq \mathcal{R}_{2}(t)/L^2$ for non-identity Pauli operators with $AB=I$. Since the early time behavior of $\mathcal{R}_{2}(t)$ factorizes and is given by
\begin{align}
\langle A(0) A^{\dagger}(t) \rangle_{\text{GUE}} \simeq \frac{J_{1}(2t)^2}{t^2}\,,
\end{align}
the time scale for the decay of $2$-point correlation functions, denoted by $t_{2}$, is $\mathcal{O}(1)$. This is consistent with our intuition from thermalization in strongly coupled systems where $t_{2}\simeq \beta$. As such, quantum information appears to be lost in $\mathcal{O}(1)$ time for local observers in systems governed by GUE Hamiltonians.

Next, let us consider $4$-point OTOCs and scrambling. To recap the relation between OTOCs and scrambling in the context of black hole physics, consider a scenario where Alice has thrown an unknown quantum state into a black hole and Bob attempts to reconstruct Alice's quantum state by collecting the Hawking radiation. Hayden and Preskill added an interesting twist to this classic setting of black hole information problem by assuming that the black hole has already emitted half of its contents and Bob has collected and stored early radiation in some quantum memory he possesses. The surprising result by Hayden and Preskill is that, if time evolution $U=e^{-iHt}$ is approximated by a Haar random unitary operator, then Bob is able to reconstruct Alice's quantum state by collecting only a few Hawking quanta~\cite{HaydenPreskill}. This mysterious phenomenon, where a black hole reflects a quantum information like a mirror, relies on scrambling of quantum information where Alice's input quantum information is delocalized over the whole system~\cite{ChaosChannels}. The definition of scrambling can be made precise and quantitative by using quantum information theoretic quantities as briefly reviewed in App.~\ref{app:QIreview} and App.~\ref{app:QI}.

The scrambling of quantum information can be probed by the decay of $4$-point OTOCs of the form $\langle A(0) B(t) A^{\dagger}(0) B^{\dagger}(t)\rangle$ where $A,B$ are some local unitary operators. An intuition is that an initially local operator $B(0)$ grows into some non-local operator under time evolution via conjugation by $e^{-iHt}$, and OTOCs measure how non-locally $B(t)$ has spread. For this reason, the time scale $t_{4}$ when OTOCs start decaying is called the scrambling time.

Having reviewed the concepts of scrambling and OTOCs, let us study scrambling in random matrices. For the GUE, we found $\langle A(0) B(t) C(0) D(t)\rangle \simeq \mathcal{R}_{4}(t)/L^4$ for non-identity Pauli operators with $ABCD=I$. Since one can approximate $\mathcal{R}_{4}$ as $\mathcal{R}_{4}(t)\simeq \mathcal{R}_{2}(t)^2$ at early times, we obtain
\begin{align}
\langle A(0) B(t) C(0) D(t)\rangle_{\text{GUE}} \simeq \frac{J_{1}(2t)^4}{t^4}\,.
\end{align}
This implies that the decay time scale of $4$-point OTOCs is $t_{4}\simeq \frac{1}{2}t_{2}$, which is $\mathcal{O}(1)$ and is faster than the decay time of $2$-point correlation functions. This behavior is in strong contrast with behaviors in chaotic systems studied in the context of black hole physics. Namely, in holographic large-$N$ CFTs with classical gravity duals, the decay times are
\begin{align}
t_{2}\simeq \beta\,, \qquad t_{4}\simeq \beta \log N^2
\end{align}
with $t_{4}\gg t_{2}$. Also, the scrambling time $t_{4}\sim \mathcal{O}(1)$ violates a bound on quantum signalling which would hold for quantum systems with local interactions~\cite{HaydenPreskill, FastScrambling}. The pathology can be also seen from the viewpoint of black hole information problems. If black hole dynamics is modeled by the time evolution of some Hamiltonian sampled from GUE random matrices, then the scrambling time for OTOC decay is $\mathcal{O}(1)$. So Bob might be able to reconstruct Alice's quantum state in $\mathcal{O}(1)$ time. If Bob jumps into the black hole after decoding Alice's quantum state, Alice can send a quantum message with $\mathcal{O}(1)$ energy to Bob and verify the quantum cloning.

Another difference between GUEs and actual chaotic systems can be seen from the behaviors of correlators of the form $\langle A(0)B(t)C(2t)D(t) \rangle$. In the previous subsection, we showed that $\langle A(0)B(t)C(2t)D(t) \rangle\simeq \langle A(0)B(t)C(0)D(t) \rangle$. In strongly chaotic large-$N$ systems, we expect the following behaviors~\cite{Kitaev15, MSSbound}:
\begin{align}
\langle A(0)B(t)A(0)B(t) \rangle &= 1 - \frac{1}{N} \, e^{\lambda t}\,,\qquad \qquad \, \beta \ll t\ll \beta \log N.
\\
\langle A(0)B(t)C(2t)B(t) \rangle &= \langle A\rangle \langle B\rangle \langle C \rangle \langle B \rangle\,, \qquad t\simeq \beta.
\end{align}
Thus these two types of correlators should behave in a rather different manner.

These discrepancies clearly highlight the failure of GUE to capture early-time quantum chaos behavior which is present in realistic strongly-coupled systems. What was wrong about random matrices? Recent developments from black hole physics teach us that the butterfly effect in chaotic systems stems from delocalization of quantum information where initially local operators grow into non-local operators. However, for the GUE, the system does not distinguish local and non-local operators. To be concrete, let $A_{\text{local}}$ be some one-qubit Pauli operator, and $A_{\text{non-local}}=UA_{\text{local}}U^{\dagger}$ be some non-local operator created by conjugating $A_{\text{local}}$ via some non-local unitary $U$. Due to the Haar invariance of the GUE measure, we have
\begin{align}
\langle  A_{\text{local}}(0) A_{\text{local}}(t) \rangle_{\text{GUE}} = \langle  A_{\text{non-local}}(0) A_{\text{non-local}}(t) \rangle_{\text{GUE}}\,.
\end{align}
As this argument suggests, the GUE is a good description of quantum systems which have no notion of locality. After the scrambling time, we expect that an initially local operator $A_{\text{local}}(0)$ will time evolve to $A_{\text{local}}(t)$ which has support on the whole system, and the notion of locality is lost (or at least obfuscated) after the scrambling time. We thus expect that $\langle A_{\text{local}}(0) A_{\text{local}}(t) \rangle_{\text{GUE}}$ will be a good description of two-point correlation functions after the scrambling time. Similarly, the GUE does not distinguish time-ordering as seen from $\langle A(0)B(t)C(2t)D(t) \rangle\simeq \langle A(0)B(t)C(0)D(t) \rangle$. This implies that, at late time scales when the GUE becomes a good description, the system forgets the locality of time. In this sense, the GUE captures physics of quantum chaos after the locality of spacetime is forgotten. We will elaborate on this issue in Sec.~\ref{sec:haarinv}.

\section{Frame potentials and random matrices}
\label{sec:FPRMT}
In discussions of black hole information loss, we often approximate the chaotic internal dynamics of a black hole as evolution by a Haar random unitary \cite{HaydenPreskill, SSbutterfly}, and talk about typical black hole states as random pure states generated by Haar unitaries \cite{Page93}. While it is impractical to generate a Haar random unitary operator -- due to its exponential quantum circuit complexity, as noted by \cite{HaydenPreskill} -- it often suffices to sample from an ensemble that only reproduces the first few moments of the Haar ensemble. \cite{ChaosDesign} made significant progress in quantifying chaos in OTOCs by relating the late-time decay of $2k$-point OTOCs to the $k$-th frame potential, measuring the distance to Haar-randomness.\footnote{Also of interest, \cite{Liu17} recently discussed scrambling and randomness and showed that the R\'enyi $k$-entropies averaged $k$-designs are typically near maximal.}

One efficient way of generating a unitary $k$-design is to employ random local quantum circuits where one applies random two-qubit unitary gates at each unit time~\cite{Dankert09, HaydenPreskill, Brandao12} and the ensemble monotonically becomes a $k$-design as time evolves. Motivated by tensor network descriptions of the AdS/CFT correspondence~\cite{HPPY15, Hayden16}, random local quantum circuits have been used as a toy model of the Einstein-Rosen bridge and the dynamics of the two-sided AdS black hole \cite{ChaosChannels}. While such toy models are successful in capturing key qualitative features such as fast scrambling and complexity growth, their dynamics is not invariant under time translations.
A natural question is to ask if systems of time-independent Hamiltonians are able to form $k$-designs or not.

In this section we study time-evolution by the ensemble of GUE Hamiltonians and quantify its approach to Haar-randomness by asking when it forms a unitary $k$-design. We consider the ensemble of unitary time evolutions at a fixed time $t$, with Hamiltonians drawn from the GUE
\begin{equation}
\CE_t^{\rm GUE} = \big\lbrace e^{-iHt}, ~{\rm for}~ H\in {\rm GUE} \big\rbrace\,.
\end{equation}
As the frame potential quantifies the ensemble's ability to reproduce Haar moments, \ie form a $k$-design, we will be interested in the time scales at which we approach ``Haar values." Making use of the spectral form factors computed for the GUE, we derive explicit expressions for the frame potentials and extract the key time scales. We find that the GUE ensemble forms an approximate $k$-design after some time scales, but then deviates from being a $k$-design.

\subsection{Overview of QI machinery}

We begin by introducing the formalism of unitary $k$-designs and defining the frame potential. Consider a finite dimensional Hilbert space $\CH$ of dimension $L$. In this paper we are primarily interested in ensembles of unitary operators $\CE = \{p_i, U_i\}$, where the unitary $U_i$ appears with some probability $p_i$. A familiar ensemble might be the Haar ensemble. The Haar ensemble is the unique left and right invariant measure on the unitary group $U(L)$, where
\begin{equation}
\int_{\rm Haar} dU = 1\,, \quad \int_{\rm Haar}  dU\, f(U) = \int_{\rm Haar}  dU\, f(VU) = \int_{\rm Haar}  dU\, f(UV)\,,
\end{equation}
for some function $f$ and for all $V\in U(L)$. Taking $k$ copies of $\CH$, we can consider an operator $O$ acting on $\CH^{\otimes k}$, \ie $O \in \CA(\CH^{\otimes k})$ the algebra of operators on the Hilbert space. The $k$-fold channel of $O$ with respect to Haar is\footnote{The $k$-fold channel of $O$ is also referred to in the literature as the $k$-fold twirl of $O$.}
\begin{equation}
\Phi_{\rm Haar}^{(k)} (O) \equiv \int_{\rm Haar} dU\, (U^{\otimes k})^\dagger O U^{\otimes k}\,.
\end{equation}

Given an ensemble of unitary operators $\CE = \{p_i, U_i\}$, we might ask how Haar-random it is. More specifically, we should ask to what extent our ensemble reproduces the first $k$ moments of the Haar ensemble, a notion quantified by unitary $k$-designs.\footnote{Note that in the quantum information literature, these are often referred to as unitary $t$-designs. But here $t$ will always denote time.}  The $k$-fold channel with respect to the ensemble $\CE$ is
\begin{equation}
\Phi^{(k)}_\CE (O) \equiv \int_{U\in \CE} dU  (U^{\otimes k})^\dagger O U^{\otimes k}\,,
\end{equation}
written here for a continuous ensemble. We say that an ensemble $\CE$ is a unitary $k$-design if and only if
\begin{equation}
\Phi^{(k)}_\CE (O) = \Phi^{(k)}_{\rm Haar} (O) \,,
\end{equation}
meaning we reproduce the first $k$ moments of the Haar ensemble. But it does not make sense to compute the $k$-fold channels and check this equality for all operators in the algebra. Thus, we want a quantity which measures how close our ensemble is to being Haar-random. The frame potential, defined with respect to an ensemble as~\cite{Scott08}
\begin{equation}
\CF^{(k)}_\CE = \int_{U,V\in \CE}dU dV\, \big| \Tr (U^\dagger V)\big|^{2k}\,,
\label{eq:FP}
\end{equation}
measures Haar-randomness in the sense that is tells us how close the ensemble is to forming a unitary $k$-design. More precisely, it measures the 2-norm distance between the $k$-fold channel $\Phi^{(k)}_\CE$ with respect to the ensemble $\CE$, and the $k$-fold twirl $\Phi^{(k)}_{\rm Haar}$ with respect to the Haar ensemble. The frame potential will be a central object of study in this section.

The $k$-th frame potential for the Haar ensemble is given by
\begin{equation}
\CF^{(k)}_{\rm Haar} = k! \for k\leq L\,.
\end{equation}
Furthermore, for any ensemble $\CE$ of unitaries, the frame potential is lower bounded by the Haar value
\begin{equation}
\CF^{(k)}_{\rm \CE} \geq \CF^{(k)}_{\rm Haar}\,,
\end{equation}
with equality if and only if $\CE$ is a $k$-design. In particular, the deviation from the Haar value $\CF^{(k)}_{\rm \CE}-\CF^{(k)}_{\rm Haar}$ corresponds to the $2$-norm distance of $2$-fold quantum channels. The notion of an approximate $k$-design is reviewed in App.~\ref{app:QIreview}.

We will also need to compute moments of the Haar ensemble, \ie the ability to integrate monomials of Haar random unitaries. The exact formula \cite{Collins02, Collins04} for evaluating these moments is given by
\begin{equation}
\int dU\, U_{k_1}^{j_1} \ldots U_{k_n}^{j_n} U^\dagger{}^{\ell_1}_{m_1} \ldots U^\dagger{}^{\ell_n}_{m_n} = \sum_{\sigma, \tau \in S_n} \delta^{j_1}_{m_{\sigma(1)}}\ldots \delta^{j_n}_{m_{\sigma(n)}} \delta^{\ell_1}_{k_{\tau(1)}}\ldots \delta^{\ell_n}_{k_{\tau(n)}} \Wg ( \tau \sigma^{-1})\,,
\label{eq:Haarint}
\end{equation}
where, for the $n$-th moment, we sum over cycles of the permutation group $S_n$. The {\it Weingarten function} $\Wg$, a function of cycles $\sigma\in S_n$, is defined in App.~\ref{app:wein}. Performing Haar integrals then simply amounts to contracting indices and computing the Weingarten functions.

\subsection{Frame potentials for the GUE}
\begin{figure}[htb!]
\centering
\includegraphics[width=0.45\linewidth]{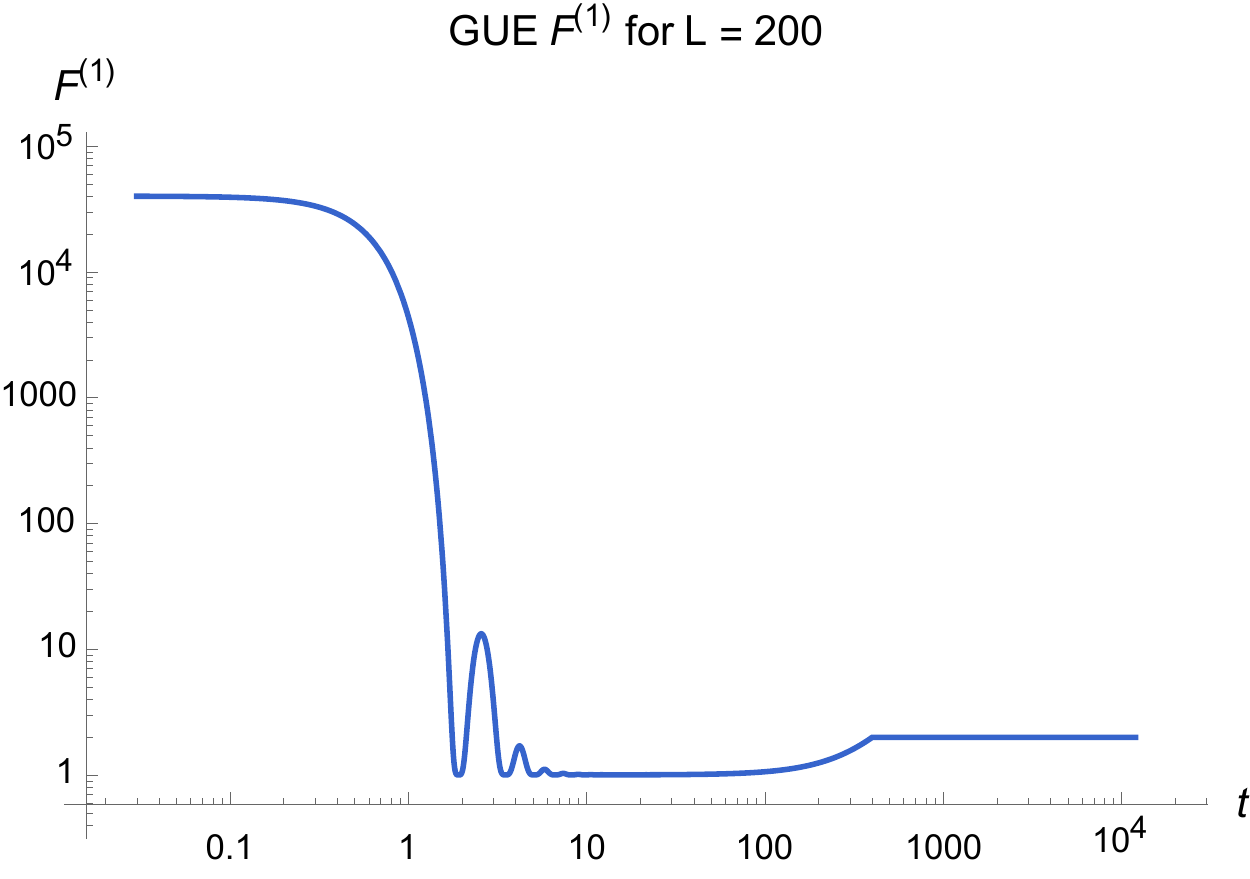}~~
\includegraphics[width=0.45\linewidth]{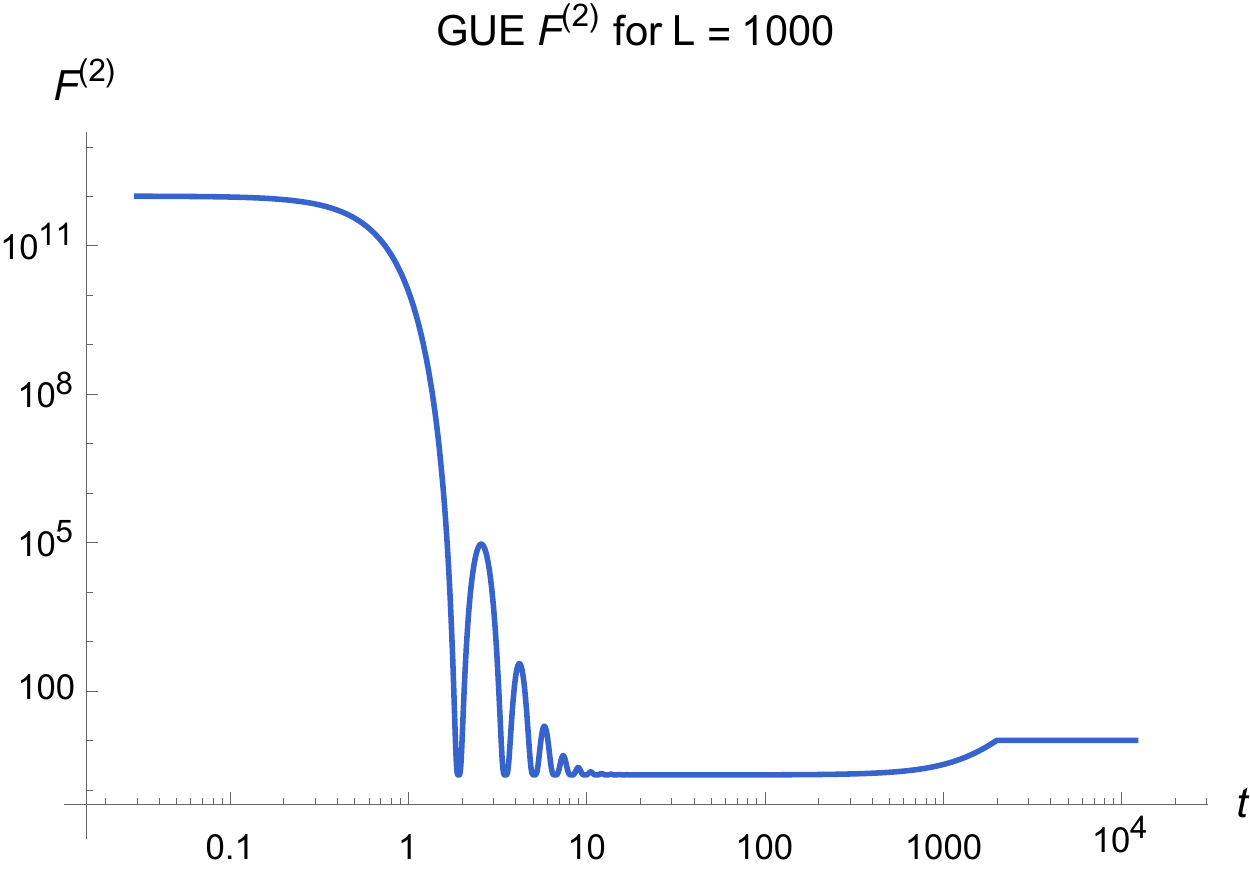}
\caption{The first and second frame potentials for the GUE, using the infinite temperature 2-point and 4-point form factors computed in Sec.~\ref{sec:FFRMT}, plotted for $L=200$ and $L=1000$, respectively. We observe the decay to the Haar value at the dip time and a subsequent rise at late times.
}
\label{fig:GUEFP}
\end{figure}

{\bf $k=1$ frame potential}\\
The first frame potential for the GUE is written as
\begin{equation}
\CF_{\rm GUE}^{(1)} = \int dH_1dH_2\, e^{-\frac{L}{2} \Tr H_1^2} e^{-\frac{L}{2} \Tr H_2^2} \Big| \Tr \big( e^{i H_1 t} e^{-iH_2 t}\big) \Big|^2\,.
\end{equation}
Noting that the GUE measure is invariant under unitary conjugation, we find
\begin{equation}
\CF_{\rm GUE}^{(1)} = \int_{\rm Haar} dU dV\int dH_1dH_2\, e^{-\frac{L}{2} \Tr H_1^2} e^{-\frac{L}{2} \Tr H_2^2} \Big| \Tr \big( U^\dagger \Lambda_1^\dagger U V^\dagger \Lambda_2 V\big) \Big|^2\,,
\end{equation}
where we define $\Lambda \equiv U e^{-i H t} U^\dagger$, \ie the matrix exponential of the GUE matrix in the diagonal basis. Going into the eigenvalue basis, we can express the GUE integral as
\begin{equation}
\CF_{\rm GUE}^{(1)} = \int D \lambda_1 D\lambda_2 \int dU \, \Tr \big( U^\dagger \Lambda_1^\dagger U \Lambda_2\big) \Tr \big( \Lambda_2^\dagger U^\dagger \Lambda_1 U \big)\,,
\end{equation}
where we have used the left and right invariance of the Haar measure to write the expression as a single Haar integral. Written out explicitly with indices,
\begin{equation}
\CF_{\rm GUE}^{(1)} = \int D \lambda_1 D\lambda_2 \int dU \, \Big( U_{k_1}^{j_1} U_{k_2}^{j_2} U^\dagger{}^{\ell_1}_{m_1} U^\dagger{}^{\ell_2}_{m_2}\, \Lambda_1^\dagger{}^{m_1}_{j_1} \Lambda_2{}^{k_1}_{\ell_1} \Lambda^\dagger_2{}^{k_2}_{\ell_2} \Lambda_1{}^{m_2}_{j_2} \Big)\,,
\end{equation}
and we can do the Haar integral using the second moment
\begin{align}
\int dU\, U_{k_1}^{j_1} U_{k_2}^{j_2} U^\dagger{}^{\ell_1}_{m_1} U^\dagger{}^{\ell_2}_{m_2}
= \frac{1}{L^2-1} \Big( &\delta_{m_1}^{j_1} \delta_{m_2}^{j_2} \delta_{k_1}^{\ell_1} \delta_{k_2}^{\ell_2} +\delta_{m_2}^{j_1} \delta_{m_1}^{j_2} \delta_{k_2}^{\ell_1} \delta_{k_1}^{\ell_2}\nn
&-\frac{1}{L} \delta_{m_1}^{j_1} \delta_{m_2}^{j_2} \delta_{k_2}^{\ell_1} \delta_{k_1}^{\ell_2} -\frac{1}{L} \delta_{m_2}^{j_1} \delta_{m_1}^{j_2} \delta_{k_1}^{\ell_1} \delta_{k_2}^{\ell_2} \Big)\,.
\end{align}
We find
\begin{equation}
\CF_{\rm GUE}^{(1)} = \int D\lambda_1 D\lambda_2\, \frac{1}{L^2-1} \bigg( \Tr \Lambda_1^\dagger \Tr\Lambda_1 \Tr \Lambda_2^\dagger \Tr \Lambda_2 + L^2 - \frac{1}{L} \Big( L\Tr \Lambda_1^\dagger \Tr\Lambda_1 + L\Tr \Lambda_2^\dagger \Tr \Lambda_2\Big) \bigg)\nonumber
\end{equation}
or equivalently
\begin{equation}
\CF_{\rm GUE}^{(1)} = \frac{1}{L^2-1} \Big( \CR_2^2 +L^2 - 2\CR_2\big)\,,
\label{eq:GUE_FP1}
\end{equation}
written in terms of the $2$-point form factor
\begin{equation}
\CR_2 = \int D\lambda \sum_{i,j} e^{i(\lambda_i-\lambda_j) t}\,.
\end{equation}
We know from the expression found in Sec.~\ref{sec:FFRMT}, that at early times $\CR_2 \sim L^2$, so the early time behavior of the frame potential is dominated by the $\CR_2^2$ term until near the dip time. At the dip time, $\CR_2 \approx \sqrt{L}$ and $\CF_{\rm GUE}^{(1)} \approx 1$, achieving the Haar value and forming a 1-design. At late times $t\ra\infty$, we take the late time limit of $\CR_2$ where only the $\delta_{ij}$ terms contribute, and find $\CR_2\approx L$, meaning that the first frame potential $\CF_{\rm GUE}^{(1)} \approx 2$ or double the Haar value. The first frame potential is plotted in Fig.~\ref{fig:GUEFP}.

A common intuition is that physical systems will become more and more uniformly random as time goes passes. Then one might expect that the frame potential, a measure of Haar randomness, would be a monotonically decreasing function with time. While it is monotonic for random local quantum circuits, we found that it is not generically monotonic for ensembles of unitaries generated by fixed Hamiltonians.\footnote{Frame potentials monotonically decrease in local random circuits and Brownian circuits~\cite{Dankert09, FastScrambling} where the time evolution is Markovian in the sense that the system samples different Hamiltonians, or infinitesimal time evolution operators, at random at each time step. In Markovian ensembles, spectral form factors are monotonically decreasing, and there is no ramp behavior. If the ensemble $\mathcal{E}$ is generated by a Markovian process and is invariant under complex transposition $\mathcal{E}=\mathcal{E}^{\dagger}$, then we have $\mathcal{F}^{(k)}(t)= \mathcal{R}_{2k}(2t)$.} In Sec.~\ref{sec:haarinv}, we propose an alternative quantity which may be monotonic at late times.

\vspace*{8pt}
\noindent {\bf $k=2$ frame potential}\\
We can similarly compute the second frame potential using the unitary invariance of the GUE measure:
\begin{align}
\CF_{\rm GUE}^{(2)} &= \int dH_1 dH_2\, e^{-\frac{L}{2} \Tr H_1^2} e^{-\frac{L}{2} \Tr H_2^2} \left| \Tr \left( e^{i H_1 t} e^{-i H_2 t} \right) \right|^{4}\\
&= \int D\lambda_1 D\lambda_2 \int dU\, \Tr \Big( U^\dagger \Lambda_1^\dagger U \Lambda_2 \Big) \Tr \Big( \Lambda_2^\dagger U^\dagger \Lambda_1 U \Big)\Tr \Big( U^\dagger \Lambda_1^\dagger U \Lambda_2 \Big) \Tr \Big( \Lambda_2^\dagger U^\dagger \Lambda_1 U \Big)\,, \nonumber
\end{align}
where again, $\Lambda $ is the exponentiated diagonal matrix. The fourth moment of the Haar ensemble that appears here generates $4!^2=576$ terms. Recalling Eq.~\eqref{eq:Haarint}, we can compute the fourth moment by computing the necessary Weingarten functions and summing over $\delta$-function contractions.

We relegate the presentation of the full expression for the $k=2$ frame potential, and the definitions of the spectral quantities on which it depends, to Appendix \ref{app:FPs}. While $\CF_{\rm GUE}^{(2)}$ depends on a number of spectral form factors, the dominant and interesting behavior is entirely captured by the $2$-point and $4$-point spectral form factors. At early times, the dominant contribution is
\begin{equation}
{\rm Early}: \quad \CF_{\rm GUE}^{(2)} \approx \frac{\CR_4^2}{L^4}\,.
\end{equation}
As we approach the dip time, the spectral quantities in the second frame potential,
\begin{align}
\CF_{\rm GUE}^{(2)} \approx 2 + \frac{\CR_4^2}{L^4} -\frac{8 \CR_4^2}{L^6} +\frac{6 \CR_4^2}{L^8} -\frac{36 \CR_2^2}{L^4} + \frac{4 \CR_2^2}{L^2}  +\frac{64 \CR_2 \CR_4}{L^6}-\frac{8 \CR_2 \CR_4}{L^4} + \ldots\,,
\label{eq:FP2dip}
\end{align}
are suppressed. From the calculation in Sec.~\ref{sec:FFRMT}, we have $\CR_2\approx \sqrt{L}$ and $\CR_4\approx L$ at the dip, meaning all terms are suppressed, with the exception of the leading constant. Thus, at the dip time, the $\mathcal{E}_t^{\text{GUE}}$ achieves the Haar value $\CF^{(2)}_{\rm Haar} \approx 2$ and forms an approximate unitary $2$-design.

At late times, in the infinite time average, we know that $\CR_2\ra L$, and $\CR_4 \ra 2L^2-L$ from the two eigenvalue pairings in the sum where the exponent vanishes, \ie $\delta_{ik}\delta_{j\ell}$ and  $\delta_{i\ell}\delta_{jk}$, and accounting for the $i=j=k=\ell$ terms. This tells us that the only terms that survive at late times, and are not suppressed in $L$, are
\begin{equation}
{\rm Late}:\quad \CF_{\rm GUE}^{(2)} \approx 2 + \frac{\CR_4^2}{L^4} + \frac{4 \CR_2^2}{L^2} \,,
\end{equation}
which gives us $\CF_{\rm GUE}^{(2)} \approx 10$, to leading order in $1/L$.

\subsection{Higher $k$ frame potentials}
Let us review what we have discussed so far.\\

\ni {\bf $k=1$ Frame Potential}\\
We computed the first frame potential for the GUE to be
\begin{equation}
\CF_{\rm GUE}^{(1)} = \frac{1}{L^2-1} \Big( \CR_2^2 +L^2 - 2\CR_2\big) \approx 1+\frac{\CR_2^2}{L^2} - \frac{2\CR_2}{L^2}
\end{equation}
for large $L$. In the late time limit, where $t\ra\infty$, we have that $\CR_2 \ra L$, and the late time behavior goes like $\CF_{\rm GUE}^{(1)} \sim 1+\CR_2^2/L^2 $, and $\CF_{\rm GUE}^{(1)} \ra 2$ or double the Haar value.
\begin{equation}
{\rm Early:}~ \CF_{\rm GUE}^{(1)} \approx \frac{\CR_2^2}{L^2} \,, \quad {\rm Dip:}~ \CF_{\rm GUE}^{(1)} \approx 1\,, \quad {\rm Late:}~ \CF_{\rm GUE}^{(1)} \approx 2\,.
\end{equation}

\ni {\bf $k=2$ Frame Potential}\\
We discussed the early and dip behaviors above. The terms unsuppressed at late times are
\begin{equation}
\CF_{\rm GUE, ~late}^{(2)} \approx 2 + \frac{\CR_4^2}{L^4} + \frac{4 \CR_2^2}{L^2}\,.
\end{equation}
Since $\CR_2\ra L$ and $\CR_4 \ra 2L^2-L$ in the late time limit, $\CF_{\rm GUE}^{(2)}$ approaches 10.
\begin{equation}
{\rm Early:}~ \CF_{\rm GUE}^{(2)} \approx \frac{\CR_4^2}{L^4} \,, \quad {\rm Dip:}~ \CF_{\rm GUE}^{(2)} \approx 2\,, \quad {\rm Late:}~ \CF_{\rm GUE}^{(2)} \approx 10\,.
\end{equation}

\ni {\bf $k=3$ Frame Potential}\\
The full expression for the third frame potential is given in App.~\ref{app:FPs}. The leading order behavior at early times is $\CR_6^2/L^6$, and at the dip time, the third frame potential approaches its Haar value. Again, the late time behavior above is better understood by looking at the dominant form factors. At late times, the terms that contribute at zeroth order in $L$ are
\begin{equation}
\CF_{\rm GUE, ~late}^{(3)} \approx 6+\frac{\CR_6^2}{L^6}+\frac{9\CR_4^2}{L^4} + \frac{18 \CR_2^2}{L^2} \ra 96\,,
\end{equation}
as $\CR_2\ra L$, $\CR_4\ra 2L^2$, and $\CR_6 \ra 6 L^3$ to leading order in $L$. In summary,
\begin{equation}
{\rm Early:}~ \CF_{\rm GUE}^{(3)} \approx \frac{\CR_6^2}{L^6} \,, \quad {\rm Dip:}~ \CF_{\rm GUE}^{(3)} \approx 6\,, \quad {\rm Late:}~ \CF_{\rm GUE}^{(3)} \approx 96\,.
\end{equation}

\ni {\bf $k=4$ Frame Potential}\\
It is not tractable to compute the $k=4$ frame potential, as the Haar integrals involved (the eighth moment of the Haar ensemble), generate $(8!)^2\sim 1.6$ billion terms. But the interesting behavior can be understood from the dominant terms at leading order in $L$ at different time scales. Recall that the $2k$-th moment of the Haar ensemble can be written as the sum of $\delta$-functions and the Weingarten function $\Wg$ (defined in App.~\ref{app:wein}) over elements of the permutation group $S_{2k}$. At large $L$, the Weingarten functions go as \cite{Weingarten78, Collins04}
\begin{equation}
\Wg(\sigma) \sim \frac{1}{L^{4k-\#{\rm cycles}}}\,,
\end{equation}
where `$\#{\rm cycles}$' denotes the number of cycles in the permutation $\sigma$. The Weingarten function contributing at leading order in $1/L$ is the one labeled by the partitioning of $2k$ into ones, \ie the trivial permutation of $S_{2k}$, which contributes as
\begin{equation}
\CW(\{1,1,\ldots\}) \sim \frac{1}{L^{2k}}\,.
\end{equation}
All other Weingarten functions, labeled by the integer partitions of $2k$, contribute at subleading order at early and late times. Thus, instead of computing the full fourth frame potential, we can compute the terms of combinations of spectral functions with this Weingarten function as their coefficient. In the sum over elements of the permutation group $\sigma,\tau\in S_{2k}$, we simply need the terms where $\tau\sigma^{-1}$ is the trivial permutation, \ie $\tau = \sigma$. Computing this we find the dominant contribution to the $k=4$ frame potential, at leading order in $1/L$. The full expression is still too large to reproduce here, but we can comment on the relevant features. The early time behavior is
\begin{equation}
\CF_{\rm GUE, ~early}^{(4)} \approx \frac{\CR_8^2}{L^8}\,.
\end{equation}
At the dip, where $\CR_n\sim L^{n/2}$, all terms are suppressed, leaving only the constant Haar value $24$. Lastly, the late time behavior is
\begin{equation}
\CF_{\rm GUE, ~late}^{(4)} \approx 24+\frac{\CR_8^2}{L^8}+\frac{16\CR_6^2}{L^8}+\frac{72\CR_4^2}{L^4} + \frac{96 \CR_2^2}{L^2} \rightarrow 1560\,,
\end{equation}
In summary,
\begin{equation}
{\rm Early:}~ \CF_{\rm GUE}^{(4)} \approx \frac{\CR_8^2}{L^8} \,, \quad {\rm Dip:}~ \CF_{\rm GUE}^{(4)} \approx 24\,, \quad {\rm Late:}~ \CF_{\rm GUE}^{(4)} \approx 1560\,.
\end{equation}

\ni {\bf $k$-th Frame Potential}\\
We are now poised to discuss the general form of the $k$-th frame potential
\begin{equation}
{\rm Early:}~ \CF_{\rm GUE}^{(k)} \approx \frac{(\CR_{2k})^{2}}{L^{2k}} \,, \quad {\rm Dip:}~ \CF_{\rm GUE}^{(k)} \approx k!\,.
\end{equation}
We can also determine what the general late time value should look like. Above, we understood that the plateau value of the $k$-th frame potential is the sum of the Haar value and the contributions of the spectral functions. It was only the squares of the spectral functions that gave contributions which were not suppressed by $1/L$ at late times. Extrapolating from above, we expect the $k$-th frame potential to have
\begin{equation}
\CF_{\rm GUE, ~late}^{(k)} \approx {\rm Haar} + {\rm spectral~functions} \approx k! + \frac{\CR_{2k}^2}{L^{2k}}+ c_{1} \frac{\CR_{2k-2}^2}{L^{2k-2}}+\ldots+ c_{k-1}\frac{\CR_{2}^2}{L^{2}}\,,
\end{equation}
with coefficients $c_\ell$. Given the way the spectral form factors are generated from Haar integration, we can understand these coefficients as the number of partial bijections of a given length. For example, for $k=3$ there are 24 partial bijections on a 3 element set of length 2, \ie 24 nonclosed cycles of length two, which gives us 24 ways of constructing the $2$-point functions for $k=3$. More generally, the coefficients above can be written as
\begin{equation}
c_\ell(k) = {k\choose \ell}^2 \ell!\,,
\end{equation}
where for $k=4$, we have the coefficients 1, 16, 72, 96, 24. The $k$-th coefficient is the Haar value $c_k(k) = k!$, \ie the number of ways to construct 0-point functions in the Haar integration. We can then write down the general late time behavior for the $k$-th frame potential
\begin{equation}
\CF_{\rm GUE, ~late}^{(k)} \approx \sum_{\ell=0}^k c_\ell(k) \frac{\CR_{2(k-\ell)}^2}{L^{2(k-\ell)}}\,.
\end{equation}

Since the late time value of the $2k$-point spectral form factor is, to leading order in $L$, $\CR_{2k} = k! L^k$, the late time floor value for the $k$-th frame potential of the GUE is
\begin{equation}
\CF_{\rm GUE, ~late}^{(k)} \approx \sum_{\ell =0}^k  {k\choose \ell}^2 \ell! \big((k-\ell)!\big)^2 = \sum_{\ell=0}^k  \frac{k!^2}{\ell!} \,.
\end{equation}
where the first few terms of this sequence are 2, 10, 96, 1560.
\vspace*{12pt}

We emphasize that while the purpose of this section is to understand GUE Hamiltonians, the derivations in this subsection where we relate the frame potential to spectral $2k$-point functions only used the unitary invariance of the measure to proceed in doing the calculations by Haar integration. Thus, if we are handed an ensemble whose measure is unitarily invariant, the same relations hold.

\subsection{Frame potentials at finite temperature}
We now generalize the discussion of the frame potential to ensembles at finite temperature and compute the thermal frame potential for the GUE. Again we consider the ensemble of unitary time evolutions at a fixed time $t$, with $H$ drawn from an ensemble $\CE$. One might consider generalizing the frame potential to finite temperature by defining the frame potential with respect to a thermal density matrix $\rho_\beta = e^{-\beta H}/\Tr(e^{-\beta H})$, and taking thermal expectation values. With this in mind, we define the frame potential at finite temperature by taking the average over all thermal $2k$-point functions, with the operator insertions $A$ and $B$ spaced equidistant on the thermal circle
\begin{equation}
\vev{AB(t)\ldots AB(t)} = \Tr \big(( e^{-\beta H/2k} A e^{-\beta H/2k} B(t) \ldots e^{-\beta H/2k} A e^{-\beta H/2k} B(t)\big) / \Tr e^{-\beta H}\,.
\end{equation}
Averaging the norm-squared $2k$-point correlation function over all operators and then averaging over the ensemble, we find
\begin{equation}
\CF_{\CE_\beta}^{(k)} = \int dH_1 dH_2 \frac{\big| \Tr \big( e^{-(\beta/2k - it) H_1}  e^{-(\beta/2k + it) H_2} \big) \big|^{2k}}{\Tr(e^{-\beta H_1})  \Tr(e^{-\beta H_2})/L^2}\,.
\end{equation}
Note that this definition differs from the one in the Appendix of \cite{ChaosDesign} by a factor of $L^2$. With this slight change in normalization, we reduce to the usual frame potential $\CF_\CE^{(k)}$ at infinite temperature. \\

\ni {\bf $k=1$ Frame Potential}\\
Let us compute the first thermal frame potential for GUE Hamiltonians:
\begin{equation}
\CF_{{\rm GUE}}^{(1)}(t,\beta) = \int D\lambda_1 D\lambda_2 \int dU \frac{\big| \Tr \big(U^\dagger e^{-(\beta/2 - it) D_1} U e^{-(\beta/2 + it) D_2} \big) \big|^{2}}{\Tr(e^{-\beta H_1})  \Tr(e^{-\beta H_2})/L^2}\,.
\end{equation}
where we use the invariance of the GUE measure under unitary conjugation, diagonalize $H$ where $D$ is the diagonalized Hamiltonian, and use the left and right invariance of the Haar measure to write a single Haar integral. Doing the Haar integral, we find
\begin{equation}
\CF_{{\rm GUE}}^{(1)}(t,\beta) = \frac{1}{L^2-1} \Big( \widetilde \CR_{2}^2(t,\beta/2) + L^2 - 2 \widetilde \CR_{2}(t,\beta/2) \Big)\,,
\end{equation}
where we define
\begin{equation}
\widetilde \CR_{2}(t,\beta) \equiv \bigg\langle \frac{Z(t,\beta)Z^*(t,\beta)}{Z(2\beta)/L}\bigg\rangle_{\rm GUE} = \int D\lambda \,\frac{\sum_{ij} e^{it(\lambda_i-\lambda_j)}e^{-\beta(\lambda_i+\lambda_j)}}{\sum_{i} e^{-2\beta \lambda_i}/L}\,,
\label{eq:R2tilde}
\end{equation}
which is normalized such that we recover the infinite temperature form factor $\CR_2(t)$ when $\beta\ra 0$. This normalization differs from $\vev{|Z(t,\beta)|^2/Z(\beta)^2}$, which gives an initial value of one. Here the thermal form factor which naturally arises from the thermal frame potential has a late time value which is $\beta$-{\it independent}. The initial value of $\CR_{2}(t,\beta)$, and thus $\CF_{{\rm GUE}}^{(1)}(t,\beta) $, depends on the $\beta$. 

In stating the time scales for the thermal frame potential, we will work with the `quenched' version of Eq.~\eqref{eq:R2tilde} where the numerator and denominator are averaged separately. As we mentioned in Sec.~\ref{sec:FFs}, the `annealed' 2-point form factor is the correct object to consider, but we opt to work with the more analytically tractable quenched form factor. Numerically, the two functions are in close agreement with each other.

\subsection{Time scales from GUE form factors}
With an understanding of the behavior of the GUE spectral form factors from Sec.~\ref{sec:FFs}, we can now look at the time scales for the dip and plateau of the first frame potential
\begin{equation}
\CF^{(1)}_{\rm GUE} = \frac{1}{(L^2-1)}\big( \CR_2^2 + L^2 - 2\CR_2 \big) \,.
\end{equation}
At $t_d \approx \sqrt{L}$, when $\CR_2 \approx\sqrt{L}$, we reach the minimal Haar value of $1$, and at the plateau time $t_p = 2L$, when $\CR_2 = L$, we reach the late time value of 2.

There is another time scale at play here which is an artifact of working at infinite temperature. We might also ask what is the first time the form factor or frame potential reaches its minimal value. This time scale can be attributed to the first zero of the Bessel function, $J_1(2t)=0$ at $t\approx 1.92$, and is universal for all values of $L$. This is the first time at which the ensemble becomes a $1$-design. Something like the scrambling time, where the frame potential begins to deviate rapidly from its initial value, occurs at $\op(1)$ time.

Using the explicit expression for the GUE $4$-point form factor, we can also verify the expected time scales in the second frame potential $\CF^{(2)}_{\rm GUE}$. At the dip time, $t_d\approx \sqrt{L}$, we have that all the form factors appearing in the $\CF^{(2)}_{\rm GUE}$ are suppressed by powers of $L$, and thus the leading term is the Haar value, $\CF^{(2)}_{\rm GUE}(t_d) \approx 2$. Further, the plateau values of the spectral form factors $\CR_2$ and $\CR_4$ give us the late time value of $\CF^{(2)}_{\rm GUE}\approx 10$.

Lastly, we can extract the time scales and values of the finite temperature frame potential from our discussion of $\CR_2(t,\beta)$. The initial value of the first frame potential is
\begin{equation}
\CF^{(1)}_{\rm GUE}(t=0,\beta)=L^2 \frac{h_1(\beta/2)^4}{h_1(\beta)^2}\,,
\end{equation}
where $h_1(\beta) = J_1(2i\beta)/i\beta$. At the dip time, $t_d \approx h_2(\beta/2) \sqrt{L}$, the thermal form factor defined above $\widetilde \CR_2(t_d,\beta/2)\approx \sqrt{L} h_3(\beta/2)/h_1(\beta)$, with the functions defined in Sec.~\ref{sec:FFs}. For $\beta \ll L$, we have
\begin{equation}
\CF^{(1)}_{\rm GUE}(t_d,\beta) \approx 1\,.
\end{equation}
Finally, as we can see from time averaging Eq.~\eqref{eq:R2tilde}, at the plateau time 
\begin{equation}
\CF^{(1)}_{\rm GUE}(t_p,\beta) = 2\,,
\end{equation}
for any $\beta$, as the late time value of the thermal frame potential does not depend on the temperature.

Let us briefly comment on the dip value of the $k$-th frame potential at infinite temperature. As we discussed, at the dip time $t_d\approx\sqrt{L}$, the frame potentials reached the Haar value and form an approximate $k$-design for some $k$.  Determining the size of $k$ requires an understanding of the corrections to the dip value.  The leading order correction to the Haar value at the dip comes from $\CR_2^2/L^2 \sim 1/L$, the coefficient of which is $c_{k-1}(k) = k!\, k$. So at the dip time
\begin{equation}
\CF^{(k)}_{\rm GUE} (t_d) \approx k! \left( 1+ \frac{k}{L}\right) \,,
\end{equation}
meaning we form an approximate $k$-design for $k\ll L$.

The claim that the GUE forms a $k$-design at intermediate times but then deviates from this behavior at late times might at first seem surprising, but the late time behavior makes sense if we consider the dephasing of GUE eigenvalues in the $t\ra \infty$ limit. Under the exponential map $\lambda \ra e^{i\lambda t}$, the GUE eigenvalues are distributed around the circle and at early times will still be correlated and logarithmically repel. However, at late times the eigenvalues will spread uniformly around the circle. Moreover, explicitly computing the level density for the GUE under the exponential map and taking the long time limit, one finds that the density becomes constant and the eigenvalues are independently and uniformly distributed. Eigenvalue statistics of Haar random unitary operators can be characterized by the following well-known relation~\cite{Diaconis94}\footnote{If one views $t$ as a discrete time and $U$ as a time evolution in a unit time with a Hamiltonian $H=i\log U$, then the above equation mimics the late-time ramp and plateau behavior. }
\begin{align}
\int_{\text{Haar}}dU \,\text{tr}(U^t) \text{tr}({U^{\dagger}}^t)=t \qquad k\leq L \,.
\end{align}
If we suppose that the eigenvalue distribution of $U$ is random, then $\int dU \, \text{tr}(U^t) \text{tr}({U^{\dagger}}^t)$ would not depend on $t$. Therefore, the late-time eigenvalue statistics of unitaries generated by fixed GUE matrices is quite different from those of Haar unitaries, which have eigenvalue repulsion.

\section{Complexity and random matrices}
\label{sec:comp}
In recent years, the notion of quantum complexity has attracted significant attention in the study of quantum many-body systems~\cite{Bravyi06, Susskind14, CompAction}. By quantum complexity of a quantum state $|\psi\rangle$, we mean the minimal number of elementary local quantum gates necessary to (approximately) create $|\psi\rangle$ from a trivial product state with no entanglement. A similar characterization applies to the quantum complexity of unitary operators constructed from the identity operator. Quantum complexity provides deep insight into what kinds of physical operations are allowed (or prohibited) in a given physical system as states or operators of very large complexity cannot be prepared or implemented in a short period of time by the evolution of local Hamiltonians with finite energy density. Quantum complexity has also proven useful in condensed matter physics where topological phases of matter can be classified in terms of the quantum complexity of ground state wavefunctions~\cite{Chen10}. More recently, it was asked whether the AMPS thought experiment can be carried out in a physically reasonable amount of time and resources by considering the computational complexity of decoding the Hawking radiation~\cite{HarlowHayden}. In the past few years, quantum complexity has been considered in holography as a possible CFT observable\footnote{At least with respect to some subspace of states of the boundary CFT.} to study the late-time dynamics of the AdS black holes~\cite{Susskind14, CompAction}.

Despite all the promises of the usefulness of quantum complexity, a precise understanding of the growth of quantum complexity in quantum many-body systems, especially in AdS/CFT, continues to elude us. While it is possible to see a hint of complexity growth from entanglement dynamics at early times before the scrambling time,\footnote{For example, from the level-statistics of the entanglement spectrum \cite{Yang17}.} the late-time complexity growth remains difficult to observe as the extremal surfaces do not go through the interior of the black hole and entanglement entropies get saturated at late times. From a mathematical perspective, it is extremely challenging to compute the quantum gate complexity of a given quantum state $|\psi\rangle$ as one essentially needs to consider all the possible quantum circuits creating $|\psi\rangle$ and find the one with the minimal number of gates. Thus it would be valuable to have an analytical toy example of Hamiltonians whose dynamics indeed makes the quantum complexity of wavefunctions increase even after the scrambling time by providing a rigorous lower bound on quantum complexity.

Here, we present analysis of complexity growth of typical Hamiltonian time evolution by GUEs and show that quantum complexity indeed grows in time. A lower bound on a typical unitary operator in an ensemble $\mathcal{E}$ can be computed from a simple counting argument. Observe that short depth quantum circuits can prepare only a small number of unitary operators which occupy a tiny fraction of the whole space of unitary operators. The idea is that, if there are so many unitary operators in $\mathcal{E}$ which are sufficiently far apart and distinguishable, then most of operators in $\mathcal{E}$ cannot be created by a short depth circuit. Furthermore, it has been found that lower bounds on the number of distinguishable unitary operators in $\mathcal{E}$ can be obtained by frame potentials, a measure of randomness in $\mathcal{E}$. Although such a counting argument often gives a rather loose lower bound, it is still possible to obtain a rigorous complexity lower bound for a system of quantum many-body Hamiltonians. See~\cite{ChaosDesign} for a rigorous treatment and details.

To be concrete, let us consider a system of qubits where we pick a pair of qubits and apply an arbitrary two-qubit gate at each step. While the circuit complexity for generating an ensemble and the circuit complexity for generating a particular unitary in the ensemble are different, the former provides an approximate lower bound for the circuit complexity of typical unitary operators in the ensemble~\cite{ChaosDesign}.  We define the number of quantum gates necessary to create an ensemble $\mathcal{E}$ by a quantum gate complexity $\mathcal{C}_{\text{gate}}$. The lower bound on the quantum gate complexity is then given by
\begin{align}
\mathcal{C}_{\text{gate}} \geq \frac{2kn - \log_{2} \CF^{(k)}}{2\log(n)}\,,
\end{align}
up to some constant multiplicative factor. Let us consider the bound for small $k$. In Sec.~\ref{sec:FPRMT}, we found that $\CF^{(k)}$ drops to its minimal value $\sim k!$ at $t\sim \mathcal{O}(1)$ (the first zero of the Bessel function). We thus have
\begin{align}
\mathcal{C}_{\text{gate}}(t) \geq \frac{2kn- \log_{2} \frac{\mathcal{R}_{2k}^2(t)}{L^{2k}}  }{2\log(n)} \simeq \frac{4kn- \log_{2} \mathcal{R}_{2k}^2(t) }{2\log(n)} \simeq\frac{4k(n- \log_{2} \mathcal{R}_{1}(t)) }{2\log(n)}
\end{align}
up to the first dip time $t_{\text{dip}}\sim \mathcal{O}(1)$ where we have used an approximation $\mathcal{R}_{2k}\simeq (\mathcal{R}_{1})^{2k}$. Thus, at $t\sim \op(1)$, the following lower bound on the complexity is obtained:
\begin{align}
\mathcal{C}_{\text{gate}}(t_{\text{dip}}) \geq \mathcal{O}\left(\frac{kn}{\log(n)}\right).
\end{align}
Converting it into a quantum circuit complexity, we obtain
\begin{align}
\mathcal{C}_{\text{circuit}}(t_{\text{dip}}) \geq \mathcal{O}\left(\frac{k}{\log(n)}\right).
\end{align}
This lower bound should be valid as long as $k\sim \mathcal{O}(1)$. As we have discussed in Sec.~\ref{sec:FFRMT} and Sec.~\ref{sec:FPRMT}, the early-time oscillations of spectral form factors and frame potentials disappear at finite temperature. It would be then useful to consider the complexity lower bound based on envelope functions of form factors and frame potentials. Since the asymptotic behavior is given by $\mathcal{R}_{1}(t)\sim 1/t^{3/2}$, we would have
\begin{align}
\mathcal{C}_{\text{gate}}(\beta,t) \geq \mathcal{O}\left(\frac{k \log t }{\log(n)}\right)
\end{align}
where $\beta$ implies that we consider the asymptotic behaviors of the envelope. Thus, the quantum circuit complexity grows at least logarithmically in $t$ up to the thermal dip time.

While the above studies are able to provide rigorous lower bounds on quantum circuit complexity, the bounds are not meaningful when $k$ is small. To obtain a meaningful lower bound on quantum complexity, we need to evaluate the frame potential and form factor for large $k$. Analytically computing $\mathcal{R}_{2k}$ and $\mathcal{F}^{(k)}$ for large $k$ seems rather challenging. Instead, we employ a certain heuristic argument to derive the decay of $\mathcal{R}_{2k}$ and $\mathcal{F}^{(k)}$. Let us begin by recalling the early-time behavior of $1$-point form factor. The $1$-point form factor $\mathcal{R}_{1}(t)$ can be analytically written via a contour integral as follows~\cite{BrezinRMT}
\begin{align}
\mathcal{R}_{1}(t) = L e^{- \frac{t^2}{2L}} \oint
\frac{du}{2\pi i} \left(\frac{1}{-it} \right)\left( 1 - \frac{it}{Lu} \right)^L e^{-itu}.
\end{align}
For $L \rightarrow \infty$, the integral gives the Bessel function:
\begin{align}
 \oint
\frac{du}{2\pi i} \left(\frac{1}{-it} \right)\left( 1 - \frac{it}{Lu} \right)^L e^{-itu}\simeq \frac{J_{1}(2t)}{t}.
\end{align}
But $J_{1}(2t)\simeq t$ for $t \ll 1$, so we have
\begin{align}
\mathcal{R}_{1}(t) \simeq L e^{- \frac{t^2}{2L}}\frac{J_{1}(2t)}{t}
\end{align}
where the Gaussian decay is dominant for $t \ll 1$ while, for $1 \ll t \ll \sqrt{L}$, the Bessel function dominates the decay. In a similar manner, the $2k$-point form factor can be analytically written as
\begin{align}
\mathcal{R}_{2k}(t) = L^{2k} e^{- \frac{k t^2}{L}} \oint
\prod_{j=1}^{2k}\frac{du_{j}}{2\pi i} \left( 1 + (-1)^j \frac{it}{Lu_{j}} \right)^L e^{(-1)^j itu_{j}}\det \left(\frac{1}{u_{j}-u_{k}+(-1)^j it/L}\right)
\end{align}
where the sign of $\pm it$ depends on the index of $u_{i}$ and the integral part is equal to unity at $t=0$. In previous sections, we have neglected the Gaussian decay because our discussions were mostly centered on small $k$ spectral form factors. But, for large $k$, the Gaussian decay part is no longer negligible. Let us bound the form factor by using the Gaussian decay part only by neglecting the decay contribution from Bessel functions in the integral part:
\begin{align}
\mathcal{R}_{2k}(t) \leq  L^{2k} e^{- \frac{k t^2}{L}}.
\end{align}
While the validity of this inequality for large $k$ remains unclear, we assume its validity up to the dip time $\sim \sqrt{L}$ when ramp behavior kicks in. The notion of unitary $k$-design and its application to complexity would be meaningful only up to $k \sim \mathcal{O}(L)$ (see~\cite{ChaosDesign} for instance). By using this approximate bound for $k= c L$ with $c \sim \mathcal{O}(1)$, we will have
\begin{align}
\mathcal{F}^{(cL)} \lesssim  L^{2k} e^{- 2c t^2}
\end{align}
up to the dip time $\sim \sqrt{L}$. This leads to the following estimate of quantum complexity growth for the GUE:
\begin{align}
\mathcal{C}_{\text{gate}} \gtrsim \frac{c t^2 }{\log (n)}
\end{align}
which predicts a quadratic growth of quantum complexity.

Let us compare our estimate with predictions from the AdS/CFT correspondence. According to the conjecture that quantum complexity is proportional to the volume in the bulk, the early-time complexity (volume) growth is quadratic in time, and then becomes linear in time. Our analysis above suggests that the complexity growth for the GUE is (at least) quadratic in $t$ for a long time until very close to the saturation of quantum complexity $\sim L$. One may find that $t^2$ complexity growth is unphysical as the system has evolved only for time $t$. The point is that the GUE Hamiltonian is generically non-local and is comprised of $\mathcal{O}(n)$-body terms whereas we measure quantum complexity by using two-local quantum gates as building blocks.

\section{Characterization of Haar-invariance}
\label{sec:haarinv}

From the perspective of operator delocalization, it is clear why the GUE fails to characterize information scrambling and dynamics in local quantum systems at early times. Recall that the GUE is Haar-invariant, meaning
\begin{equation}
\int_{U \in \text{Haar}} dU \int_{H \in \text{GUE}} dH \, f(UHU^\dagger) = \int_{H \in \text{GUE}} dH \, f(H)
\end{equation}
where $U$ is integrated over the unitary group $U(L)$ and where $f(H)$ is an arbitrary function. As a consequence, a typical GUE Hamiltonian is non-local (or $\mathcal{O}(n)$-local), so local operators are delocalized essentially immediately. Indeed, the Haar-invariance of the GUE ensemble and non-locality of its Hamiltonians resulted in unusual behaviors of OTOCs whose decay time was shorter than that of $2$-point correlation functions. It thus appears that \textit{local} chaotic Hamiltonians and a typical Hamiltonian from a Haar-invariant ensemble behave in a dramatically different way.

However, previous studies on chaotic Hamiltonians suggest that at late times, Haar-invariant Hamiltonian ensembles, such as the GUE, GOE and GSE, capture behaviors of correlation functions remarkably well. This apparent tension between early time and late time behaviors may be resolved in the following manner. Initially, any ensemble of local Hamiltonians is not Haar-invariant because Hamiltonians are made of local terms. This can be clearly seen from the fact that the OTOC, $\langle A(0)B(t)A(0)B(t)\rangle$, behaves rather differently depending on the sizes of operators $A, B$. Yet, after the scrambling time when local operators become delocalized by Hamiltonian evolution, it becomes harder to tell whether the original operators $A(0),B(0)$ were local or not, and we expect that the unitary ensemble becomes `approximately' Haar-invariant.

With this observation in mind, we are naturally led to consider a fine-grained characterization of Haar-invariance which we shall call \textit{$k$-invariance}. Intuitively, $k$-invariance refers to an ensemble of unitary operators which appear to be Haar-invariant up to $k$-th moments. More precisely, let $\mathcal{E}$ be an ensemble of unitary operators. We define a Haar-invariant extension $\widetilde{\mathcal{E}}$ of this ensemble by:
\begin{align}
\int_{U \in \widetilde{\mathcal{E}}} dU = \int_{W \in \text{Haar}} dW\int_{U\in \mathcal{E}} d(WUW^{\dagger})\,.
\end{align}
From the construction, we can easily see $W\widetilde{\mathcal{E}} W^{\dagger}=\widetilde{\mathcal{E}}$ for any unitary operator $W$, and so the \emph{Haar'ed} ensemble is independent of any basis. Let us consider the $k$-fold twirl superoperator:
\begin{align}
\Phi^{(k)}_{\mathcal{E}}(\cdot) = \int_{U \in \mathcal{E}} dU \, U^{\otimes k}  (\cdot) {U^{\dagger}}^{\otimes k}\,.
\end{align}
Then, $\mathcal{E}$ is said to be $k$-invariant if and only if
\begin{align}
\Phi^{(k)}_{\mathcal{E}}(\cdot)  = \Phi^{(k)}_{\widetilde{\mathcal{E}}}(\cdot)\,.
\end{align}
An ensemble of unitaries is Haar-invariant if and only if it is $k$-invariant for all $k \geq 1$. Similar definitions apply to Haar-invariance with respect to orthogonal and symplectic groups. 

The utility of $k$-invariance can be seen from an explicit relation between correlation functions and spectral statistics. Recall that we have derived the following relation in the GUE by using the Haar-invariance of the GUE measure:
\begin{align}
\langle A_{1}(0)B_{1}(t)\ldots A_{k}(0)B_{k}(t) \rangle_{\text{GUE}}\simeq  \langle A_{1}B_{1}\ldots A_{k}B_{k} \rangle \frac{\mathcal{R}_{2k}(t)}{L^{2k}}\,.
\end{align}
It is clear that the same derivation applies to any ensemble which is $k$-invariant. The implication is that, after the \emph{$k$-invariance time}, the behavior of $2k$-point OTOCs can be completely determined by the spectral statistics alone. The physical significance of the $k$-invariance time is that it is the time scale when OTOCs behave in a similar way regardless of the locality or non-locality of the operators $A_{j},B_{j}$ (as well as their time-ordering). A similar conclusion holds for $k$-th frame potentials which can be written only in terms of spectral form factors for $k$-invariant ensembles. Thus, $k$-invariance and its associated time scale will be a useful notion to characterize the loss of locality from the perspective of $2k$-point OTOCs and the onset of random matrix behavior.

How can one verify that some ensemble $\mathcal{E}$ is $k$-invariant?
One formal approach is to use frame potentials. Let us define the following operator
\begin{align}
S = \int_{\mathcal{E}} dU U^{\otimes k}  \otimes {U^{\dagger}}^{\otimes k} - \int_{\widetilde{\mathcal{E}}} dU U^{\otimes k}  \otimes {U^{\dagger}}^{\otimes k}
\end{align}
which corresponds to the difference between tensor expanders from $\mathcal{E}$ and its Haar-invariant extension $\widetilde{\mathcal{E}}$. Then we have
\begin{equation}
\begin{split}
0 \leq \text{tr}(S^{\dagger}S) = &\int_{U,V\in\mathcal{E}} dUdV |\text{tr}(U^{\dagger}V)|^{2k} \\
&-   \int_{U,V\in\mathcal{E}} dUdV \int_{W \in \text{Haar}} dU\, |\text{tr}(U^{\dagger}WVW^{\dagger})|^{2k} \\
&-   \int_{U,V\in\mathcal{E}} dUdV \int_{W \in \text{Haar}} dU\, |\text{tr}(WU^{\dagger}W^{\dagger}V)|^{2k} \\
&+ \int_{U,V\in\mathcal{E}} dUdV \int_{W,Y \in \text{Haar}} dWdY\, |\text{tr}(WU^{\dagger}W^{\dagger}YVY^{\dagger})|^{2k}  \\
&= \CF^{(k)}_{\mathcal{E}} - \CF^{(k)}_{\widetilde{\mathcal{E}}}
\end{split}
\end{equation}
where $\CF^{(k)}_{\mathcal{E}}$ is the $k$-th frame potential for an ensemble $\mathcal{E}$. Here we used the fact that the Haar unitary ensemble is left and right invariant. Therefore, we arrive at the following inequality
\begin{align}
\CF^{(k)}_{\mathcal{E}} \geq  \CF^{(k)}_{\widetilde{\mathcal{E}}}
\end{align}
with equality if and only if $\mathcal{E}$ being $k$-invariant. The difference $\CF^{(k)}_{\mathcal{E}} - \CF^{(k)}_{\widetilde{\mathcal{E}}}$ measures the $2$-norm distance to being $k$-invariant.\footnote{For a more rigorous analysis, the diamond distance should be considered. While the diamond norm is difficult to compute in general, there are some examples of ensembles of realistic Hamiltonians where the diamond norm can be analytically computed. We hope to address this in a future publication.} The above derivation is a straightforward generalization of a method used in~\cite{Scott08}.

\subsubsection*{Haar-invariance in a spin system}
\begin{figure}[htb!]
\centering
\includegraphics[width=0.45\linewidth]{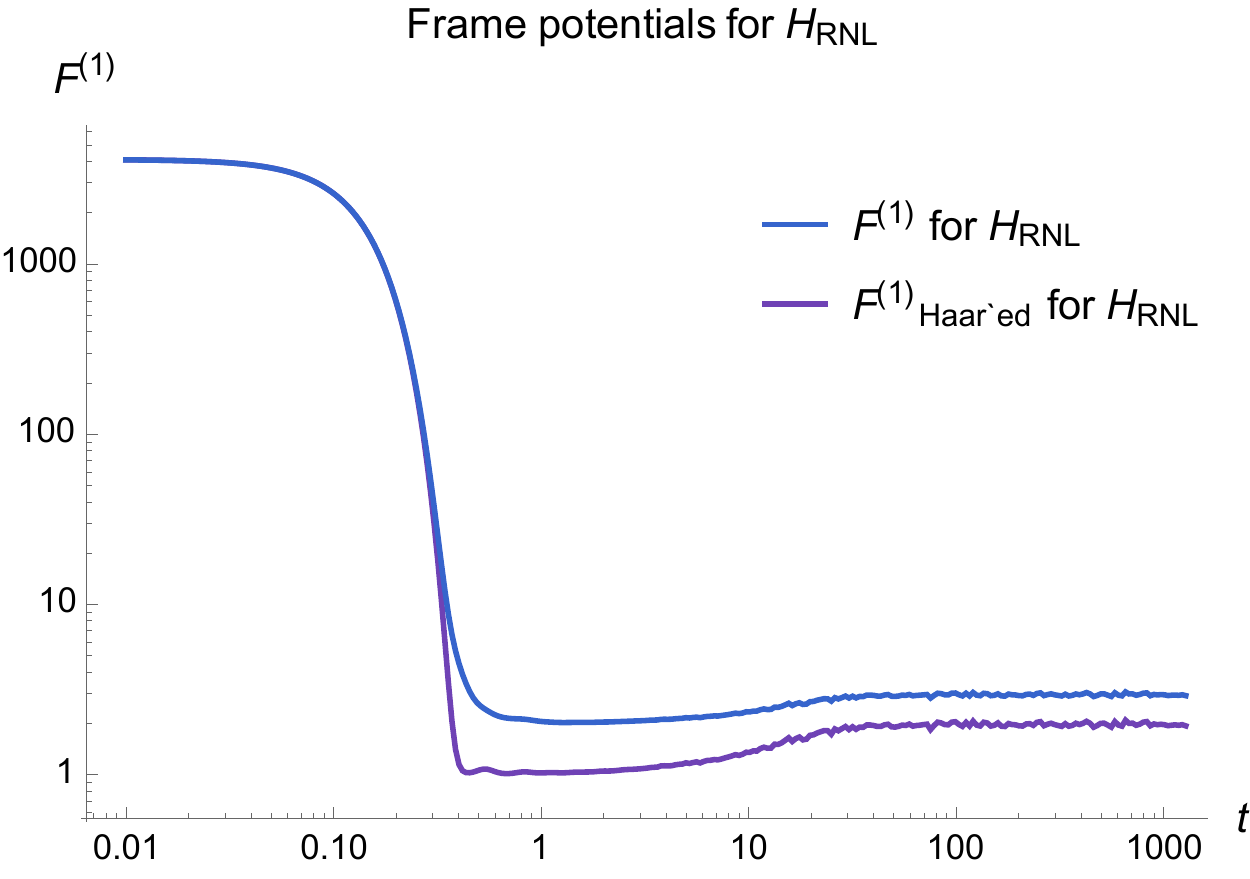}
\includegraphics[width=0.45\linewidth]{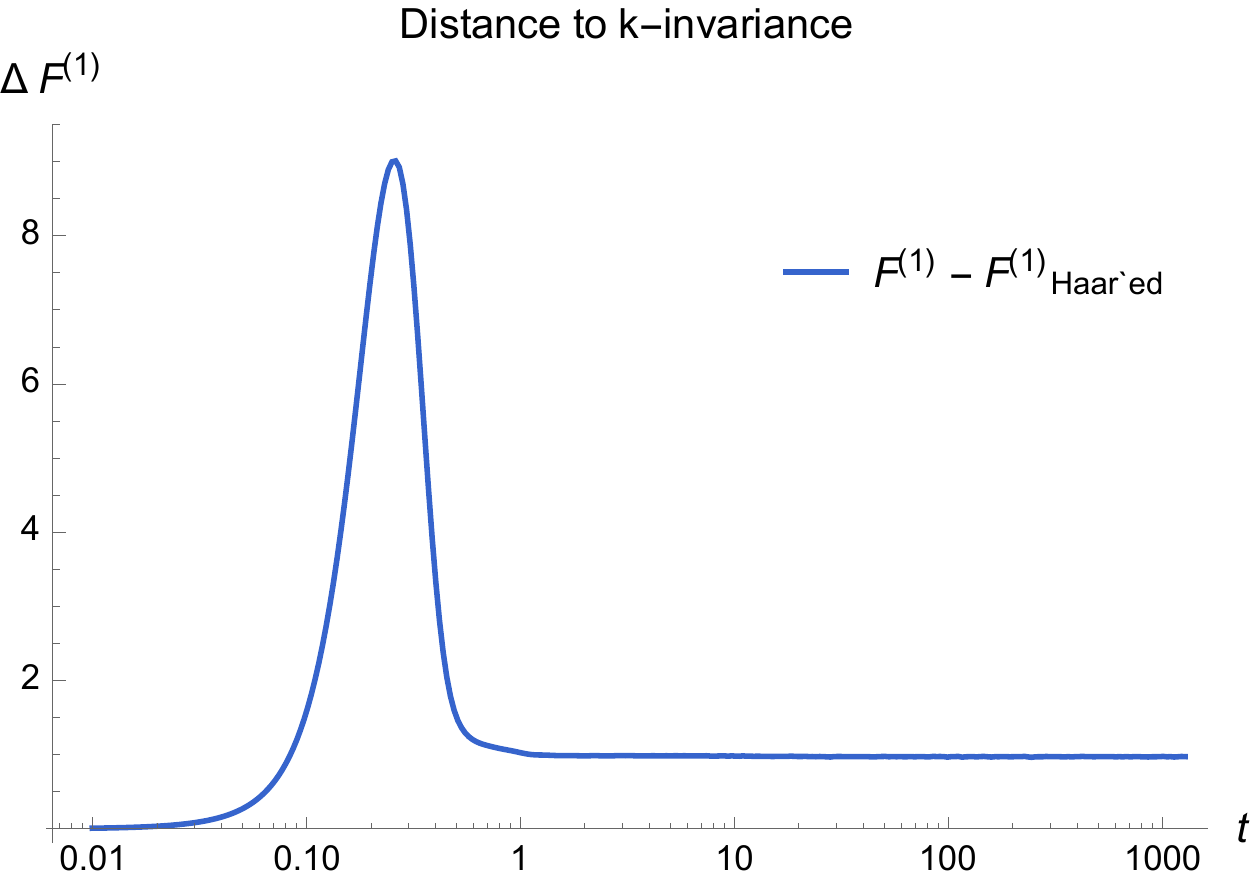}
\caption{On the left we plot the first frame potential $\CF^{(1)}_{\CE_{\rm RNL}}$ for $H_{\rm RNL}$ along side the first frame potential for its Haar-invariant extension $\CF^{(1)}_{\widetilde\CE_{\rm RNL}}$, computed numerically using the 2-point form factor as in Eq.~\eqref{eq:GUE_FP1}. On the right we plot the difference, measuring the 2-norm distance to $1$-invariance and observe approximate 1-invariance at late times.
}
\label{fig:HRNL_fp}
\end{figure}

Let us examine $k$-invariance for the random non-local (RNL) spin system discussed in Sec.~\ref{sec:OTOCFF} where we defined the Hamiltonian in Eq.~\eqref{eq:HRNL} as the sum over all 2-body operators with random Gaussian couplings $J_{ij\alpha \beta}$. The time evolution of the first frame potential for this ensemble as well as its Haar-conjugated generalization are shown in Fig.~\ref{fig:HRNL_fp} along side the difference $\CF^{(1)}_{\mathcal{E}} - \CF^{(1)}_{\widetilde{\mathcal{E}}}$, measuring the distance to $1$-invariance. We only report numerics for a modest spin system of $n=6$ spins. The difficulty of performing frame potential numerics is mentioned in App.~\ref{app:num}.

We find that in this chaotic spin system, at early times we quickly deviate from $1$-invariance, but after evolution by the system's chaotic dynamics, we observe an approach to approximate $1$-invariance at late times. For this system, we see that the frame potential approaches, but does not equal, its Haar-invariant counterpart at later times. But we found numerically that increasing the number of sites makes this late time difference smaller. Thus we expect that at large $N$ for chaotic systems, we reach $k$-invariance at late times.

\subsubsection*{Comments on $k$-invariance}
While frame potentials provide a quantitative way of judging if an ensemble $\mathcal{E}$ is $k$-invariant or not, it would be beneficial to relate it to some physical observables such as correlation functions. It is perhaps not a big surprise that $k$-invariance can be verified by $2k$-point OTOCs. The following statement holds:
\begin{align}
&\langle A_{1}(0)B_{1}(t)\ldots  A_{k}(0)B_{k}(t) \rangle_{\mathcal{E}}\nn
&\qquad\qquad = \langle \tilde{A}_{1}(0)\tilde{B}_{1}(t)\ldots  \tilde{A}_{k}(0)\tilde{B}_{k}(t) \rangle_{\mathcal{E}}\quad \forall \tilde{A}_{j},\tilde{B}_{j} \quad \Longleftrightarrow \quad \text{$\mathcal{E}$ is $k$-invariant}
\end{align}
where $A_{j},B_{j}$ are Pauli operators, and $\tilde{A}_{j},\tilde{B}_{j}$ are some transformations from $A_{j},B_{j}$ such that
\begin{align}
\tilde{A}_{j} = W A_{j} W^{\dagger} \qquad \tilde{B}_{j} = W B_{j} W^{\dagger}
\end{align}
where $W$ is an arbitrary element of unitary $2k$-design. The proof is straightforward and thus is skipped.

Motivated by late-time random matrix universality of chaotic quantum systems, we have introduced a novel quantum information theoretic concept, $k$-invariance, as a possible way of bridging early-time and late-time physics.  We would like to comment on a few caveats. First, consider an ensemble of unitary operators $\mathcal{E}$ generated by some Hamiltonians. Since $\mathcal{E}_{t=0}=\{I\}$, the ensemble is Haar-invariant at time $t=0$. Thus, an ensemble is initially $k$-invariant and is expected to immediately deviate at $t>0$ and then eventually become approximately $k$-invariant. Therefore $\CF^{(k)}_{\mathcal{E}} -  \CF^{(k)}_{\widetilde{\mathcal{E}}}$, which quantifies $k$-invariance, is not a monotonic quantity under time evolution. However, we expect that it is monotonically decreasing at late times. We observe these features in the non-local spin system described above. Depending on the symmetries of the system of interest, we would need to consider the Haar measure with respect to an appropriate Lie group $G \subset U(L)$.

Second, for realistic physical systems with local Hamiltonians, it is not likely that an ensemble $\mathcal{E}_{t}$ becomes $k$-invariant in an exact sense even at very late times. This can be seen from a recent work which shows that the late-time value of infinite temperature OTOCs $\langle A(0)B(t)A(0)B(t) \rangle$ of $q$-local Hamiltonians is $\mathcal{O}(1/N)$ if operators $A,B$ are local and have overlaps with the Hamiltonian~\cite{Huang17}, based on an Eigenstate Thermalization Hypothesis (ETH) argument. A similar argument applies to late-time values of two-point correlators. On the other hand, the Haar average of OTOCs is $\mathcal{O}(1/L^2)$ (or $\mathcal{O}(1/L)$ for an average of absolute values). Thus, OTOCs for local operators and OTOCs for non-local operators may have significantly different late-time values. However, it should be noted that a prediction from the AdS/CFT seems to suggest that correlation functions may become exponentially small $e^{- \mathcal{O}(S)}$ even if $A,B$ are local operators. This may suggest a subtle but important distinction between ordinary strongly interacting systems and gravitational systems which leads to a far-reaching question concerning the universality of gravity and the universality of random matrix theory, seen from the lens of $k$-invariance. 

Let us conclude the section with a brief remark on the Eigenstate Thermalization Hypothesis (ETH). The notion of $k$-invariance may be viewed a dynamical analog of Berry's conjecture about random eigenvectors, which was the motivation behind ETH~\cite{Berry77, Srednicki94, ChaosETH}. A basic assumption of ETH is that matrix elements of a local operator $O$, with respect to energy eigenstates, look ``random'' inside some sufficiently small energy window $\Delta E$. A system achieving $k$-invariance roughly tells us that energy eigenstates may be treated as random vectors 
after sufficiently long times for studying dynamics via OTOCs.\footnote{The related notion of quantum ergodicity and randomness of eigenstates was recently discussed in \cite{Ho17}.} Given the prevalence of eigenstate thermalization in strongly correlated many-body systems,\footnote{See \cite{ChaosETH} and references therein. Interestingly, evidence for ETH has also been discussed recently both in the SYK model \cite{Sonner17} as well as in its free fermion counterpart \cite{Magan15}.} a precise relation between $k$-invariance, ETH and OTOCs would provide clarity on defining what it means for a quantum system to be chaotic.

\section{Discussion}
\label{sec:discussion}

Random matrix theory provides a powerful paradigm for studying late-time chaos. We have leveraged the technology of random matrix theory and Haar-invariance to study correlation functions like OTOCs which diagnose early-time chaos, and frame potentials which diagnose randomness and complexity. The salient feature of the GUE which gave us computational traction is its Haar-invariance, namely that the ensemble looks the same in any basis.  As a result, the dynamics induced by GUE Hamiltonians is non-local ($\mathcal{O}(N)$-local) with respect to any tensor factor decomposition of the Hilbert space, and so the dynamics immediately delocalizes quantum information and other more subtle forms of correlations. Accordingly, the GUE captures features of the long-time physics of a local system that has been delocalized.

In a chaotic quantum system described by a local Hamiltonian, there are two temporal regimes of interest: times before the system scrambles and thus has mostly local correlations, and times after the system scrambles when correlations have effectively delocalized.  We suggested that the transition between these two regimes may be due to the onset of approximate Haar-invariance, and we defined $k$-invariance as a precise characterization. A careful understanding of Haar-invariance for ensembles of local quantum systems could yield precise insights into the apparent breakdown of locality, and tell us in what time regimes we can use Haar-invariance to calculate late-time physics (i.e., correlation functions, frame potentials, complexity, etc.) A concrete way of studying delocalization of operators and the emergence of $k$-invariance would be to compare connected pieces of OTOCs with local and non-local operators and observe their eventual convergence. Of particular interest is to find the $2$-invariance time when all the $4$-point OTOCs, regardless of sizes of operators, start to behave in a similar manner. This time scale must be at least the scrambling time since OTOCs with local operators start to decay only around the scrambling time while OTOCs with non-local operators decay immediately.  Relatedly, we would like to draw attention to an upcoming work~\cite{ShenkerTBA} which studies the onset of random matrix behavior at early times.

In this paper, we computed correlation functions averaged over an ensemble of Hamiltonians. Chaotic systems described by disordered ensembles tend to have small variance in their correlators, and their averaged correlation functions are close to those computed for a simple instance of the ensemble. Even in regimes where replica symmetries are broken, performing time bin averaging reproduces the averaged behaviors very well.  We find in App. \ref{app:timebin} that the time bin-averaged frame potential in the large $L$ limit for two samples agrees with averaging over the whole ensemble. 

We conclude by mentioning a far reaching goal, but one that provides the conceptual pillars for these ideas, namely understanding black holes as quantum systems. While black holes are thermodynamic systems whose microscopic details remain elusive, questions about information loss can be precisely framed by late-time values of correlation functions within AdS/CFT~\cite{MaldacenaEternal}, where unitary evolution can be discussed in terms of the boundary CFT. Ultimately, we would like to use random matrix theory to characterize chaos and complexity in local quantum systems and identify late-time behaviors which are universal for gravitational systems. An interesting future question is to see if gravitational systems are described by random matrices in the sense of $k$-invariance and pinpoint some late-time behavior which results from gravitational universality.

\section*{Acknowledgments}
We thank Yoni Bentov, Fernando Brand\~ao, Clifford Cheung, Patrick Hayden, Alexei Kitaev, John Preskill, Daniel Ranard, Daniel Roberts, Lukas Schimmer, and Steve Shenker for valuable comments and insights.  JC is supported by the Fannie and John Hertz Foundation and the Stanford Graduate Fellowship program. JC and NHJ would like to thank the Perimeter Institute for their hospitality during the completion of part of this work.  BY and NHJ acknowledge support from the Simons Foundation through the ``It from Qubit'' collaboration. NHJ is supported the Institute for Quantum Information and Matter (IQIM), an NSF Physics Frontiers Center (NSF Grant PHY-1125565) with support from the Gordon and Betty Moore Foundation (GBMF-2644). JL is
supported in part by the U.S. Department of Energy, Office of Science, Office of High Energy Physics, under Award Number DE-SC0011632. Research at Perimeter Institute is supported by the Government of Canada through Industry Canada and by the Province of Ontario through the Ministry of Research and Innovation.

\appendix

\section{Scrambling and $2$-designs}
\label{app:QIreview}

Recently there has been growing interest in scrambling and unitary designs from the high energy and quantum information communities. Here we provide a short summary of different ways of quantifying them for infinite temperature cases.

\subsection{Scrambling}

We begin with scrambling. Consider a system of qubits and non-overlapping local ($\mathcal{O}(1)$-body) Pauli operators $V, W$ and compute $\text{OTOC}=\langle VW(t)VW(t)\rangle $ where $W(t)=UWU^{\dagger}$. The initial value of OTOC at $t=0$ is $1$. Scrambling is a phenomenon where the OTOC becomes $ \mathcal{O}(\epsilon)$ with $\epsilon\ll 1$ being a small but finite constant:
\begin{align}
\text{$\langle VW(t)VW(t)\rangle = \mathcal{O}(\epsilon)$ \qquad for all pairs of local operators $V,W$}
\end{align}
It is often the case that OTOCs with local operators are the slowest to decay. This can be seen from our analysis on $4$-point spectral form factors. So, by the scrambling time, OTOCs with non-local operators are already $\mathcal{O}(\epsilon)$ or smaller. The scrambling time is lower bounded by $\mathcal{O}(\log(n))$ in the case of $0$-dimensional $\mathcal{O}(1)$-local systems due to a Lieb-Robinson--like argument~\cite{FastScrambling}.

Scrambling has caught significant attention from the quantum gravity community since it is closely related to the Hayden-Preskill thought experiment on black hole information problems~\cite{HaydenPreskill}. Assume that $V,W$ act on qubits on some local regions $A,D$ respectively, and define their complements by $B=A^{c}, C=D^c$. Imagine that $A$ is an unknown quantum state $|\psi\rangle$ thrown into a ``black hole'' $B$, and the whole system evolves by some time-evolution operator $U=e^{-iHt}$. At time $t$, we collect the ``Hawking radiation'' $D$ and attempt to reconstruct (an unknown) $|\psi\rangle$ from measurement on $D$. Such a thought experiment was considered by Page who argued that, if a black hole's dynamics $U$ is approximated by a random unitary operator, then reconstructing $|\psi\rangle$ is not possible unless we collect more than $n/2$ qubits of the Hawking radiation \cite{PageTime}. As we shall show in Appendix~\ref{app:QI}, the impossibility of reconstruction of $A$ from $D$ is reflected in the smallness of the $2$-point correlation functions:
\begin{align}
|\langle VW(t) \rangle| = \mathcal{O}(\epsilon) \qquad \text{for local $V,W$} \quad \longrightarrow \quad \text{no reconstruction of $A$ from $D$.}
\end{align}
The famous calculations by Hawking and Unruh imply that these two-point correlators are thermal, and quickly become small.

Hayden and Preskill considered a situation where a black hole $B$ has already emitted half of its contents, and we have collected its early radiation and stored it in some secure quantum memory $M$. The quantum memory $M$ is maximally entangled with $B$, and the question is whether we can reconstruct $|\psi\rangle$ by having access to $M$. It has been shown that scrambling, as defined above, implies that we can reconstruct $|\psi\rangle$ with some good average fidelity by collecting the Hawking radiation on $D$ at time $t$:
\begin{align}
\langle VW(t)VW(t)\rangle = \mathcal{O}(\epsilon) \quad \longrightarrow \quad \text{reconstruction of $A$ from $D$ and $M$.}
\end{align}
Therefore, scrambling implies the possibility of recovering local quantum information via local measurements on the Hawking radiation. A random unitary operator $U$ typically gives very small OTOCs which enables reconstruction of $A$ in the Hayden-Preskill thought experiment. 

Reconstruction problems in the Hayden-Preskill setting are closely related to the problem of decoupling. A crucial difference between scrambling and decoupling is that decoupling typically considers $A,D$ to be some finite fraction of the whole system and concerns the reconstruction of unknown many-body quantum states supported on a big region $A$. Since we quantify the reconstruction via fidelity for many-body quantum states, the requirement tends to be more stringent. The relation between scrambling and decoupling is discussed in~\cite{Brown15} in the context of local random circuits.

\subsection{Unitary designs}

Next let us discuss unitary $2$-designs. Consider an ensemble of time evolution operators $U_{j}$ with probability distributions $p_{j}$; $\mathcal{E}=\{ U_{j},p_{j} \}$ with $\sum_{j}p_{j}=1$. The 2-fold channels of $\CE$ and the Haar ensemble are
\begin{align}
\Phi_{\mathcal{E}}(\rho)= \sum_{j}p_{j} U_{j}\otimes U_{j}(\rho) U_{j}^{\dagger}\otimes U_{j}^{\dagger}\qquad
\Phi_{\text{Haar}}(\rho)= \int_{\text{Haar}} dU \ U \otimes U (\rho) U^{\dagger}\otimes U^{\dagger}.
\end{align}
If $\Phi_{\mathcal{E}}(\rho)=\Phi_{\text{Haar}}(\rho)$ for all $\rho$, then we say $\mathcal{E}$ is $2$-design. One can check if $\mathcal{E}$ is $2$-design or not by looking at OTOCs. Consider the OTOC $\langle VW(t)VW(t)\rangle $ for arbitrary Pauli operators $V,W$ which are not necessarily local operators. We will be interested in the ensemble averages of OTOCs:
\begin{align}
\langle VW(t)VW(t)\rangle_{\mathcal{E}} \equiv \sum_{j}p_{j} \langle VU_{j}WU_{j}^{\dagger}VU_{j}WU_{j}^{\dagger}\rangle.
\end{align}
If $\langle VW(t)VW(t)\rangle_{\mathcal{E}}= \langle VW(t)VW(t)\rangle_{\text{Haar}}$ for all pairs of Pauli operators $V,W$, then the ensemble forms a unitary $2$-design~\cite{ChaosDesign}.

A typical unitary operator from a $2$-design achieves scrambling because
\begin{align}
|\langle VW(t)VW(t)\rangle|_{\text{Haar}} \simeq \frac{1}{L}
\qquad
\langle VW(t)VW(t)\rangle_{\text{Haar}} \simeq \frac{1}{L^2}
\end{align}
for any (possibly non-local) Pauli operators $V,W$. The first equation implies that the OTOC value for a single instance from the ensemble is typically $1/L$ in absolute value while the second equation implies that the OTOC, after ensemble averaging, is $1/L^2$. Since OTOCs are small, a typical $2$-design unitary operator $U$ implies scrambling, but the converse is not always true. Recall that scrambling only requires $\text{OTOC}= \mathcal{O}(\epsilon)$. There is thus a big separation in the smallness of the OTOC, and the scrambling time may be much shorter than the $2$-design time. Also, scrambling requires $\text{OTOC}= \mathcal{O}(\epsilon)$ only for local operators while a $2$-design unitary makes the OTOC small for all pairs of Pauli operators. The lower bound for the exact $2$-design time is $\mathcal{O}(\log(n))$, but no known protocol achieves this time scale.

One important distinction between scrambling and the $2$-design time is how small the OTOCs becomes. The phenomena of scrambling concerns the deviation of OTOC values from the maximal value $1$. The concept of a $2$-design concerns the deviation of OTOC values from the minimal value $\mathcal{O}(1/L)$. The former is related to early-time chaos and the latter is related to late-time chaos. 

\subsection{Approximate $2$-designs}

Finally, let us briefly discuss the notion of approximate $2$-design. When two quantum operations $\Phi_{\mathcal{E}}$ and $\Phi_{\text{Haar}}$ are close to each other, we say that $\mathcal{E}$ is an approximate $2$-design. In order to be quantitative, however, we need to pick appropriate norms with which two quantum operations can be compared. The $2$-norm distance can be defined in a simple way via
\begin{equation}
\begin{split}
&\text{$2$-norm}= \sqrt{\text{tr}(SS^{\dagger})} \\
&S = \int \sum_{j}p_{j} U_{j}\otimes U_{j}\otimes U_{j}^{\dagger}\otimes U_{j}^{\dagger}-  \int_{\text{Haar}} dU \ U \otimes U \otimes U^{\dagger}\otimes U^{\dagger}.
\end{split}
\end{equation}
If $S=0$, then $\Phi_{\mathcal{E}}$ and $\Phi_{\text{Haar}}$ would be the same. We say that $\mathcal{E}$ is a $\delta$-approximate $2$-design in the $2$-norm if $\sqrt{\text{tr}(SS^{\dagger})}\leq \delta$.

Frame potentials are closely related to the $2$-norm distance because $\text{tr}(SS^{\dagger}) = \CF_{\mathcal{E}} - \CF_{\text{Haar}}\geq 0$. In~\cite{ChaosDesign}, a relation between the frame potential and OTOCs has been derived
\begin{align}
\int dAdBdCdD |\langle AB(t)CD(t) \rangle_{\mathcal{E}}|^2 = \frac{\CF_{\mathcal{E}}^{(2)}}{L^6}.
\end{align}
In practice, the main contribution to the left-hand side comes from OTOCs of the form $\langle AB(t)AB(t) \rangle_{\mathcal{E}}$. For simplicity of discussion, let us assume that $\langle AB(t)CD(t) \rangle_{\mathcal{E}}=0$ when $C\not=A$ or $D\not=B$ (where $A,B,C,D$ are non-identity Pauli operators). Then, a simple analysis leads to
\begin{align}
|\langle AB(t)AB(t) \rangle_{\mathcal{E}}|^2 \simeq \delta^2
\end{align}
for typical non-identity Pauli operators $A,B$. Thus, being a $\delta$-approximate $2$-design in the $2$-norm implies that OTOCs are typically small. However, this does not necessarily imply scrambling because OTOCs with local operators are often the slowest to decay. In order to guarantee scrambling, we would need a $\frac{\delta}{L}$-approximate design in the $2$-norm (under an assumption on $\langle AB(t)CD(t) \rangle_{\mathcal{E}}=0$ for $C\not=A$ or $D\not=B$). For this reason, an alternative distance measure called the diamond norm is often used in quantum information literature. See~\cite{Low_Thesis} for relations between different norms.

\section{Information scrambling in black holes}\label{app:QI}

In this Appendix, we discuss behaviors of $2$-point correlators and $4$-point OTOCs from the viewpoint of information scrambling in black holes. We begin by deriving a formula which relates two-point autocorrelation functions and mutual information. We will be interested in the following quantity
\begin{align}
\big|\langle O_{A} O_{D}(t) \rangle_{\rm avg}\big|^2 \equiv \frac{1}{L_{A}^2 L_{D}^2} \sum_{O_{A}\in \mathcal{P}_{A}} \sum_{O_{D} \in \mathcal{P}_{D}} |\langle O_{A}O_{D}(t) \rangle|^2
\end{align}
where $\langle O_{A}O_{D}(t)\rangle = \frac{1}{L} \text{Tr} (O_{A}UO_{D}U^{\dagger})$ and $U$ is the time-evolution operator of the system, and $\mathcal{P}_{A}$ and $\mathcal{P}_{D}$ are sets of Pauli operators on $A$ and $D$. There are $L_{A}^{2}$ and $L_{D}^2$ Pauli operators.

The relation between apparent information loss and two-point correlators can be understood by using the state representation $|U\rangle$ of a unitary operator $U$. Given a unitary operator $U$ acting on an $n$-qubit Hilbert space $\mathcal{H}$, one can view $U$ as a pure quantum state $|U\rangle$ defined on a $2n$-qubit Hilbert space $\mathcal{H}\otimes \mathcal{H}$:
\begin{align}
|U\rangle \equiv U \otimes I |\text{EPR}\rangle, \qquad |\text{EPR}\rangle =\frac{1}{\sqrt{2^n}}\sum_{j=1}^{2^n} |j\rangle \otimes |j\rangle.\label{eq:state_rep}
\end{align}
Or equivalently, $|U\rangle \equiv \frac{1}{\sqrt{2^n}}\sum_{i,j}U_{i,j}|i\rangle\otimes |j\rangle$ where $U = \sum_{i,j}U_{i,j}|i\rangle \langle j|$. One easily see that the quantum state $|U\rangle$ is uniquely determined by a unitary operator $U$. The state representation allows us to view $|U\rangle_{ABCD}$ as a four-partite quantum state:
\begin{align}
\ket{U} = \frac{1}{\sqrt{2^n}}\begin{gathered}\includegraphics{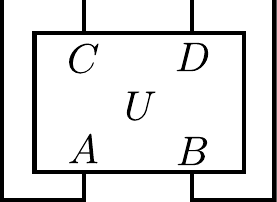}\end{gathered}
\end{align}
where $B=A^c$ and $D=C^c$ in the original system of qubits. Given the state representation $|U\rangle$ of a unitary operator, we can derive the following formula
\begin{align}
\big|\langle O_{A} O_{D}(t) \rangle_{\rm avg}\big|^2 = \frac{1}{L_{A}^2L_{D}^2}2^{I^{(2)}(A,D)}
\end{align}
where $I^{(2)}(A,D)$ is the R\'{e}nyi-$2$ mutual information between $A$ and $D$ for $|\Psi\rangle$, defined by $I^{(2)}(A,D)\equiv S_{A}^{(2)}+S_{D}^{(2)}-S_{AD}^{(2)}$.

To derive the formula, let $\rho_{AD}$ be the reduced density matrix of $|U\rangle$ on $AD$. Its graphical representation is
\begin{align}
\rho_{AD} = \frac{1}{L}~ \begin{gathered}\includegraphics{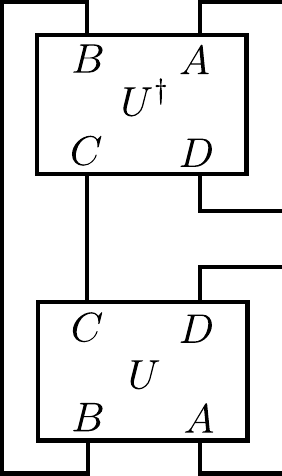}\end{gathered}
\end{align}
The averaged $2$-point correlator is given by
\begin{align}
\big| \vev{O_A O_D(t)}_{\rm avg}\big|^2 = \frac{1}{L^2} \begin{gathered}\includegraphics{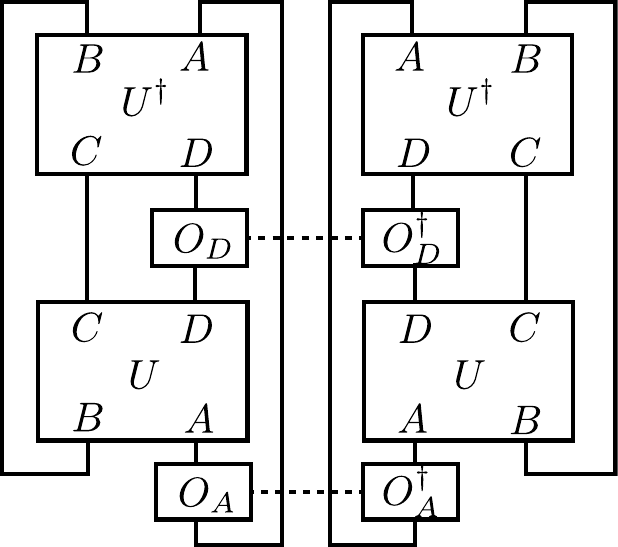}\end{gathered}
\end{align}
where dotted lines represent averaging over Pauli operators. By using $\frac{1}{L}\sum_{O\in \mathcal{P}} O\otimes O^{\dagger}=\text{SWAP}$, we obtain
\begin{align}
|\langle O_{A} O_{D}(t) \rangle_{\text{ave}}|^2 = \frac{\text{Tr}(\rho_{AD}^2)}{L_{A}L_{D}}=  \frac{1}{L_{A}^2L_{D}^2}2^{I^{(2)}(A,D)}.
\end{align}

Let us further ponder this formula. For strongly interacting systems, it is typically the case that
\begin{align}
\langle O_{A}O_{D}(t) \rangle \simeq 0 \qquad \text{if}\quad\text{Tr}(O_{A}O_{D})=0.
\end{align}
So, the following relation for the autocorrelation functions holds approximately:
\begin{align}
\sum_{O_{A}\in \mathcal{P}_{A}} |\langle O_{A}O_{A}(t) \rangle|^2 \simeq 2^{I^{(2)}(A,D)}
\end{align}
where we took $A$ and $D$ to be the same subset of qubits.

The above formula has an interpretation as information retrieval from the early Hawking radiation. Consider scenarios where Alice throws a quantum state $|\psi\rangle$ into a black hole and Bob attempts to reconstruct it from the Hawking radiation. In accordance with such thought experiments, let $A$ be qubits for Alice's quantum state, $B$ be the black hole, $C$ be the remaining black hole and $D$ be the Hawking radiation. Then, the averaged $2$-point correlation functions have an operational interpretation as Bob's strategy to retrieve Alice's quantum state. Let us assume that the initial state of the black hole is unknown to Bob and model it by a maximally mixed state $\rho_{B}=\frac{I_{B}}{L_{B}}$. Alice prepares an EPR pair $|\text{EPR}\rangle_{AR}$ on her qubits and her register qubits. Notice the difference from the Hayden-Preskill setup where Bob had access to some reference system $B'$ which is maximally entangled with the black hole $B$. In this decoding problem, we do not grant such access to Bob. He just collects the Hawking radiation $D$ and tries to reconstruct Alice's quantum state.

The most obvious strategy is to apply the inverse $U^{\dagger}$. However, Bob does not have an access to qubits on $C$. So, he applies $U_{CD}^{\dagger}\otimes I_{R}$ to $\rho_{C}\otimes \rho_{DR}$ where $\rho_{C}=\frac{I_{C}}{L_{C}}$. Graphically, this corresponds to
\begin{align}
\ket\Psi = \frac{L}{\sqrt{L_A L_B L_C}}\begin{gathered}\includegraphics{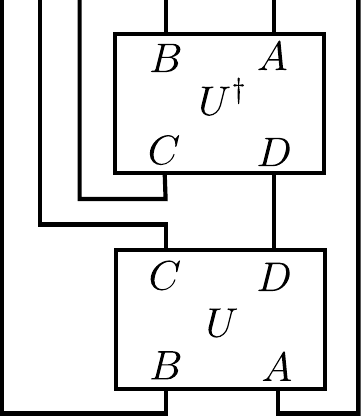}\end{gathered}
\end{align}
The success of decoding is equivalent to distillation of an EPR pair between $A$ and $R$. So, we compute the EPR fidelity. Namely, letting $\Pi$ be a projector onto an EPR pair between $A$ and $R$, we have
\begin{align}
F = \braket{\Psi|\Pi| \Psi} = \frac{1}{L^2}\begin{gathered}\includegraphics{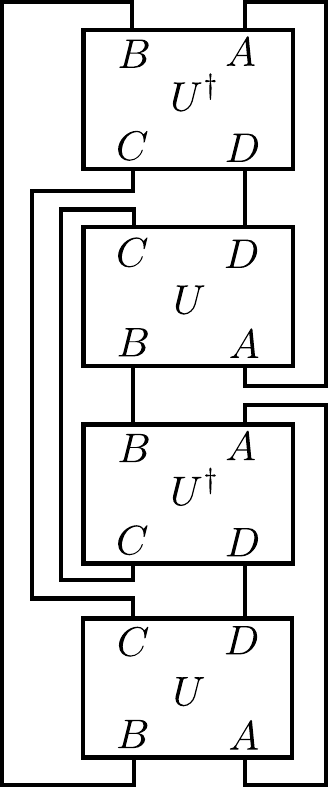}\end{gathered}
\end{align}
which leads to
\begin{align}
F = \text{Tr}(\rho_{BC}^2) = \text{Tr}(\rho_{AD}^2) = L_{A}L_{D} |\langle O_{A} O_{D}(t)\rangle_{\rm avg}|^2.
\end{align}
Therefore, the decay of $2$-point correlation functions indeed implies that Bob cannot reconstruct Alice's quantum state.

Finally, let us summarize the known relations between correlation functions and mutual information:
\begin{align}
2^{-I^{(2)}(A,BD)} = \langle O_{A}O_{D}(t)O_{A}O_{D}(t) \rangle_{\rm avg} \\
2^{I^{(2)}(A,D)} = |\langle O_{A}O_{D}(t) \rangle_{\rm avg}|^{2}\cdot L_{A}^2 L_{D}^2.
\end{align}
Note that the first formula proves that the decay of OTOCs leads to large $I^{(2)}(A,BD)$ which implies the possibility of Bob decoding Alice's quantum state by accessing both the early radiation $B$ and the new Hawking radiation $D$. These two formulae allow us to formally show that a black hole can be viewed as a quantum error-correcting code. Let $A,D$ be degrees of freedom corresponding to incoming and outgoing Hawking radiation, and $B,C$ be degrees of freedom corresponding to other exotic high energy modes at the stretched horizon. Since a black hole is thermal, we know that $|\langle O_{A}O_{D}(t) \rangle_{\rm avg}|$ decays at $t\sim \mathcal{O}(\beta)$. Also, due to the shockwave calculation by Shenker and Stanford \cite{SSbutterfly}, we know that $\langle O_{A}O_{D}(t)O_{A}O_{D}(t) \rangle_{\rm avg}$ decays at $t\sim \mathcal{O}(\beta\log N)$. These results imply that after the scrambling time:
\begin{align}
I^{(2)}(A,D)\simeq 0 \qquad I^{(2)}(A,C)\simeq 0.
\end{align}
The implication is that quantum information injected from $A$ gets delocalized and non-locally is hidden between $C$ and $D$. The error-correction property can be seen by
\begin{align}
I^{(2)}(A,BD)\simeq 2a \qquad I^{(2)}(A,BC)\simeq 2a \qquad I^{(2)}(A,CD)\simeq 2a
\end{align}
where $a$ is the number of qubits on $A$. Namely, if we see the black hole as a quantum code which encodes $A$ into $BCD$, then the code can tolerate erasure of any single region $B,C,D$. In other words, accessing any two of $B,C,D$ is enough to reconstruct Alice's quantum state. Thus, black hole dynamics, represented as a four-partite state $|U\rangle_{ABCD}$, can be interpreted as a three-party secret sharing quantum code.

\section{Spectral correlators and higher frame potentials}
\label{app:FFFP}
In this Appendix we will present formulas for form factors from random matrix theory. For GUE$(L,0,1/\sqrt{L})$, $L\times L$ matrices with off-diagonal complex entries and real diagonal entries chosen with variance $\sigma^2=1/L$, the joint probability of eigenvalues for GUE, with normalizing factors, is
\begin{align}
\label{jointprobwithfactors}
P(\lambda_1,\ldots,\lambda_L) = \frac{L^{L^2/2}}{(2\pi)^{L/2} \prod_{p=1}^L p!} \, e^{-\frac{L}{2} \sum_i \lambda_i^2} \prod_{i<j} (\lambda_i-\lambda_j)^2\,
\end{align}
and the joint probability distribution of $n$ eigenvalues (i.e., the $n$-point spectral correlation function), defined as
\begin{align}
\rho^{(n)}(\lambda_1,\ldots,\lambda_n) = \int d\lambda_{n+1}\ldots d\lambda_L P(\lambda_1,\ldots,\lambda_L)\,.
\end{align}
We can compactly express $\rho^{(n)}(\lambda_1,\ldots,\lambda_n)$ in terms of a kernel $K$ \cite{TaoRMT,MehtaRMT} as
\begin{align}
\label{deteq1}
\rho^{(n)}(\lambda_1,\ldots ,\lambda_n)=\frac{(L-n)!}{L!}\det \big( K(\lambda_i ,\lambda_j) \big)_{i,j=1}^{n}
\end{align}
In the large $L$ limit, the kernel $K$ is approximately
\begin{align}\label{sinekernel}
K(\lambda_i,\lambda_j)\equiv \begin{cases}
\dfrac{L}{\pi }\dfrac{\sin (L({{\lambda }_{i}}-{{\lambda }_{j}}))}{L({{\lambda }_{i}}-{{\lambda }_{j}})} &\mbox{for } i \not = j \\ \\
\dfrac{L}{2\pi }\sqrt{4-\lambda _{i}^{2}} & \mbox{for } i=j
\end{cases}
\end{align}
where the $i \not = j$ case is called the sine kernel, and the $i = j$ case is simply the Wigner semicircle.  In the large $L$ limit, the basic approach for computing spectral form factors will be expanding the determinant in Eq.~\eqref{deteq1} using the kernel in Eq.~\eqref{sinekernel}, and computing the Fourier transform of the resulting sums of product of kernels. Thus we will have sums of integrals of the form \cite{MehtaRMT}
\begin{align}\label{finiteL}
&\int{\prod\limits_{i=1}^{m}{d{{\lambda }_{i}}}}{K}({{\lambda }_{1}},{{\lambda }_{2}}){K}({{\lambda }_{2}},{{\lambda }_{3}})\ldots {K}({{\lambda }_{m-1}},{{\lambda }_{m}}){K}({{\lambda }_{m}},{{\lambda }_{1}}) \,e^{i \sum_{i=1}^{m} k_i \lambda_i} \nn
&\qquad = \frac{L}{\pi }\int d\lambda \,e^{i \sum_{i=1}^{m} k_i \lambda_i} \int{dk}\,g(k)g\Big(k+\frac{{{k}_{1}}}{2L}\Big) g\Big(k+\frac{{{k}_{2}}}{2L}\Big) \ldots g\Big(k+\frac{{{k}_{m-1}}}{2L}\Big)
\end{align}
where we define the Fourier transform of the sine kernel
\begin{align}
g(k)\equiv \int dr \, {{{e}^{2\pi ikr}}\,\frac{\sin(\pi r)}{\pi r}}=\left\{ \begin{matrix}
   1 &\mbox{for } \left| k \right|<\frac{1}{2}  \\
   0 &\mbox{for } \left| k \right|>\frac{1}{2}  \\
\end{matrix} \right.\,.
\end{align}

The delta function singularity from the $\int{d{{\lambda }}\,{{e}^{\sum\nolimits_{i=1}^{m}{i{{k}_{i}}{{\lambda }}}}}}$ integral in Eq.~\eqref{finiteL} is an artifact of our expansion around infinite $L$, namely that $\frac{L}{\pi} \frac{\sin(L(\lambda_i - \lambda_j))}{L(\lambda_i - \lambda_j)}$ is not regulated in the $(\lambda_i + \lambda_j)$ direction.  The most direct method to soften this divergence is to impose a cutoff
\begin{align}
\frac{L}{\pi}\int d\lambda \, e^{i \sum_{i=1}^{m} k_i \lambda_i} \longrightarrow \frac{L}{\pi}\int_{-\pi/2}^{\pi/2} d\lambda\, e^{i \sum_{i=1}^{m} k_i \lambda_i}
\end{align}
which is fixed by the normalization condition
\begin{equation}
\frac{L}{\pi }\int_{-\pi/2}^{\pi/2}d \lambda \, e^{i \sum_{i=1}^{m} k_i \lambda_i} \int{dk}\,g(k)g\Big(k+\frac{{{k}_{1}}}{2L}\Big) g\Big(k+\frac{{{k}_{2}}}{2L}\Big)\ldots g\Big(k+\frac{{{k}_{m-1}}}{2L}\Big)\bigg|_{k_1, \ldots, k_m = 0} = L\,.
\end{equation}
While the `box approximation' of applying the cutoff allows us to compute higher-point spectral correlators in the large $L$ limit, it does lead to errors relative to an exact answer whose closed form is not tractable.\footnote{For instance, the Fourier transform of the semicircle distribution decays as $t^{-3/2}$, whereas the Fourier transform of a box decays as $t^{-1}$.} Thus we must be careful to keep track of these errors and compare with numerics.  However, we find that at infinite temperature, the box approximation of the spectral form factors is analytically controlled at early times like $\mathcal{O}(1)$ and late times greater than $\mathcal{O}(\sqrt{L})$.

To understand the errors of the box approximation, we first consider various cases heuristically: when we have $\sum_i k_i=0$, the $\lambda$ integral in Eq.~\eqref{finiteL} is directly fixed by normalization.  When $\sum_i k_i \ne 0$, the $\lambda$ integral in Eq.~\eqref{finiteL} dephases and so decays when $|\sum_i k_i|$ is large, and thus the induced error is unimportant at long times.  At small, $\mathcal{O}(1)$ values of the $|k_i|$'s (assuming that $m$ is $\mathcal{O}(1)$), the error induced by the box approximation is also small and the value is still close to the $\sum_i k_i=0$ value.

For instance, carefully keeping track of factors of $L$ tells us that in $\mathcal{R}_4$, for early times like $\mathcal{O}(1)$ the error is suppressed by $\mathcal{O}(1/L)$ relative to largest order terms, while for late times after $\mathcal{O}(\sqrt{L})$ the error is suppressed by $\mathcal{O}(1/\sqrt{L})$ relative to the largest order terms.

In this discussion, particularly for $\sum_i k_i=0$, we assumed simple sine kernel correlations and found $r_2$ to be a pure linear function. However, a more delicate treatment shows some other transition time scale at early times, which likely complicates the functional form of $r_2$ and gives a different slope for the ramp. We briefly address this issue for our numerics in App.~\ref{app:num}.

Since the dephasing of the $\lambda$ integral at large $|\sum_i k_i|$ is suppressed at finite temperature, to better capture long-time finite temperature eigenvalue correlations we use a modified kernel $\widetilde{K}$ which is valid in the short distance limit $|\lambda_a-\lambda_b|\sim \mathcal{O}(1/L)$ \cite{BrezinHikami2,BrezinRMT},
\begin{align}
\widetilde{K}(\lambda_i, \lambda_j)=\frac{\sin \big(L\pi ({{\lambda }_{i}}-{{\lambda }_{j}}){{\rho }^{(1)}}(({{\lambda }_{i}}+{{\lambda }_{j}})/2)\big)}{\pi ({{\lambda }_{i}}-{{\lambda }_{j}})}
\end{align}
which naturally provides a cutoff in the $(\lambda_i + \lambda_j)$ direction. However, this approximation assumes the continued domination of the regulated integral in the short distance limit, which may not be true for large $\beta$. However, for small $\beta$ the modified kernel is reliable. In the generic case, one should consider the full expression of Hermite polynomials as the sine kernel, and correctly take the limit. A complicated formula has been derived in \cite{BrezinHikami2,BrezinRMT} from a saddle point approximation.

\subsection{Expressions for spectral correlators}
\label{app:FF}
Using the analysis above, it is straightforward to compute form spectral correlation functions for the GUE. It is convenient to define
\begin{align}
r_1(t) \equiv \frac{J_1(2 t)}{ t}\,, \quad r_2(t) \equiv \begin{cases} 1-\frac{t}{2L} &~{\rm for}~ t< 2L\\ 0 &~{\rm for}~ t> 2L \end{cases} ~\,, \quad r_3(t) \equiv \frac{\sin(\pi  t/2)}{\pi  t/2}\,.
\end{align}
as mentioned earlier.  The infinite temperature form factors which appear in the calculation of the first and second frame potentials are
\begin{align}
&\CR_2(t) = \int{D\lambda} \sum_{i,j=1}^{L} e^{i(\lambda_i-\lambda_j)t}\,,
&\CR_{4,1}(t) &= \int D\lambda \sum_{i,j,k=1}^{L} e^{i(\lambda_i+\lambda_j - 2\lambda_k )t}\,,\nn
&\CR_4 (t) =\int D\lambda \sum_{i,j,k,\ell=1}^{L} e^{i(\lambda_i+\lambda_j-\lambda_k-\lambda_\ell)t}\,,
&\CR_{4,2} (t) &= \int D\lambda \sum_{i,j=1}^{L} e^{2i(\lambda_i- \lambda_j)t}\,.
\label{eq:FFs}
\end{align}
As $\mathcal{R}_{4,2}$ is simply $\mathcal{R}_2(2t)$, we only need to compute the first three spectral correlation functions. We will also investigate the finite temperature version of $\mathcal{R}_2$, which we defined as
\begin{align}
\CR_2(t,\beta) \equiv \int{D\lambda} \sum_{i,j=1}^{L} e^{i(\lambda_i-\lambda_j)t}e^{-\beta (\lambda_i+\lambda_j)}\,.
\end{align}
\subsubsection*{$\mathcal{R}_2$ at infinite temperature}
We start by computing $\mathcal{R}_2$ at infinite temperature:
\begin{align}
\mathcal{R}_2(t) & =L+\int d\lambda_1 \, d\lambda_2 \, \Big(  {K} (\lambda_1, \lambda_1)  {K}(\lambda_2, \lambda_2) -  {K}^2 (\lambda_1, \lambda_2) \Big) e^{i (\lambda_1-\lambda_2)t}\,.
\end{align}
Evaluating the first term in the integral, we find
\begin{align}
\int d\lambda_1  {K}(\lambda_1,\lambda_1) e^{i\lambda_1 t} \int d\lambda_2  {K}(\lambda_2,\lambda_2) e^{-i\lambda_2 t} = L^2  {r}_1^2(t)\,.
\end{align}
The second term can be evaluated using Eq.~\eqref{finiteL}, and we find
\begin{equation}
\int d\lambda_1 d\lambda_2  {K}^2(\lambda_1,\lambda_2) e^{i(\lambda_1-\lambda_2) t} = L  {r}_2(t)\,.
\end{equation}
The final result is
\begin{align}
{{\mathcal {R}}_{2}}(t)=L+{{L}^{2}} {r}_{1}^{2}(t)-L{ {r}_{2}}(t).
\end{align}
\subsubsection*{$\mathcal{R}_2$ at finite temperature}
As explained above, to better capture long-time correlations at finite temperature we will use the short-distance-limit kernel $\widetilde{K}$. Firstly, for $i=j$, we have
\begin{align}
L\int D\lambda \,{{e}^{-2\beta {{\lambda }_{1}}}}=L{ {r}_{1}}(2i\beta )\,.
\end{align}
For $i\ne j$ we have
\begin{align}
&L(L-1)\int D\lambda \, e^{i(\lambda_1 -\lambda_2)t-\beta (\lambda_1+\lambda_2)}\nn
 & \qquad=\int d\lambda_1 d\lambda_2 \Big( \widetilde{K} (\lambda_1, \lambda_1) \widetilde{K}(\lambda_2, \lambda_2) - \widetilde{K}^2(\lambda_1, \lambda_2)\Big) e^{i(\lambda_1-\lambda_2)t-\beta( \lambda_1 + \lambda_2)} \nn
& \qquad = {{L}^{2}}{ {r}_{1}}(t+i\beta ){ {r}_{1}}(-t+i\beta )-L{{ {r}}_{1}}(2i\beta ){{ {r}}_{2}}(t)\,.
\end{align}
Putting everything together, we obtain
\begin{align}
&{{\mathcal {R}}_{2}}=L{ {r}_{1}}(2i\beta )+{{L}^{2}}{ {r}_{1}}(t+i\beta ){ {r}_{1}}(-t+i\beta )-L{ {r}_{1}}(2i\beta ){ {r}_{2}}(t)\,.
\end{align}
\subsubsection*{$\mathcal{R}_4$ at infinite temperature}

We now compute ${\mathcal{R}_{4}}(t)$, again by separately considering coincident eigenvalues, using the determinant of kernels, and Fourier transforming to find
\begin{align}
{\mathcal{R}_{4}}(t) =\,\,&{{L}^{4}} {r}_{1}^{4}(t) -2{{L}^{3}} {r}_{1}^{2}(t){{ {r}}_{2}}(t){{ {r}}_{3}}(2t)-4{{L}^{3}} {r}_{1}^{2}(t){{ {r}}_{2}}(t)+2{{L}^{3}}{{ {r}}_{1}}(2t) {r}_{1}^{2}(t)+4{{L}^{3}} {r}_{1}^{2}(t) \nn
&+2{{L}^{2}} {r}_{2}^{2}(t)+{{L}^{2}} {r}_{2}^{2}(t) {r}_{3}^{2}(2t)+8{{L}^{2}}{{ {r}}_{1}}(t){{ {r}}_{2}}(t){{ {r}}_{3}}(t)-{2{L}^{2}}{{ {r}}_{1}}(2t){{ {r}}_{2}}(t){{ {r}}_{3}}(2t)\nn
&-4{{L}^{2}}{{ {r}}_{1}}(t){{ {r}}_{2}}(2t){{ {r}}_{3}}(t)  +{{L}^{2}} {r}_{1}^{2}(2t)-4{{L}^{2}} {r}_{1}^{2}(t) -4{{L}^{2}}{{ {r}}_{2}}(t)+2{{L}^{2}}\nn &-7L{{ {r}}_{2}}(2t)+4L{{ {r}}_{2}}(3t)+4L{{ {r}}_{2}}(t)-L\,.
\end{align}
We can simplify this formula at early times of $\mathcal{O}(1)$ and late times greater than $\mathcal{O}(\sqrt{L})$ by dropping subdominant terms and find
\begin{align}\label{main}
{\mathcal{R}_{4}}\approx{{L}^{4}}r_{1}^{4}(t)+2{{L}^{2}}r_{2}^{2}(t)-4{{L}^{2}}{{r}_{2}}(t)+2{{L}^{2}}-7{{L}^{}}{{r}_{2}}(2t)+4L{{r}_{2}}(3t)+4L{{r}_{2}}(t)-L\,,
\end{align}
where the $2{{L}^{2}}r_{2}^{2}$ term gives a quadratic rise at late times, akin to the ramp in $\mathcal{R}_2$.
\subsubsection*{$\mathcal{R}_{4,1}$ at infinite temperature}
We find that
\begin{align}
{\mathcal{R}_{4,1}}(t)=&\,\,{{L}^{3}}{{ {r}}_{1}}(2t) {r}_{1}^{2}(t) -{{L}^{2}}{{ {r}}_{1}}(2t){{ {r}}_{2}}(t){{ {r}}_{3}}(2t)-2{{L}^{2}}{{ {r}}_{1}}(t){{ {r}}_{2}}(2t){{ {r}}_{3}}(t)\nn
&+{{L}^{2}} {r}_{1}^{2}(2t)+2{{L}^{2}} {r}_{1}^{2}(t)+2L{{ {r}}_{2}}(3t)-L{{ {r}}_{2}}(2t)-2L{{ {r}}_{2}}(t)+L\,.
\end{align}
Just as above, we can approximate $\mathcal{R}_{4,1}$ at early and late times by
\begin{align}\label{main2}
\mathcal{R}_{4,1}\approx {{L}^{3}}{{r}_{1}}(2t)r_{1}^{2}(t)+2L{{r}_{2}}(3t)-L{{r}_{2}}(2t)-2L{{r}_{2}}(t)+L\,.
\end{align}

\subsection{Expressions for higher frame potentials}
\label{app:FPs}
\subsubsection*{$k=2$ frame potential}
We computed the second frame potential for the GUE to be
\begin{align*}
\CF_{\rm GUE}^{(2)} = & \bigg( \left(L^4-8 L^2+6\right) \CR_4^2 + 4 L^2 \left(L^2-9\right) \CR_4+ 4 \left(L^6-9 L^4+4 L^2+24\right) \CR_2^2\\
&-8 L^2 \left(L^4-11 L^2+18\right) \CR_2 + 2 \left(L^4-7 L^2+12\right) \CR_{4,1}^2 -4 L^2 \left(L^2-9\right) \CR_{4,2}  \\
&+\left(L^4-8 L^2+6\right) \CR_{4,2}^2-8 \left(L^4-8 L^2+6\right) \CR_2 \CR_4 -4 L \left(L^2-4\right) \CR_4 \CR_{4,1}\\
&+16 L \left(L^2-4\right) \CR_2 \CR_{4,1} -8 \left(L^2+6\right) \CR_2 \CR_{4,2}+2 \left(L^2+6\right) \CR_4 \CR_{4,2}\\
&-4 L \left(L^2-4\right) \CR_{4,1} \CR_{4,2}+2 L^4 \left(L^4-12 L^2+27\right) \bigg)\\
&\Big/\Big((L - 3) (L - 2) (L - 1) L^2 (L + 1) (L + 2) (L + 3)\Big)\,.
\end{align*}
with form factors as defined in Eq.~\eqref{eq:FFs}. Let us try and extract the interesting behavior encoded in the expression.
We know the maximal value of the spectral $n$-point functions defined above at early times, $\CR_2\sim L^2$, $\CR_4 \sim L^4$, $\CR_{4,1}\sim L^3$, and $\CR_{4,2}\sim L^2$. From the expression for the frame potential above, we keep the terms that are not suppressed in $1/L$, \ie can contribute at least at zeroth order:
\begin{align*}
\CF_{\rm GUE}^{(2)} \sim 2 &-\frac{8 \CR_2}{L^2} -\frac{36 \CR_2^2}{L^4} + \frac{4 \CR_2^2}{L^2} + \frac{4 \CR_4}{L^4}+\frac{6 \CR_4^2}{L^8} -\frac{8 \CR_4^2}{L^6}+\frac{\CR_4^2}{L^4} +\frac{\CR_{4,2}^2}{L^4}  -\frac{14 \CR_{4,1}^2}{L^6}\\
&+\frac{2 \CR_{4,1}^2}{L^4}+\frac{16 \CR_2 \CR_{4,1}}{L^5}+\frac{16 \CR_4 \CR_{4,1}}{L^7} -\frac{4 \CR_4 \CR_{4,1}}{L^5}+\frac{2 \CR_4 \CR_{4,2}}{L^6}-\frac{4 \CR_{4,1} \CR_{4,2}}{L^5}\\
&+\frac{64 \CR_2 \CR_4}{L^6}-\frac{8 \CR_2 \CR_4}{L^4}\,,
\end{align*}
with the Haar value appearing at the beginning. At early times, the leading order behavior is $\CF_{\rm GUE}^{(2)} \sim \CR_4^2/L^4$. From our calculation of the $n$-point form factors, we know that at the dip time all form factor terms above are suppressed in $L$, meaning the frame potential goes like the Haar value.
Knowing the late time value of the $2$-point and $4$-point form factors, the terms above that will contribute at late times are
\begin{equation}
{\rm Late:}\quad \CF_{\rm GUE}^{(2)} \approx 2 + \frac{\CR_4^2}{L^4} + \frac{4 \CR_2^2}{L^2}\,,
\end{equation}
which gives $\approx 10$ in the large $L$ limit. In the strict $t \ra \infty $ limit, where $\CR_2\ra L$, $\CR_4 \ra 2L^2-L$, and $\CR_{4,1} ,\CR_{4,2}\ra L$, we have
\begin{equation}
\CF_{\rm GUE}^{(2)} = \frac{10 L^2+22 L-20}{L^2+5 L+6} \and \CF_{\rm GUE}^{(2)} \approx 10 {~\rm for~} L\gg 1\,.
\end{equation}
As the left-hand side expression is valid for any $L$ at late times, in doing the numerics and taking the sample size to be large, this is the value for $L$ we should converge to.

\subsubsection*{$k=3$ frame potential}
The full expression for the third frame potential of the GUE is

\vspace*{8pt}
\noindent $\CF_{\rm GUE}^{(3)}= $\vspace*{-21pt}
{\fontsize{4}{4}
$$\renewcommand*{\arraystretch}{.4}
\begin{array}{l}
\Big(6 L^{14}+18 \CR_2^2 L^{12}-36 \CR_2 L^{12}-318 L^{12}-846 \CR_2^2 L^{10}+9 \CR_4^2 L^{10}+18 \CR_{4,1}^2 L^{10}+9 \CR_{4,2}^2 L^{10}+1836 \CR_2 L^{10}-72 \CR_2 \CR_4 L^{10}+36 \CR_4 L^{10}-36 \CR_{4,2} L^{10}+5550 L^{10}\\
+144 \CR_2 \CR_{4,1} L^9-36 \CR_4 \CR_{4,1} L^9-36 \CR_{4,1} \CR_{4,2} L^9+11574 \CR_2^2 L^8-369 \CR_4^2 L^8+\CR_6^2 L^8-828 \CR_{4,1}^2 L^8+9 \CR_2^2 \CR_{4,2}^2 L^8-18 \CR_2 \CR_{4,2}^2 L^8 -441 \CR_{4,2}^2 L^8+6 \CR_{6,1}^2 L^8\\
+4 \CR_{6,2}^2 L^8+12 \CR_{6,3}^2 L^8+4 \CR_{6,4}^2 L^8-29772 \CR_2 L^8+3276 \CR_2 \CR_4 L^8-1728 \CR_4 L^8+36 \CR_2 \CR_6 L^8-18 \CR_4 \CR_6 L^8 -12 \CR_6 L^8-36 \CR_2^2 \CR_{4,2} L^8+18 \CR_4 \CR_{4,2} L^8
\\+1800 \CR_{4,2} L^8-36 \CR_{4,1} \CR_{6,1} L^8-24 \CR_{6,4} L^8-37158 L^8-6192 \CR_2 \CR_{4,1} L^7+1332 \CR_4 \CR_{4,1} L^7+36 \CR_6 \CR_{4,1} L^7 +108 \CR_2 \CR_{4,1} \CR_{4,2} L^7+1548 \CR_{4,1} \CR_{4,2} L^7\\
-144 \CR_2 \CR_{6,1} L^7+108 \CR_4 \CR_{6,1} L^7-12 \CR_6 \CR_{6,1} L^7-36 \CR_2 \CR_{4,2} \CR_{6,1} L^7+36 \CR_{4,2} \CR_{6,1} L^7 +72 \CR_{4,1} \CR_{6,2} L^7-24 \CR_{6,1} \CR_{6,2} L^7+144 \CR_2 \CR_{6,3} L^7-72 \CR_2 \CR_{4,2} \CR_{6,3} L^7\\+72 \CR_{4,2} \CR_{6,3} L^7-24 \CR_{6,2} \CR_{6,3} L^7-48 \CR_{6,3} \CR_{6,4} L^7-39978 \CR_2^2 L^6+3726 \CR_4^2 L^6-41 \CR_6^2 L^6+11610 \CR_{4,1}^2 L^6-297 \CR_2^2 \CR_{4,2}^2 L^6+594 \CR_2 \CR_{4,2}^2 L^6+6750 \CR_{4,2}^2 L^6\\
-204 \CR_{6,1}^2 L^6-156 \CR_{6,2}^2 L^6-348 \CR_{6,3}^2 L^6-148 \CR_{6,4}^2 L^6+169812 \CR_2 L^6-42768 \CR_2 \CR_4 L^6+24732 \CR_4 L^6-1512 \CR_2 \CR_6 L^6+738 \CR_4 \CR_6 L^6+528 \CR_6 L^6\\
+1512 \CR_2^2 \CR_{4,2} L^6-432 \CR_2 \CR_{4,2} L^6-162 \CR_2 \CR_4 \CR_{4,2} L^6-486 \CR_4 \CR_{4,2} L^6+18 \CR_2 \CR_6 \CR_{4,2} L^6-18 \CR_6 \CR_{4,2} L^6-27972 \CR_{4,2} L^6+1224 \CR_{4,1} \CR_{6,1} L^6+144 \CR_2 \CR_{6,2} L^6\\
-144 \CR_4 \CR_{6,2} L^6+16 \CR_6 \CR_{6,2} L^6+72 \CR_2 \CR_{4,2} \CR_{6,2} L^6-72 \CR_{4,2} \CR_{6,2} L^6-48 \CR_{6,2} L^6-360 \CR_{4,1} \CR_{6,3} L^6+120 \CR_{6,1} \CR_{6,3} L^6-144 \CR_2 \CR_{6,4} L^6+72 \CR_2 \CR_{4,2} \CR_{6,4} L^6\\
-72 \CR_{4,2} \CR_{6,4} L^6+32 \CR_{6,2} \CR_{6,4} L^6+1032 \CR_{6,4} L^6+89040 L^6+72576 \CR_2 \CR_{4,1} L^5-11232 \CR_4 \CR_{4,1} L^5-1188 \CR_6 \CR_{4,1} L^5-3132 \CR_2 \CR_{4,1} \CR_{4,2} L^5-18792 \CR_{4,1} \CR_{4,2} L^5\\
+5040 \CR_2 \CR_{6,1} L^5-3564 \CR_4 \CR_{6,1} L^5+396 \CR_6 \CR_{6,1} L^5+1044 \CR_2 \CR_{4,2} \CR_{6,1} L^5-1044 \CR_{4,2} \CR_{6,1} L^5-2232 \CR_{4,1} \CR_{6,2} L^5+744 \CR_{6,1} \CR_{6,2} L^5-5040 \CR_2 \CR_{6,3} L^5\\
+432 \CR_4 \CR_{6,3} L^5-48 \CR_6 \CR_{6,3} L^5+2088 \CR_2 \CR_{4,2} \CR_{6,3} L^5-2088 \CR_{4,2} \CR_{6,3} L^5+648 \CR_{6,2} \CR_{6,3} L^5+288 \CR_{4,1} \CR_{6,4} L^5-96 \CR_{6,1} \CR_{6,4} L^5+1488\CR_{6,3} \CR_{6,4} L^5-522 \CR_4^2 L^4\\
-52128 \CR_2^2 L^4+458 \CR_6^2 L^4-55692 \CR_{4,1}^2 L^4 +2430 \CR_2^2 \CR_{4,2}^2 L^4-4860 \CR_2 \CR_{4,2}^2 L^4-35190 \CR_{4,2}^2 L^4+1794 \CR_{6,1}^2 L^4+1660 \CR_{6,2}^2 L^4+2388 \CR_{6,3}^2 L^4+1440 \CR_{6,4}^2 L^4\\
-274320 \CR_2 L^4+146412 \CR_2 \CR_4 L^4 +17172 \CR_2 \CR_6 L^4-8244 \CR_4 \CR_6 L^4-6276 \CR_6 L^4-15876 \CR_2^2 \CR_{4,2} L^4+18144 \CR_2 \CR_{4,2} L^4+3078 \CR_2 \CR_4 \CR_{4,2} L^4+324 \CR_4 \CR_{4,2} L^4\\
-342 \CR_2 \CR_6 \CR_{4,2} L^4 +342 \CR_6 \CR_{4,2} L^4+141408 \CR_{4,2} L^4-10764 \CR_{4,1} \CR_{6,1} L^4-4608 \CR_2 \CR_{6,2} L^4+3672 \CR_4 \CR_{6,2} L^4-408 \CR_6 \CR_{6,2} L^4-1368 \CR_2 \CR_{4,2} \CR_{6,2} L^4\\
+1368 \CR_{4,2} \CR_{6,2} L^4+1968 \CR_{6,2} L^4+7200 \CR_{4,1} \CR_{6,3} L^4-2400 \CR_{6,1} \CR_{6,3} L^4+3312 \CR_2 \CR_{6,4} L^4-288 \CR_4 \CR_{6,4} L^4+32 \CR_6 \CR_{6,4} L^4-1368 \CR_2 \CR_{4,2} \CR_{6,4} L^4\\
+1368 \CR_{4,2} \CR_{6,4} L^4-752 \CR_{6,2} \CR_{6,4} L^4-11568 \CR_{6,4} L^4-96000 L^4-199728 \CR_2 \CR_{4,1} L^3-4392 \CR_4 \CR_{4,1} L^3+9144 \CR_6 \CR_{4,1} L^3+26352 \CR_2 \CR_{4,1} \CR_{4,2} L^3\\
+51552 \CR_{4,1} \CR_{4,2} L^3-37296 \CR_2 \CR_{6,1} L^3+27432 \CR_4 \CR_{6,1} L^3-3048 \CR_6 \CR_{6,1} L^3-8784 \CR_2 \CR_{4,2} \CR_{6,1} L^3+8784 \CR_{4,2} \CR_{6,1} L^3+17928 \CR_{4,1} \CR_{6,2} L^3-5976 \CR_{6,1} \CR_{6,2} L^3\\
+37296 \CR_2 \CR_{6,3} L^3-1080 \CR_4 \CR_{6,3} L^3+120 \CR_6 \CR_{6,3} L^3-17568 \CR_2 \CR_{4,2} \CR_{6,3} L^3+17568 \CR_{4,2} \CR_{6,3} L^3-190512 \CR_2 \CR_{4,2} L^2-100800 \CR_4 L^4-5736 \CR_{6,2} \CR_{6,3} L^3\\
-720 \CR_{4,1} \CR_{6,4} L^3+240 \CR_{6,1} \CR_{6,4} L^3-11952 \CR_{6,3} \CR_{6,4} L^3+141840 \CR_2^2 L^2-49284 \CR_4^2 L^2-1258 \CR_6^2 L^2+111852 \CR_{4,1}^2 L^2+1098 \CR_2^2 \CR_{4,2}^2 L^2-2196 \CR_2 \CR_{4,2}^2 L^2\\
+53712 \CR_{4,2}^2 L^2-3756 \CR_{6,1}^2 L^2-3188 \CR_{6,2}^2 L^2+108 \CR_{6,3}^2 L^2-2736 \CR_{6,4}^2 L^2+288000 \CR_2 L^2+5472 \CR_2 \CR_4 L^2-47376 \CR_2 \CR_6 L^2+22644 \CR_4 \CR_6 L^2+14400 \CR_6 L^2\\
+14400 \CR_2^2 \CR_{4,2} L^2-9396 \CR_2 \CR_4 \CR_{4,2} L^2+49824 \CR_4 \CR_{4,2} L^2+1044 \CR_2 \CR_6 \CR_{4,2} L^2-1044 \CR_6 \CR_{4,2} L^2-115200 \CR_{4,2} L^2+22536 \CR_{4,1} \CR_{6,1} L^2+24624 \CR_2 \CR_{6,2} L^2\\
 -16488 \CR_4 \CR_{6,2} L^2+1832 \CR_6 \CR_{6,2} L^2+4176 \CR_2 \CR_{4,2} \CR_{6,2} L^2-4176 \CR_{4,2} \CR_{6,2} L^2-19200 \CR_{6,2} L^2-45720 \CR_{4,1} \CR_{6,3} L^2+15240 \CR_{6,1} \CR_{6,3} L^2+8352 \CR_2 \CR_{6,4} L^2\\
 -8352 \CR_4 \CR_{6,4} L^2+928 \CR_6 \CR_{6,4} L^2+4176 \CR_2 \CR_{4,2} \CR_{6,4} L^2-4176 \CR_{4,2} \CR_{6,4} L^2+5520 \CR_{6,2} \CR_{6,4} L^2+19200 \CR_{6,4} L^2 +133200 \CR_2 \CR_{4,1} L+53208 \CR_4 \CR_{4,1} L\\
-12312 \CR_6 \CR_{4,1} L-62208 \CR_2 \CR_{4,1} \CR_{4,2} L+4608 \CR_{4,1} \CR_{4,2} L+32400 \CR_2 \CR_{6,1} L-36936 \CR_4 \CR_{6,1} L+4104 \CR_6 \CR_{6,1} L+20736 \CR_2 \CR_{4,2} \CR_{6,1} L-20736 \CR_{4,2} \CR_{6,1} L\\
-33048 \CR_{4,1} \CR_{6,2} L+11016 \CR_{6,1} \CR_{6,2} L-32400 \CR_2 \CR_{6,3} L-25272 \CR_4 \CR_{6,3} L+2808 \CR_6 \CR_{6,3} L+41472 \CR_2 \CR_{4,2} \CR_{6,3} L-41472 \CR_{4,2} \CR_{6,3} L+16632 \CR_{6,2} \CR_{6,3} L\\
-16848 \CR_{4,1} \CR_{6,4} L+5616 \CR_{6,1} \CR_{6,4} L+22032 \CR_{6,3} \CR_{6,4} L-216000 \CR_2^2-2160 \CR_4^2+240 \CR_6^2-105840 \CR_{4,1}^2-12960 \CR_2^2 \CR_{4,2}^2+25920 \CR_2 \CR_{4,2}^2-34560 \CR_{4,2}^2\\
-2160 \CR_{6,1}^2-2160 \CR_{6,2}^2-19440 \CR_{6,3}^2-960 \CR_{6,4}^2+43200 \CR_2 \CR_4 +14400 \CR_2 \CR_6-4320 \CR_4 \CR_6+172800 \CR_2 \CR_{4,2}+25920 \CR_2 \CR_4 \CR_{4,2}-69120 \CR_4 \CR_{4,2}\\
-2880 \CR_2 \CR_6\CR_{4,2}+2880 \CR_6 \CR_{4,2}+12960 \CR_{4,1} \CR_{6,1}+14400 \CR_2 \CR_{6,2}+4320 \CR_4 \CR_{6,2}-480 \CR_6 \CR_{6,2}-11520 \CR_2 \CR_{4,2} \CR_{6,2}+11520 \CR_{4,2} \CR_{6,2}+90720 \CR_{4,1} \CR_{6,3}\\
-30240 \CR_{6,1} \CR_{6,3}-28800 \CR_2 \CR_{6,4}-2880 \CR_6 \CR_{6,4}+25920 \CR_4  \CR_{6,4}-11520 \CR_2 \CR_{4,2} \CR_{6,4}+11520 \CR_{4,2} \CR_{6,4}-6720 \CR_{6,2} \CR_{6,4} \Big)\\
\Big/\Big((L-5) (L-4) (L-3) (L-2) (L-1) L^2 (L+1) (L+2) (L+3) (L+4) (L+5)\Big)\,.
\end{array}
$$
}
The expression is best appreciated from a distance.

\subsection{Expressions for Weingarten}
\label{app:wein}
Lastly, we give the definition of the unitary Weingarten function, which appeared in the integration of Haar random unitaries in Eq.~\eqref{eq:Haarint}. The $2k$-th moment of the Haar ensemble appeared in the $k$-th frame potential. For the $n$-th moment, the Weingarten function is a function of an element $\sigma$ of the permutation group $S_n$ and presented as defined in \cite{Collins04},
\begin{equation}
\Wg(\sigma) = \frac{1}{(n!)^2} \sum_\lambda \frac{\chi_\lambda(e)^2 \chi_\lambda(\sigma)}{s_\lambda (1)}\,,
\end{equation}
where we sum over integer partitions of $n$ (recall that the conjugacy classes of $S_n$ are labeled by integer partitions of $n$). $\chi_\lambda$ is an irreducible character of $S_n$ labeled by $\lambda$ (as each irrep of $S_n$ can be associated to an integer partition) and $e$ is the identity element. $s_\lambda (1)=s_\lambda (1,\ldots,1)$ is the Schur polynomial evaluated on $L$ arguments and indexed by the partition $\lambda$. For instance, the Weingarten functions needed to compute the first frame potential were
\begin{equation}
\Wg(\{1,1\}) = \frac{1}{L^2-1} \and \Wg(\{2\}) = -\frac{1}{L(L^2-1)}\,.
\end{equation}

\section{Additional numerics}
\label{app:num}
We conclude with a few numerical checks on the formulae we derived for the form factors and frame potentials.

\subsection{Form factors and numerics}
As we mentioned in Sec.~\ref{sec:FFs} and discussed in App.~\ref{app:FF}, in order to derive expressions for the form factors for the GUE we had to make approximations which should be compared to numerics for the GUE.
\begin{figure}
\centering
\includegraphics[width=0.5\linewidth]{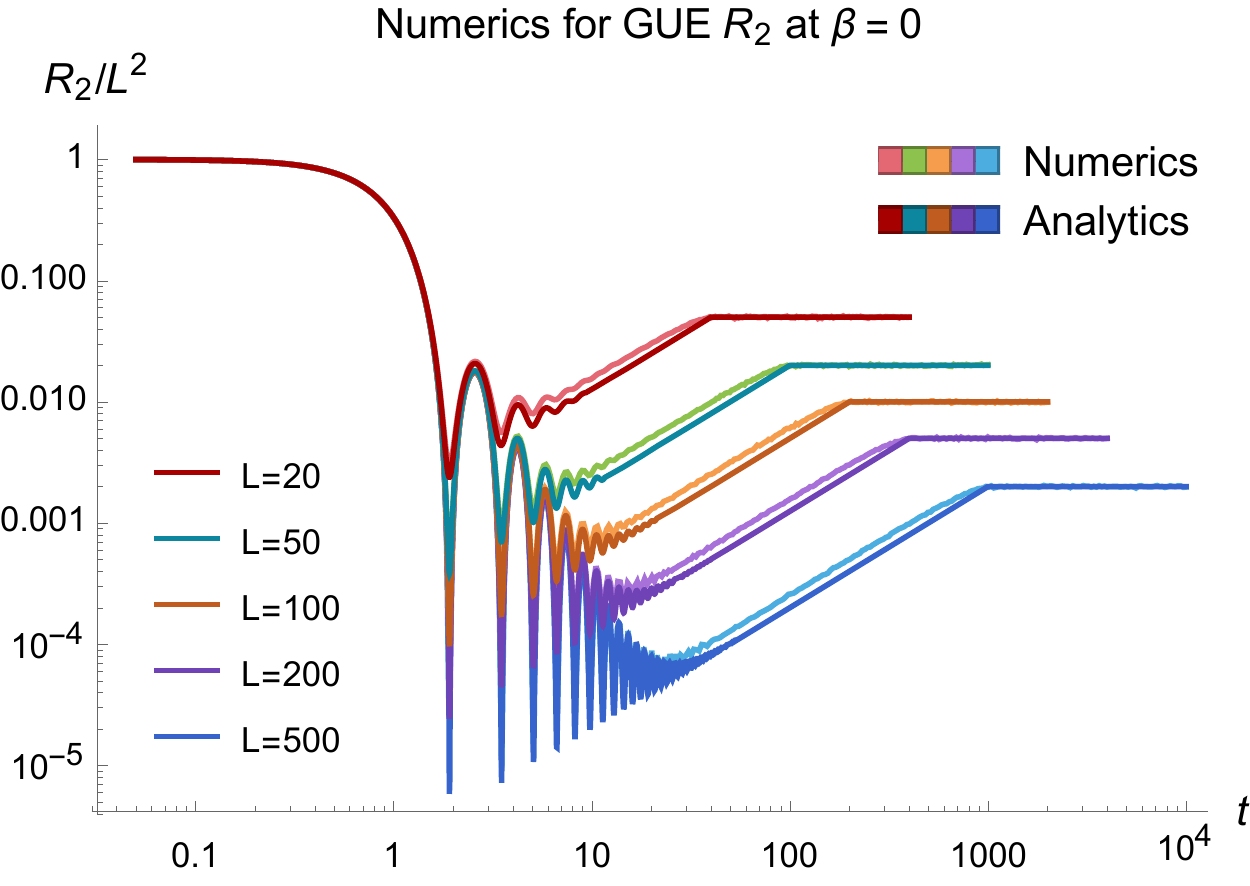}\quad
\includegraphics[width=0.3\linewidth]{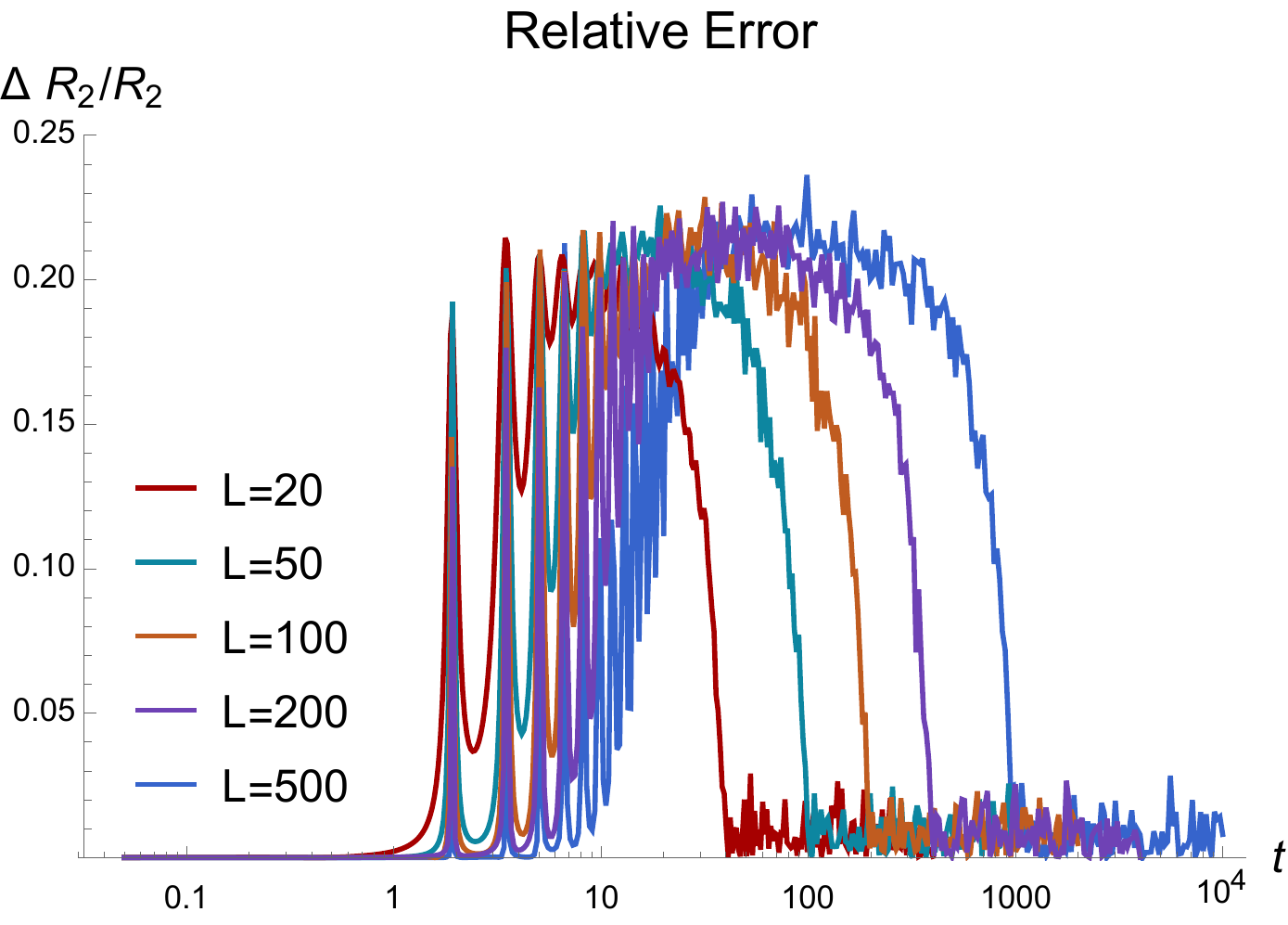}
\caption{Numerical checks of the GUE $2$-point spectral form factor at infinite temperature for various values of $L$ and normalized by $L^2$. The analytic expressions derived in Sec.~\ref{sec:FFRMT} are in the lighter shades and the numerics for GUE are in darker shades. Numerics were done $10000$ samples from the GUE. On the right we plot the relative error between the numerics and analytic predictions. We observe good agreement at early and late times, and see deviations around the ramp.
}
\label{fig:GUEnum}
\end{figure}

We briefly remind the reader that at infinite temperature, we derived the expression
\begin{equation}
\CR_2(t) = L^2 r_1^2(t) - Lr_2 (t) +L\,.
\end{equation}
Numerical checks of this expression are shown in Fig.~\ref{fig:GUEnum}. We see that the approximations employed work well at $\beta=0$, reproducing the early time oscillations, dip, plateau, and ramp features. But there is some discrepancy in the ramp behavior which merits discussion. As we take $L\ra \infty$, the difference between the predicted ramp and numerical ramp is not suppressed. In Fig.~\ref{fig:GUEnum}, we see that the relative error between the numerics and analytic prediction does not decrease as we increase $L$, indicating that this difference in the ramp prediction is not an artifact of finite $L$ numerics. On a log-log plot, this shift from the numerics suggests that we capture the correct linear behavior, but with a slightly different slope for the ramp.

The $r_2(t) = 1 - t/2L$ function which controls the slope behavior comes from the Fourier transform of the square of the sine kernel. Recall that in our approximation, we integrated over the entire semicircle. A phenomenological observation is that the modified ramp function defined by $\tilde r_2(t) \equiv 1- 2 t/\pi L$, where we change the slope to $2/\pi$, does a much better job of capturing the ramp behavior. Working in the short-distance limit of the 2-point correlator $\rho^{(2)}(\lambda_1,\lambda_2)$ (as in \cite{BrezinHikami1}) and integrating the sine kernel over the entire semicircle, we obtain $\tilde r_2$ whose behavior we only trust near the dip. 

\begin{figure}[htb!]
\centering
\includegraphics[width=0.5\linewidth]{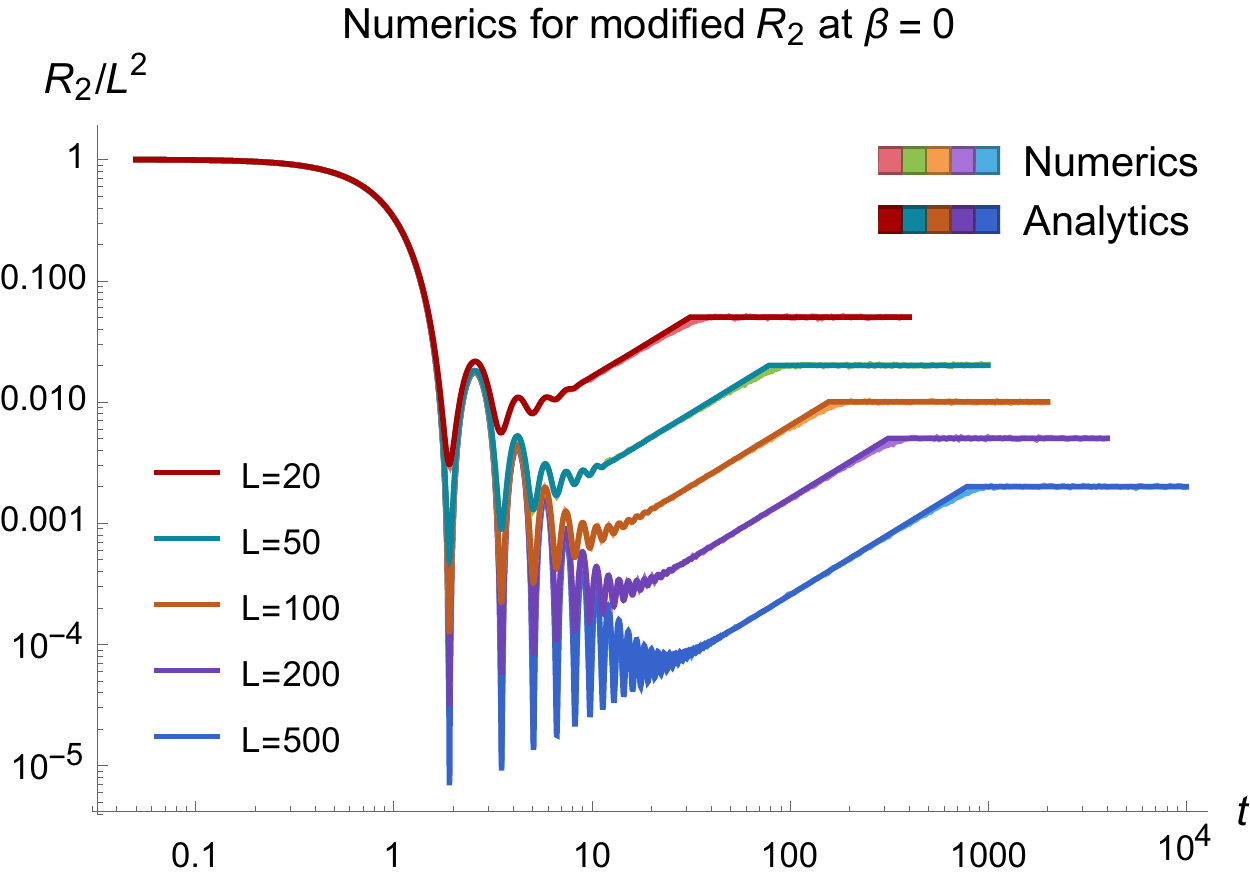}\quad
\includegraphics[width=0.3\linewidth]{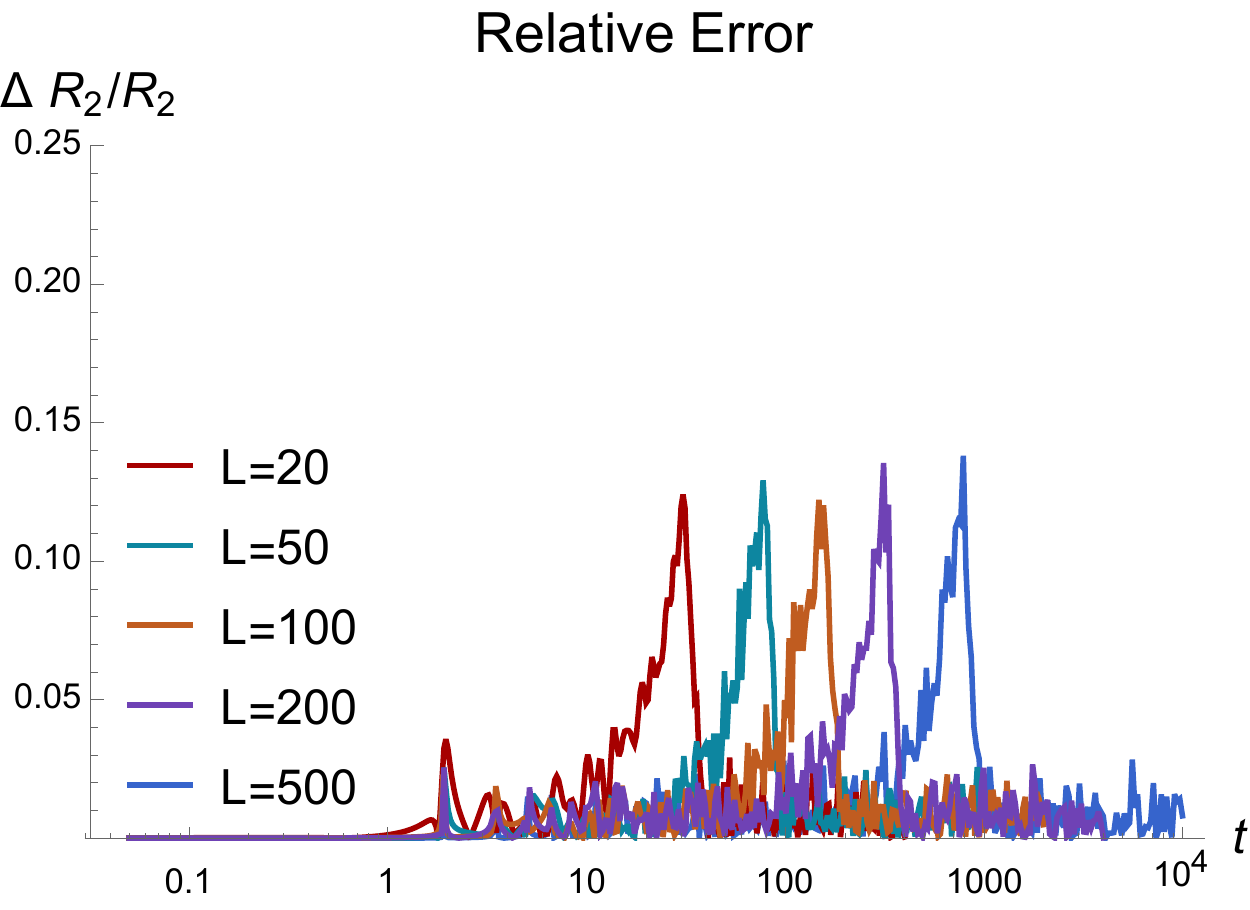}
\caption{The same numerics as reported in Fig.~\ref{fig:GUEnum}, but now compared to the analytic expression with the modified ramp behavior $\tilde r_2(t)$.
}
\label{fig:GUEnummod}
\end{figure}

Numerically, we find that this modified slope of $2\pi/L$ better captures the $r_2$ function near the dip, with error that is suppressed as we take $L\ra \infty$. The same numerics are reported in Fig.~\ref{fig:GUEnummod}, but with the modified ramp behavior. There is still some discrepancy near the plateau time when we transition to the constant plateau value, but the ramp behaviors near the dip are in much better agreement.

\begin{figure}[htb!]
\centering
\includegraphics[width=0.45\linewidth]{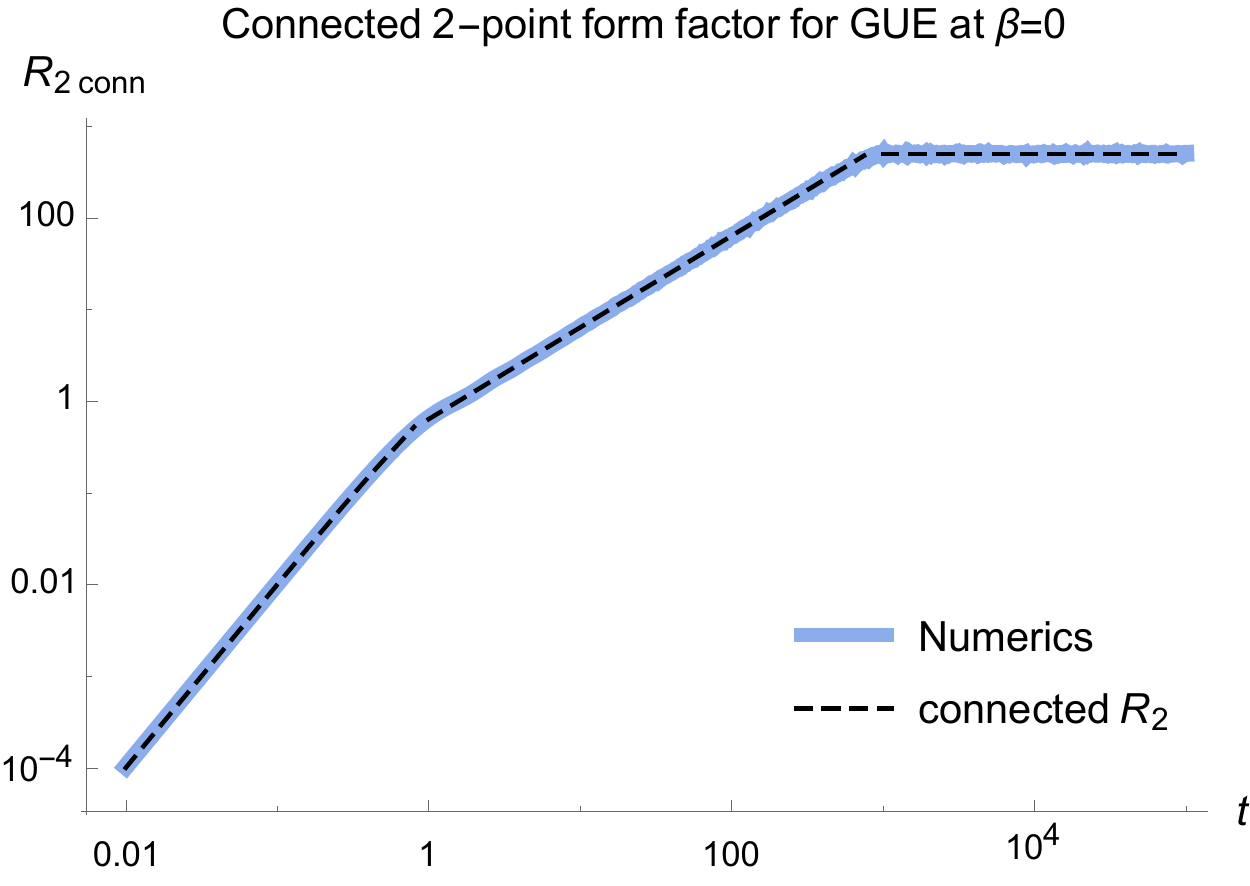}
\caption{Numerics for the connected $2$-point spectral form factor for GUE at infinite temperature plotted for $L=500$ with 10000 random samples. The dashed line is the expression Eq.~\eqref{eq:R2con} approximating the three regimes of the connected form factor.
}
\label{fig:R2con}
\end{figure}

We understand the Bessel function contribution to $\CR_2(t)$, which arises from 1-point functions. The subtlety above is really in the connected piece of the 2-point function
\begin{equation}
\CR_2(t)_{\rm conn} \equiv \CR_2(t) - L^2 r_1^2(t)\,.
\end{equation}
Numerically, we see that the connected 2-point form factor for the GUE exhibits three different behaviors: an early time quadratic growth, an intermediate linear growth, and then a late-time constant plateau. The closed form expression we derived in Sec.~\ref{sec:FFRMT} should be viewed as a coarse approximation before the plateau, approximately capturing the linear regime. The modified ramp function $\tilde r_2(t) = 1- 2 t/\pi L $ appears to capture the linear behavior near the dip with the correct slope. In \cite{BrezinRMT}, a more detailed treatment of the connected correlator is given at early times. From the integral representation of the connected $2$-point form factor, they find that
\begin{equation}
{\rm Early}:~~\CR_2(t)_{\rm conn} \approx t^2 - \frac{1}{2} \, t^4 + \frac{1}{3} \, t^6 +\ldots
\end{equation}
to leading order in $L$ (Eq.(2.28) in \cite{BrezinRMT}). The three behaviors are compared with numerics in Fig.~\ref{fig:R2con}.

\noindent In summary, the three regimes of the connected 2-point form factor are roughly captured by
\begin{equation}
\CR_2(t)_{\rm conn} = \begin{cases}\sim t^2 &\for t\lesssim 1\,,\\ \sim \frac{2}{\pi}t &\for 1\lesssim t \lesssim 2L\,,\\ L &\for t \gtrsim 2L\,.  \end{cases}
\label{eq:R2con}
\end{equation}
The early time quadratic behavior does not play an important role in our analysis of GUE correlation functions and frame potentials, but is of independent physical interest. This intruiging early-time behavior of the connected 2-point form factor will be explored in \cite{ShenkerTBA}.

\begin{figure}[htb!]
\centering
\includegraphics[width=0.4\linewidth]{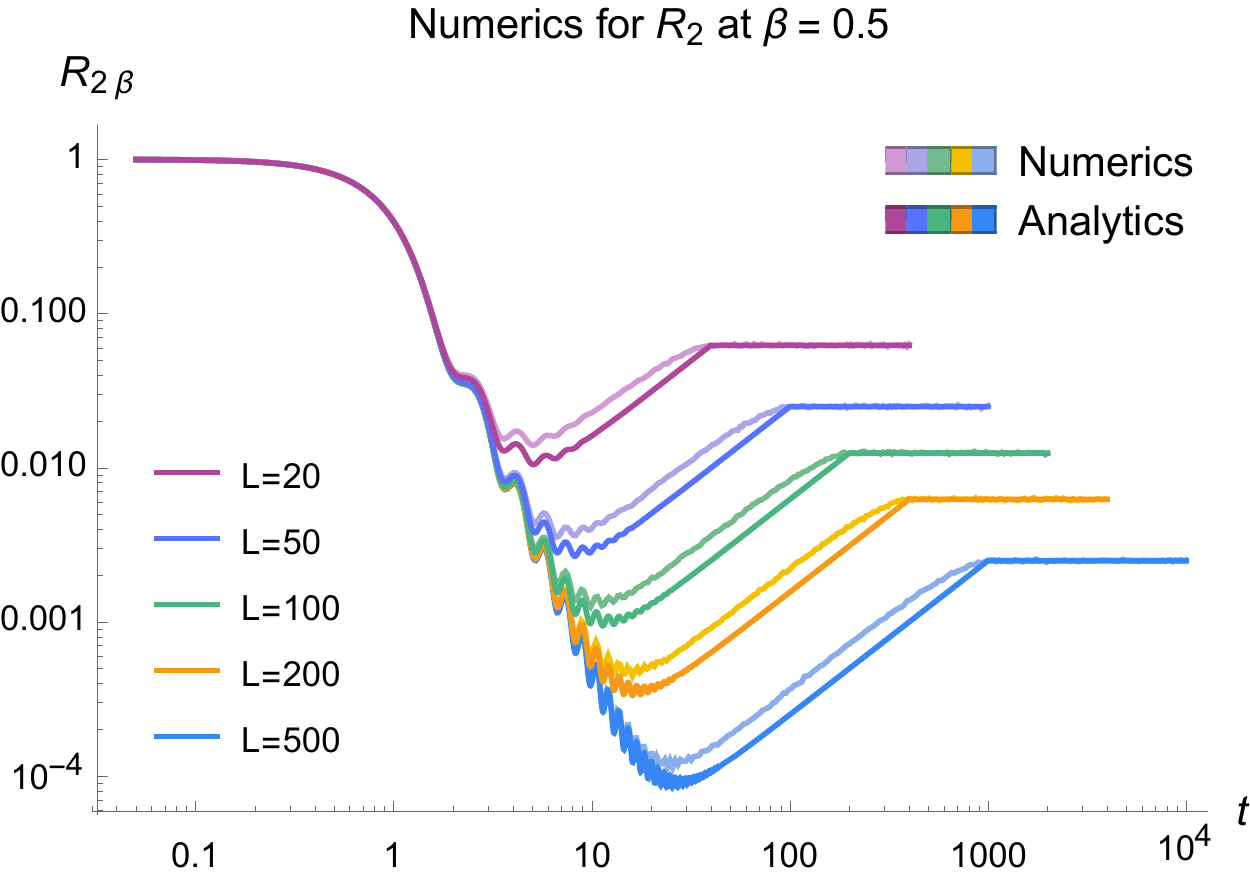}
\includegraphics[width=0.4\linewidth]{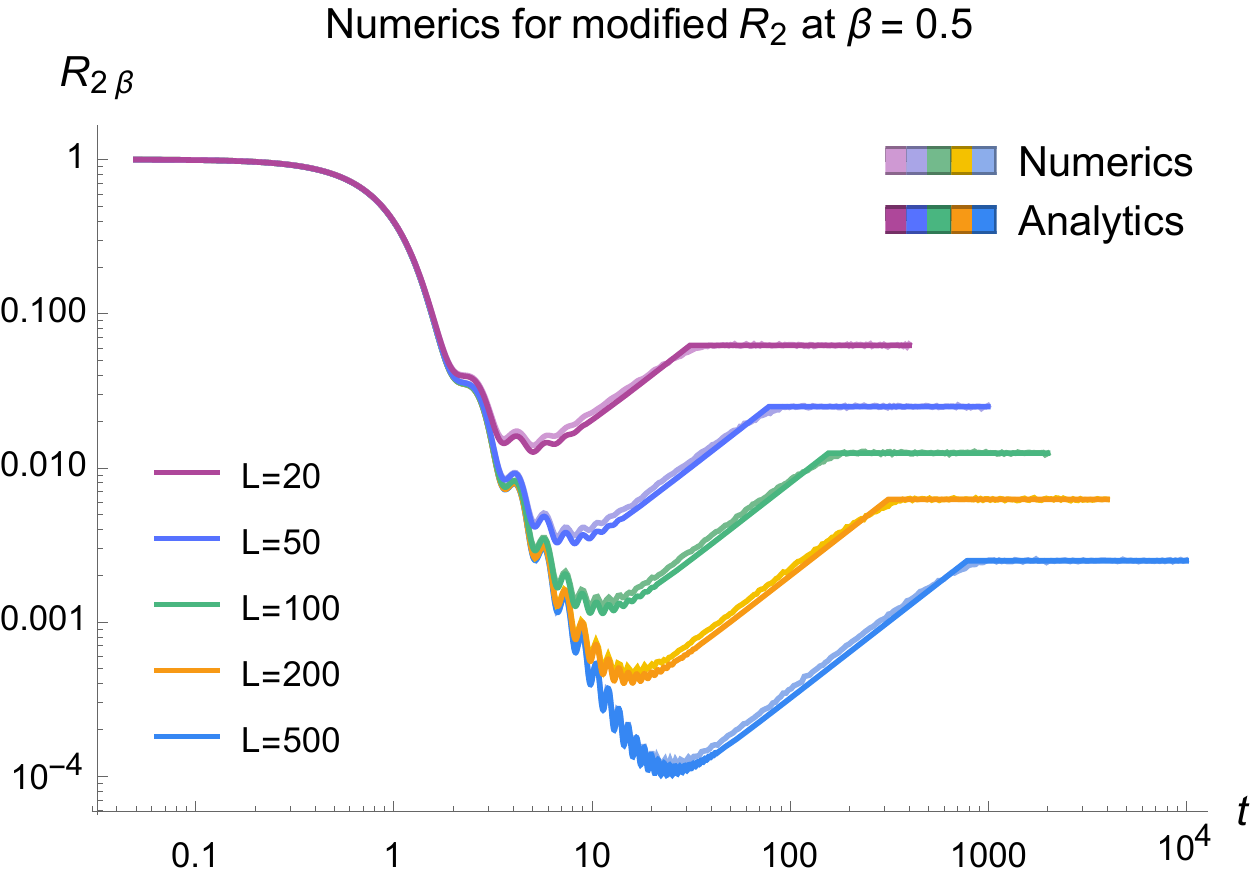}
\caption{Numerical checks of the finite temperature $2$-point spectral form factor for GUE at $\beta=0.5$, plotted for various values of $L$ and normalized by their initial values. Numerics were done with a GUE sample size of $10000$. The left figure uses the expression for $\CR_2(t,\beta)$ derived in Sec.~\ref{sec:FFs} and \ref{app:FF}, whereas the right figure uses the modified ramp $\tilde r_2$ discussed above.
}
\label{fig:GUEnumbeta}
\end{figure}

\begin{figure}[htb!]
\centering
\includegraphics[width=0.4\linewidth]{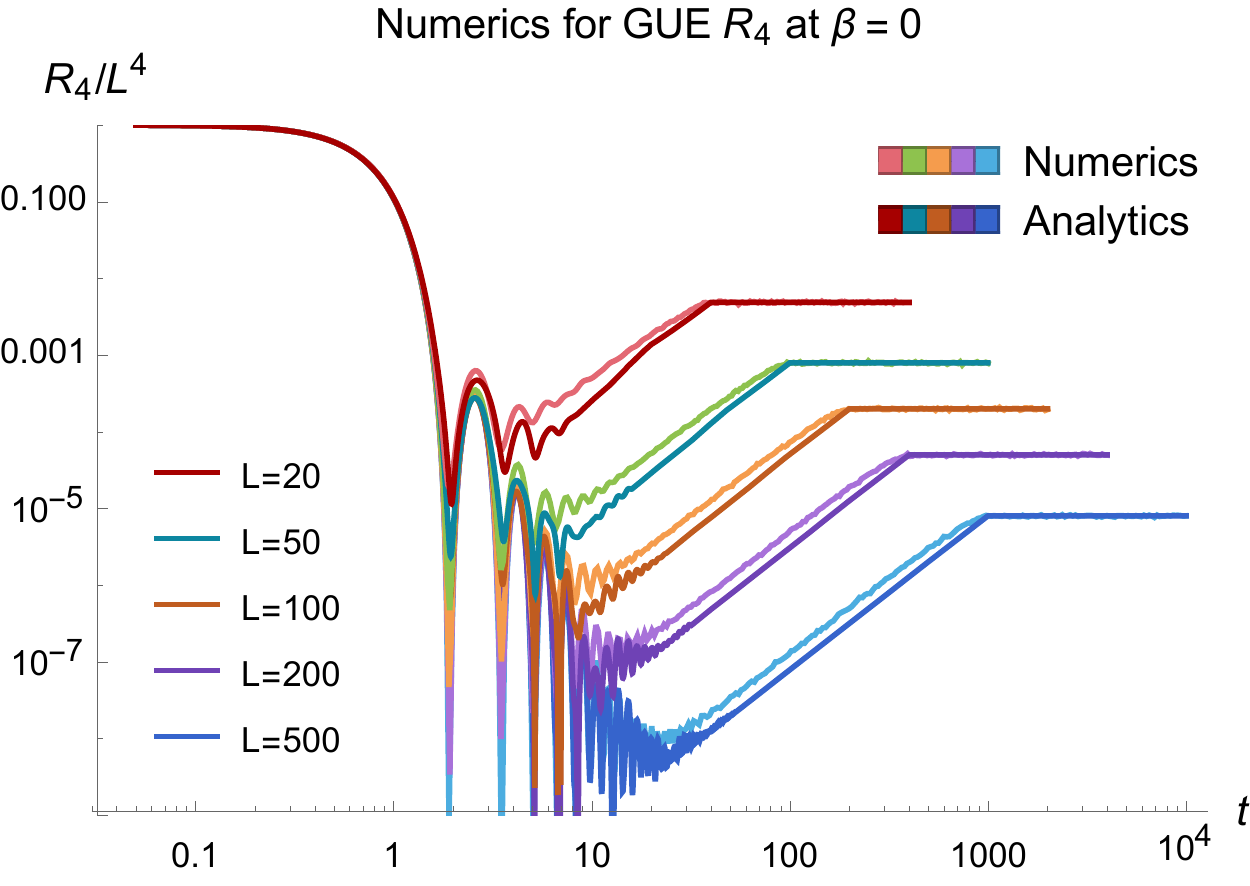}
\includegraphics[width=0.4\linewidth]{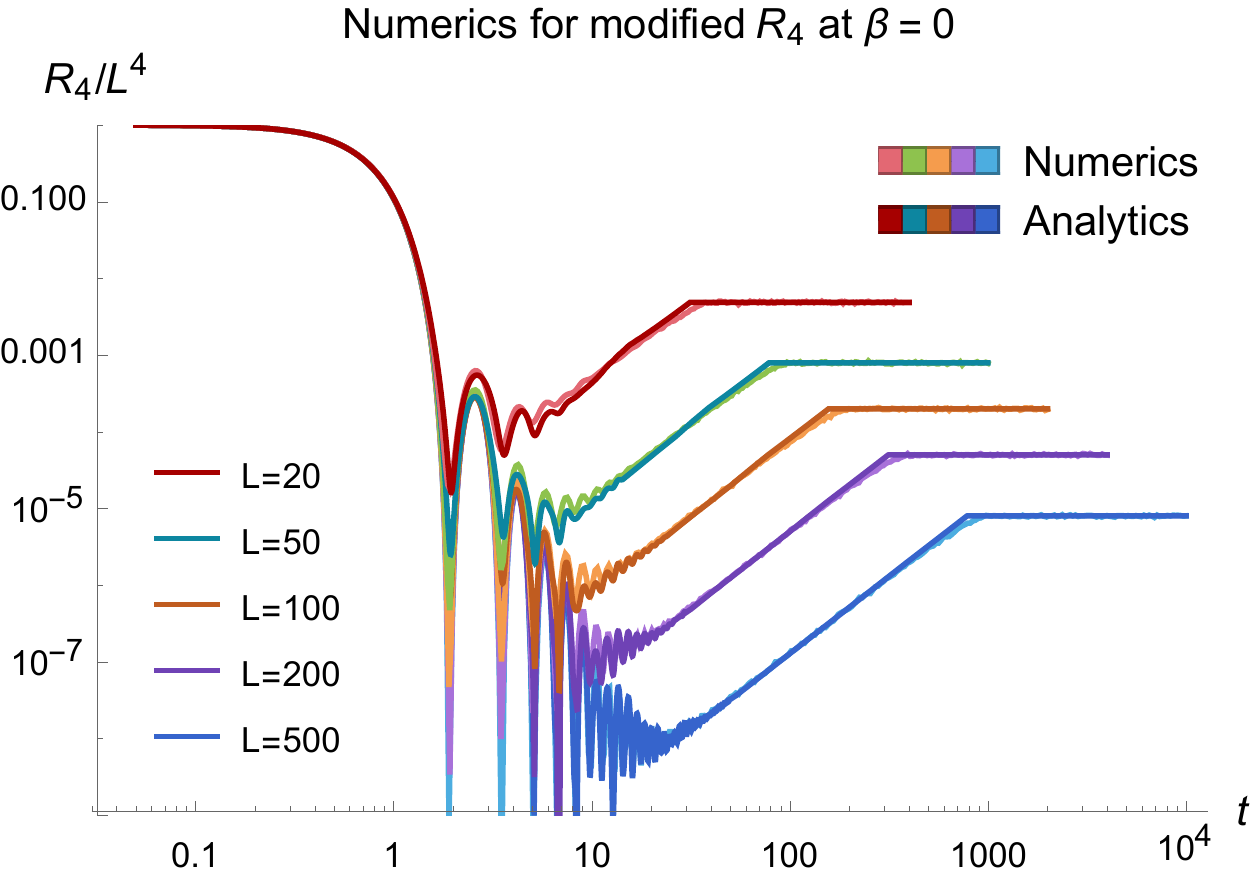}
\caption{Numerical checks the infinite temperature 4-point spectral form factor for GUE with 10000 samples, plotted for various values of $L$ and normalized by their initial values. The left figure uses the $\CR_4$ expression derived in App.~\ref{app:FF}, and the right figure uses $\tilde r_2$.
}
\label{fig:GUEnumR4}
\end{figure}

At finite temperature we find good agreement between the expression $\CR_2(t,\beta)$ and numerics at early and late times, but again see a deviation of the dip and ramp behaviors from the analytic prediction, as shown in Fig.~\ref{fig:GUEnumbeta}. Using the modified ramp $\tilde r_2$ we find closer agreement at small $\beta$, but as we increase $\beta$ the predicted ramp behavior again starts to deviate from the numerics, indicating that there is a $\beta$-dependence to the slope that we do not fully understand. But as we discussed in App.~\ref{app:FF}, we only trust the short-distance approximation at finite temperature, and thus $\CR_2(t,\beta)$, for small $\beta$. We also report numerics for the $\CR_4$ expression in Fig.~\ref{fig:GUEnumR4}.

\subsection{Frame potentials and numerics}
\begin{figure}[htb!]
\centering
\includegraphics[width=0.55\linewidth]{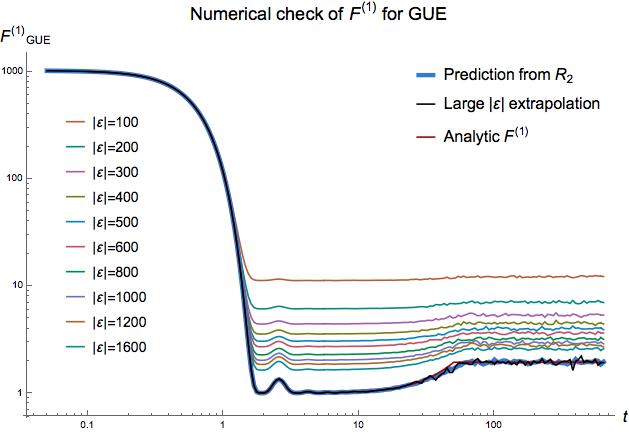}
\includegraphics[width=0.40\linewidth]{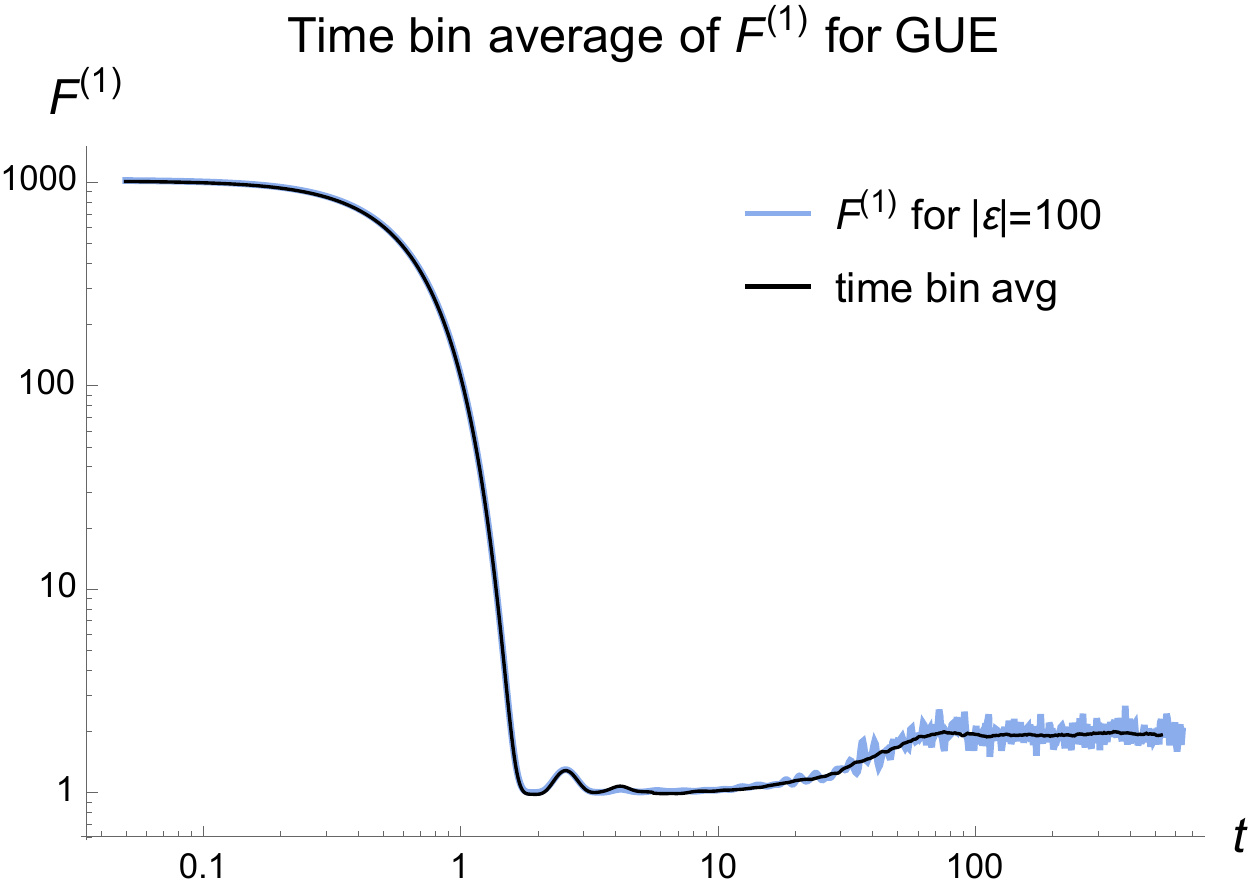}
\caption{Numerical computation of first frame potential for the GUE at $L=32$. On the left, we sequentially increase the number of samples and extrapolate to large sample size (red line), which agrees with the both the frame potential computed from $\CR_2$ numerics as in Eq.~\eqref{eq:GUE_FP1} (blue line) and the analytic expression we derived for $\CF^{(1)}_{\rm GUE}$. On the right, we time bin average $\CF^{(1)}_{\rm GUE}$ as described above and, for $L=32$ and 100 samples, we find good agreement with the quantities on the left.
}
\label{fig:FP1}
\end{figure}

As the frame potential depends on the eigenvectors of the elements in the ensemble (and not just the eigenvalues as per the form factors) and requires a double sum over the ensemble, numerical simulation of the frame potential is harder than for the form factors. For an ensemble of $L\times L$ matrices, we need to consider sample sizes greater than $L^{2k}$ for the $k$-th frame potential, which amounts to summing over many samples for fairly modest Hilbert space dimension. Instead, for a given $L$, we can sequentially increase the sample size and extrapolate to large $|\CE_{\rm GUE}|$. In Fig.~\ref{fig:FP1} we consider the first frame potential for the GUE at $L=32$ and, in the limit of large sample size, find good agreement with the analytic expression computed from $\CR_2$. Alternatively, we can numerically compute the frame potentials by ignoring the coincident contributions to the double sum in $\CF^{(k)}$, \ie when $U=V$. For a finite number of samples, these terms contribute $L^2/|\CE|$ to the sum, meaning we must look at large ensembles before their contribution does not dominate entirely. Ignoring these terms, we can time average over a sliding window to compute the frame potential with only a few samples, as shown in Fig.~\ref{fig:FP1}.

\subsection{Minimal realizations and time averaging}
\label{app:timebin}

\begin{figure}[htb!]
\centering
\includegraphics[width=0.43\linewidth]{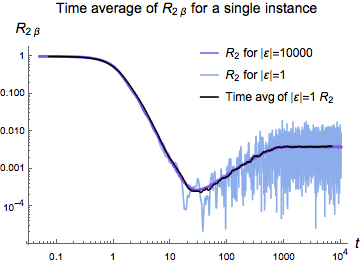}
\includegraphics[width=0.44\linewidth]{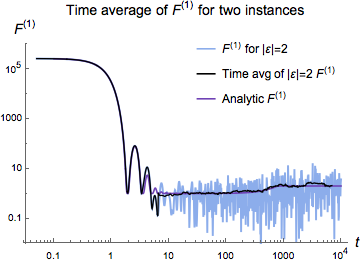}
\caption{On the left: the time average of the thermal 2-point form factor at $\beta=5$ and $L=500$. On the right: the time average of the first frame potential for $L=500$ computed for two instances. In both figures, the time average of the minimal number of instances agrees with the ensemble average.
}
\label{fig:timebin}
\end{figure}

Given an ensemble of disordered systems, one can ask whether a quantity averaged over the ensemble is the same as for a single random instance of the ensemble. It is known that up until the dip time, the spectral form factor is self-averaging, meaning that single instance captures the average for large $L$ \cite{SFFnotavg}.  However, the spectral form factor is not self-averaging at late times.  We can try to extract the averaged behavior from a single instance in regimes dominated by large fluctuations by averaging over a moving time window. In Fig.~\ref{fig:timebin}, we see that for a single instance of the GUE, the time average of the spectral form factor at finite $\beta$ gives the same result as the ensemble average for sufficiently large $L$. For the frame potential, we can consider two instances, the smallest ensemble for which the frame potential makes sense. Ignoring the coincident terms in the sum, we see that the frame potential is also self-averaging at early times and that the time average at late times agrees with the ensemble average and analytic expression.

\newpage
\bibliographystyle{utphys}
\bibliography{chaos_rmt}

\end{document}